\documentclass[a4paper,12pt]{article}
\usepackage{latexsym}%
\usepackage{amsfonts}%
\usepackage{amsthm} 
\usepackage{lscape}
\newcommand{\one}{\mathbf{1}}
\newcommand{\lap}{\bigtriangleup}
\newcommand{\hr}{\mathcal{H}}

\newcommand{\C}{\mathbb C}
\newcommand{\R}{\mathbb R}
\newcommand{\N}{\mathbb N}
\newcommand{\Z}{\mathbb Z}

\newcommand{\rkl}{\rangle}
\newcommand{\lkl}{\langle}
\newcommand{\D}{\mathcal{D}}

\newcommand{\Hel}{\mathcal{H}_{el}}
\newcommand{\F}{\mathcal{F}}
\newcommand{\cO}{\mathcal{O}}
\newcommand{\co}{o}
\newcommand{\cR}{\mathcal{R}}
\newcommand{\cS}{\mathcal{S}}
\newcommand{\hel}{H_{\mathrm{el}}}

\newcommand{\Ran}{\mathrm{Ran}}  
  
\newcommand{\supp}{\mathrm{supp}}             
\newcommand{\gs}{\Phi_{\mathrm{g}}}
\newcommand{\go}{\Phi_{0}}

\newcommand{\slim}{{\mathrm{s}}\hspace{-2pt}-\hspace{-2pt}\lim}
\newcommand{\fou}{\mathbb{F}}

\newcommand{\cM}{\mathcal{M}}

\newcommand{\fG}{\mathbf{G}}

\newcommand{\ol}{\overline}
\newcommand{\feps}{\mbox{\boldmath $\varepsilon$}}

\newtheorem{hypothesis}{Hypothesis}
\newtheorem{theorem}{Theorem}[section]        
\newtheorem{lemma}[theorem]{Lemma}            
\newtheorem{corollary}[theorem]{Corollary}    
  
\newtheorem{remark}[theorem]{Remark}           
 
\theoremstyle{plain}
\begin{document}
\title{Ionisation by quantised electromagnetic \\ fields  :
The photoelectric effect}
\author{ 
Heribert Zenk \\
Mathematisches Institut, Ludwig-Maximilians-Universit\"at\\
Theresienstra{\ss}e 39, 80333 M\"unchen\\
{\em email\,:}\,Heribert.Zenk@mathematik.uni-muenchen.de }
\maketitle
\begin{abstract}
In this paper we explain the photoelectric effect in a variant of the
standard model of non relativistic quantum electrodynamics, which is in
some aspects more closely related to the physical picture, than the
one studied in \cite{BKZ}: Now we can apply our results to an electron with 
more than one bound state and to a larger class of electron-photon
interactions. We will specify a situation, where ionisation probability 
in second order is a weighted sum of single photon terms. Furthermore
we will see, that Einstein's equality
\[E_{kin} = h \nu - \lap E >0\]
for the maximal kinetic energy $E_{kin}$ of the electron, 
energy $h \nu$ of the 
photon and ionisation gap $\lap E$ is the crucial condition, 
for these single photon terms to be nonzero.
\vspace{2mm}

\noindent
{\bf MSC:} 81Q10, 81V10, 47N50.

\vspace{2mm}

\noindent
{\bf Keywords:} Photoelectric Effect, Scattering Theory, QED. 
\end{abstract}

\section{A mathematical model for the photoelectric effect} \label{sec1}
\setcounter{equation}{0}
\subsection{Introduction}
In the first years after the discovery of the photoelectric effect it has been
a big challenge to obtain more and finer experimental results. Parallel to 
experiment, there were changes in the theoretical interpretation, which had to
be verified in the experiment:
\begin{itemize}
\item In 1887 Heinrich Hertz \cite{He} observed, that the length of a
flame in a "Funkenstrecke" depends on the light falling on the apparatus. Most
remarkable is his intuition, that this effect depends on the ultraviolet part
of the incident light.
\item A year later Wilhelm Hallwachs \cite{Ha} saw 
that an isolated, negatively charged metal plate loses its charge,
when it is enlighted with ultraviolet light. This is the simplest
setup for the photoeffect we know it already from our physics lessons.
\item Although there were a couple of experimental results in the next years,
every attempt for a theoretical description failed. There was no
theory based on classical physics, which could explain the existence of a
minimal frequency $\nu_0$ of the incoming light needed for the photoelectric 
effect to take place.
\item A turning point in our way of describing nature, is Einstein's paper
\cite{Ei} from 1905, where he takes a look at Wien's radiation formula
from the viewpoint of statistical mechanics and thermodynamic. He concludes: 
\begin{quote}
Monochromatische Strahlung von geringer Dichte (innerhalb des
G\"ul\-tig\-keits\-bereiches der Wienschen Strahlungsformel) ver\-h\"alt sich
in w\"armetheoretischer Beziehung so, wie wenn sie aus voneinander
unabh\"angigen Energiequanten von der Gr\"o{\ss}e $h\nu$ best\"unde.
\end{quote}
and applies this conclusion for the photoelectric effect. 
The ``Energie\-quanten'' are nowadays called photons and the photoelectric 
effect is in this picture a consequence of the absorption of photons 
by electrons in the metal.
An electron inside the metal needs a minimum amount of energy $\lap E$ to
leave the the metal. If one electron is allowed to absorb only one photon,
then due to conservation of energy it may escape from the metal, provided
\begin{equation} \label{eqein1}
h\nu > \lap E.
\end{equation}
This explains the minimal frequency $\nu_0=\frac{\lap E}{h}$, but at
the same time this model proposed the bound
\begin{equation} \label{eqein2}
E_{kin} = h \nu-\lap E
\end{equation}
for the maximal kinetic energy $E_{kin}$ of an escaping electron.
\item The experimental verification of (\ref{eqein2}) was done by
Robert Millikan \cite{M1}, \cite{M2} in 1916. It confirms that
Einstein's model is appropriate to describe the photoelectric effect.
\end{itemize}
\noindent The goal of this article is to explain the photoelectric
effect in some variants of 
the standard model of non relativistic quantum electrodynamics, which
are more closely related to the physical picture than
the model studied in \cite{BKZ}. Now we can apply our results to an 
electron with more than one bound state and to a larger class of 
electron-photon interactions.
The paper is organised as follows:
We start with a short overview of the photoelectric effect, motivate
the definition of zeroth and second order of ionisation probability and 
describe our results.
In Chapter \ref{ch2} we introduce the model(s) under
consideration stating all definitions and model assumptions. In particular
this includes a description of the electron and the
photon subsystems and the total interacting systems in terms of
Hamiltonians generating the dynamics. A description
of the photoelectric effect needs some special initial
states, which model a bound state plus some incoming 
photons. For this initial states we derive an asymptotic expansion of
the full interacting time evolution in terms of free Heisenberg time
evolutions in chapter \ref{sec2}. This asymptotic expansion is the key
ingredient in the definition of the zeroth and second order terms of the 
ionisation probability. 
This definition is a modification of the transported charge 
in \cite{BKZ}. The following results for ionisation probability are proven in 
Chapter \ref{sec3}:
\begin{itemize}
\item The zeroth order of the ionisation probability vanishes. 
\item If the photon wave functions are
orthonormal, then the second order term of ionisation probability is 
a weighted sum of one photon terms. This decoupling property shows, 
that the effect 
(at least in second order) does not depend on some multi-photon-phenomenon, 
hence this is a first justification for Einstein's
effective one-electron / one-photon model coming from quantum electrodynamics.
\item Theorem \ref{thm3.4} gives an explicit expression for the 
ionisation probability of a single photon. The energy conservation condition
(\ref{eqein2}) is hidden in the integration of photon momentum in 
(\ref{eq279}).
\end{itemize}
Finally the appendix contains some of the often used technical tools.

\subsection{Ionisation probability, photon clouds and photoelectric effect} 
\label{sec1.4}
For a Pauli-Fierz operator
\[H_g=H_0+gW^{(1)}+g^2W^{(2)}=(-\lap+V) \otimes \one + \one \otimes H_f
+gW^{(1)}+g^2W^{(2)}\]
with ground state $\gs$ and ground state energy $E_g$ (see Chapter \ref{ch2}
for precise definitions and model assumptions),
we want to see the relationship of
this model to the photoelectric effect we know from standard physics
textbooks. The experimental setup
consists in the simplest form of a source, emitting a beam of
photons, which are absorbed in a ``target''. A detector 
measures the current of the electrons emitted from the target. 
If there is any effect
at all, it is seen ``immediately'', which is within about 
$10^{-9} s$, see \cite{No}, p 48. 
How can we relate this experiment with theory? 
The quantum mechanical model under
consideration should cover all effects of non relativistic quantum
electrodynamics, especially Compton scattering. The borderline between
Compton scattering and photoeffect in this model is hard to define; it
depends on the initial state: In Compton scattering an electron, which
is not bound to an atom is scattered in the presence of the photon
field. On the other side, a bound state, which is
ionised by photons is the starting point of the photoelectric
effect. Hence for a description of the photoeffect, we have to choose
some initial states, which model a bound state plus some photons. 
The following points motivate our choice of the initial state:
\begin{itemize}
\item $\gs$ as bound state:
In a similar model, where the interacting Hamiltonian is also called $H_g$
and under some conditions specified in \cite{BFS1} Theorem I.2 and 
Corollary III.5
the spectrum of $H_g$ is purely absolute continuous outside a
$\cO(g)$-neighbourhood of all eigenvalues and thres\-holds of
$\hel$. Moreover the spectrum is absolutely continuous in those
neighbourhoods of the energies $e_1, e_2,...$ corresponding to the
exited states of $\hel$ below the ionisation thres\-hold. 
So the ground state $\gs$ is the only eigenstate of $H_g$ 
below a $\cO(g)$ neighbourhood of the ionisation threshold in this
slightly different model.
\item We want to decide, if the photoeffect is either
\begin{itemize}
\item a collective effect of many photons and depends e.g. on the sum of all
photon energies
\item or if it can be explained as a result of some single photon 
processes and depends e.g. on the maximum of all photon energies.
\end{itemize}
{For} this purpose, we have a look at $N>1$ incoming photons, 
otherwise we would not see any difference in the two cases.
\item In \cite{BKZ} we have seen, that a single photon result like
(\ref{eqein2}) is a result of the preparation of the initial state:
In Einstein's model the interaction is essentially turned on and off by hand,
hence the energy balance is the noninteracting one. 
If we add a photon cloud at time zero to the ground state
by just applying creation operators
\[A \gs= \prod_{j=1}^N a_{\lambda_j}^* (f_j) \gs, \]
then due to the interaction, we would expect a modified energy balance
compared to the noninteracting case. On the other hand, 
if we observe exponential decay of $\gs$,
then we could hope to mimic such an almost free energy balance by adding 
the photons wide inside the exponential tail of $\gs$, where the 
wavefunction is tiny and the interaction may be negligible. 
A way to write this vague idea in precise formulas is to use an 
incoming scatting state
\begin{equation} A(t) \gs = e^{-itH_g} e^{itH_0} A e^{-itH_0} e^{itH_g} \gs =
e^{-itH_g} \prod_{j=1}^N a_{\lambda_j}^*(e^{-it\omega}f_j) e^{itH_g} \gs 
\label{gl1.3} \end{equation}
in the limit $t \to \infty$. We will see in Section \ref{sec2.3}, 
as a little corollary of sections \ref{sec2.1} and \ref{sec2.2}, that this
limit actually exists.
\end{itemize}
Note, that the ionisation probability is not just simply a function of
$H_g$ alone, as for example $\one_{]0,\infty[}(H_g)$, because by adding enough
photons of positive energy (but too low energy according to (\ref{eqein2})) 
to the electron ground state $\varphi_0$, 
we get an overlap with $\one_{]0,\infty[}(H_g) \hr$. This would be in contrast 
to the experiments supporting (\ref{eqein2}).
So we start differently and 
introduce the orthogonal projection
\begin{equation} \label{eq222} 
F_R:= \one_{\{|x|\geq R\}} \otimes \one_{\F}
\end{equation}
onto the functions in the electron space $\Hel$
with support outside the ball of radius $R>0$.
As a first guess and with the huge distance between target and detector
and the $10^{-9}$ seconds in mind one is
probably tempted to define the ionisation probability as
\[\lim_{R \to \infty} \lim_{t \to \infty} \|F_R e^{-i\tau H_g} A(t) \gs\|^2 \]
for some fixed $\tau$ (inspired by the $10^{-9}s$).
But as $\displaystyle \lim_{t \to \infty} A(t) \gs$ exists and $F_R$
converges strongly to $0$, this expression is  for some fixed $\tau$ just $0$
and in contrast to our definition of $Q^{(0)}(A)$, there is no chance to 
see this as a zeroth order quantity in $g$.
Another idea is to choose a $g$-dependent $\tau$, such that 
$\tau(g) \nearrow \infty$ as $g \searrow 0$ and to have a look at
\begin{equation} \label{gl15}
Q^{(0)}(A):= \lim_{R \nearrow \infty} \, \lim_{g \searrow 0} \, 
\lim_{t \to \infty} \|F_R e^{-i\tau(g) H_g} A(t) \gs\|^2 ,
\end{equation}
the zeroth order of the ionisation probability. The choice 
$\tau(g) \nearrow \infty$ as $g \searrow 0$ should be seen as a weak coupling
limit: The weaker the interaction (smaller $g$) the longer you will have 
to wait until you see an effect (larger $\tau$).
In fact, we will see in Theorem
\ref{thm3.1}, that $Q^{(0)}(A)=0$ provided $\tau(g) \nearrow \infty$ as 
$g \searrow 0$. Additionally in the proof of Theorem \ref{thm3.1}
we get a decomposition of the vector (expressed in terms of the free 
time evolution $A_{\tau}:=e^{-i\tau H_0} A e^{i\tau H_0}$, see (\ref{gl3.1}))
\begin{eqnarray}
\lefteqn{ F_R e^{-i\tau(g) (H_g-E_g)} A(t) \gs= \label{gl16}} \\
&=& F_R A_{\tau(g)} \gs 
-ig F_R \int\limits_0^{t+\tau(g)} ds e^{-is(H_g-E_g)} 
[W^{(1)}+gW^{(2)}, A_{\tau(g)-s}]
\gs, \nonumber
\end{eqnarray}
which has the same norm square as $ F_R e^{-i\tau(g) H_g} A(t) \gs$.
In (\ref{gl16}) the first term does not depend on $t$ and vanishes in the limit
$\displaystyle \limsup_{R \to \infty} \lim_{g \searrow 0}$, see Lemma \ref{l2.9}
for details. The second term carries an explicit prefactor $g$, so in order
to see the contributions in second order of $g$, we subtract 
$F_R A_{\tau(g)} \gs$ and eliminate the prefactor dividing by $g$, i.e.
we define
\begin{eqnarray}
Q^{(2)}(A)&:= &
\lim_{R \nearrow \infty} \, \lim_{g \searrow 0} g^{-2} \lim_{t \to \infty} 
\|F_R e^{-i\tau(g) (H_g-E_g)} A(t) \gs-F_R A_{\tau(g)} \gs\|^2  \\
&=& \lim_{R \nearrow \infty} \, \lim_{g \searrow 0} \, \lim_{t \to \infty} 
\Big\|F_R \hspace{-0.3cm} \int\limits_0^{t+\tau(g)} \hspace{-0.3cm}
ds \,e^{-is(H_g-E_g)} 
[W^{(1)}+gW^{(2)}, A_{\tau(g)-s}] \gs \Big \|^2 
\hspace{0.15cm} 
\nonumber
\end{eqnarray}
as ionisation probability in second order. $Q^{(2)}(A)$ is the object of studies 
in sections \ref{sec3.2} and \ref{sec3.3}. Assuming
$g^{-\alpha} < \tau(g) < g^{-1}$ for some $\alpha \in ]0,1[$, we will then 
prove:
\begin{itemize}
\item a decoupling property for orthonormal photon wave functions: \\
If  $m_1,...,m_{\eta},n_1,...,n_{\eta} \in \N_0$ and 
$\varphi_1,...,\varphi_{\eta} \in C_0^{\infty}(\R^3 \backslash \{0\})$ are
orthonormal, then
\begin{eqnarray*}
\lefteqn{\frac{ Q^{(2)}(a^*_+(\varphi_1)^{m_1} a^*_-(\varphi_1)^{n_1} \cdots
a^*_+(\varphi_{\eta})^{m_{\eta}}a^*_-(\varphi_{\eta})^{n_{\eta}})}
{m_1! \cdots m_{\eta}! n_1! \cdots n_{\eta}!}= } \\
&=& \sum_{j=1}^{\eta} \Big( n_j Q^{(2)}_-(\varphi_j)+
m_j Q^{(2)}_+(\varphi_j)\Big) 
\end{eqnarray*}
for some one photon quantities $Q^{(2)}_{\lambda}(\varphi_j)$ depending on the
photon polarisation $\lambda$ and the momentum wave functions $\varphi_j$.
\item The choice $g^{-\alpha} < \tau(g)$ and the preparation of the initial
state allows us to prove in (\ref{eq279}) the expression
\[Q^{(2)}_{\lambda}(\varphi_j)
=\int\limits_{\R^3} dp \left|\; \int\limits_{S^2(p^2-e_0)} d\mu_{p^2-e_0}(k)
\widehat{\rho}_{\lambda}(p,k) \varphi_j(k) \right|^2 \]
for the contribution of a single photon with wave function $\varphi_j$ in
momentum space and polarisation $\lambda$. $p$ is the electron momentum
and $\widehat{\rho}_{\lambda}(p,k)$ can be calculated from electron Hamiltonian,
electron ground state and electron-photon interaction. 
The restriction of the photon
momentum integration to the sphere $S^2(p^2-e_0)$ of radius $p^2-e_0$
encodes the energy conservation condition $\omega(k)=p^2-e_0$ between
photon energy $\omega(k)$, free electron energy $p^2$ and the binding energy 
$|e_0|$ of $\hel$ and is therefore an analog of (\ref{eqein2}).
\end{itemize}
%
\section{Definitions, model assumptions and first conclusions} 
\label{ch2}
\setcounter{equation}{0}
Now we give a precise definition of the model(s) under consideration
including all the model assumptions and give references to literature.
\subsection{The subsystem of the electron} \label{sec1.1}
We start with a non relativistic, spinless electron whose dynamics is 
given by a Schr\"odinger operator
\begin{equation} \label{eq200}
\hel=-\lap+V
\end{equation}
in $\Hel= L^2(\R^3)$. 
\begin{hypothesis} \label{H-1}
$V$ is relatively $-\lap$-bounded with bound $<1$, thus
$\hel=-\lap +V$ defines a self-adjoint operator on the domain $\D(-\lap)$ of 
$-\lap$. $\hel$ has a non degenerate ground state
$\varphi_0 \in \Hel$ with energy $e_0<0$:
\begin{equation} \label{eq201}
\hel \varphi_0 = e_0 \varphi_0.
\end{equation}
The singular continuous spectrum $\sigma_{sc}(\hel)= \emptyset$ is
empty.
\end{hypothesis}
\begin{remark}
There is a big amount of literature about Schr\"odinger operators studying
these properties:
\begin{itemize}
\item 
\cite{RS4} chapters XIII.6, XIII.7, XIII.8 and XIII.10 are devoted to ``absence
of singular continuous spectrum'': \\
In particular $\sigma_{sc}(\hel)=\emptyset$
if $V(x)=\frac{1}{|x|}$ (Theorem XIII.36) or if $V$ is  bounded, measurable
with compact support (Theorem XIII.33).
\item
\cite{RS4} chapter XIII.12 treats ``nondegeneracy of the ground state'':\\
In particular, if for some bounded measurable potential $V$ there is 
an eigenvalue at the bottom of the spectrum of $\hel$, then it is 
nondegenerate.
\end{itemize}
So the Coulomb potential and finite potential wells, which have at least 
one bound state are some examples, that satisfy Hypothesis \ref{H-1}.
\end{remark}
\subsection{Photons}
We couple the electron described above to a quantised
photon field. The Hilbert space $\F$ carrying the photon degrees of freedom
is the bosonic Fock space 
$\F=\F_b(L^2(\R^3 \times  \Z_2))$ over the one-photon
Hilbert space $L^2(\R^3 \times \Z_2)$. $\R^3 \times \Z_2$ is viewed as 
photon momentum space, the two components describe the two independent
transversal polarisations of the photon (in radiation gauge).
\begin{equation} \label{eq202}
\F = \bigoplus_{n \in \N_0} \F^{(n)},
\end{equation}
where the vacuum sector $\F^{(0)}$ is a one-dimensional subspace
spanned by the normalised Fock vacuum $\Omega$ and the $n$-photon
sectors $\F^{(n)}$ are the subspaces of $L^2((\R^3 \times \Z_2)^{n})$ 
containing totally symmetric vectors. The Hamiltonian in $\F$ representing
the energy of the free photon field is given by
\begin{equation} \label{eq203}
H_f=\sum_{\lambda \in \Z_2}\; \int\limits_{\R^3} dk \omega(k) 
a^*_{\lambda}(k) a_{\lambda}(k),
\end{equation}
where
\begin{equation} \label{eq203.1}
\omega(k):=|k|
\end{equation}
is the photon dispersion and $a^*_{\lambda}$ and $a_{\lambda}$ are
the standard creation- and annihilation operators in $\F$, which
fulfil the canonical commutation relations
\begin{eqnarray}
{[a_{\lambda}(k),a_{\mu}(k')]}={[a^*_{\lambda}(k),a^*_{\mu}(k')]}&=&0  
\label{eq204.1}\\ 
{[a_{\lambda}(k),a^*_{\mu}(k')]}&=&\delta_{\lambda,\mu} \delta(k-k') 
 \label{eq204.2}\\ 
a_{\lambda}(k) \Omega&=&0  \label{eq204.3}
\end{eqnarray}
in the sense of operator valued distributions. In other words, $H_f$
is the second quantisation of the multiplication operator with the
photon dispersion $\omega(k)=|k|$ restricted to $\F$.
For some of the estimates, we introduce cutoff parameters
$0 \leq \tilde{r}< r \leq \infty$ and define the regularised dispersion
$\omega_{(\tilde{r},r)}(k):=\omega(k) \one_{\{\tilde{r} \leq \omega(k)\leq r\}}(k)$ 
and regularised free field
\begin{equation} \label{eq203b}
H_{f,(\tilde{r},r)}=\sum_{\lambda \in \Z_2}\; \int\limits_{\R^3} dk \,
\omega_{(\tilde{r},r)}(k) a^*_{\lambda}(k) a_{\lambda}(k).
\end{equation}
\subsection{The interaction between electron and photons}
The Hilbert space of states for the electron-photon system is the
Hilbert space tensor product 
$\hr=\Hel \widehat{\otimes} \F$. In $\hr$ the dynamics is given by
\begin{equation} \label{eq503}
H_g=H_0+W,
\end{equation}
introducing the non-interacting dynamics 
\begin{equation} \label{eq205}
H_0= \hel \otimes \one_{\F}+ \one_{\Hel} \otimes H_f.
\end{equation}
The spectral measure of $H_0=\hel \otimes \one+ \one \otimes H_f$ can be 
described very explicit in terms of the spectral measures of $\hel$ and $H_f$, 
see e.g. \cite{We}, chap. 8.5, in particular 
$\Phi_0=\varphi_0 \otimes \Omega$ is the
ground state of $H_0$ with ground state energy 
$E_0=\inf \sigma(H_0)=e_0=\inf \sigma(\hel)$.
In the interaction
\begin{eqnarray}
W&=&gW^{(1)}+g^2W^{(2)}= \label{eq700}\\
&=&gW^{(1,0)}+gW^{(0,1)}+g^2W^{(2,0)}+g^2W^{(0,2)}+g^2W^{(1,1)},
\nonumber
\end{eqnarray}
with
\begin{eqnarray}
W^{(1,0)}&=&\sum_{\lambda \in \Z_2} \; \int\limits_{\R^3} 
dk w^{(1,0)}(k,\lambda) a^*_{\lambda}(k) \label{eq504}\\
W^{(0,1)}&=&\sum_{\lambda \in \Z_2} \; \int\limits_{\R^3} 
dk w^{(0,1)}(k,\lambda) a_{\lambda}(k) \label{eq505}\\
W^{(2,0)}&=&\sum_{\lambda_1,\lambda_2 \in \Z_2} \; \int\limits_{\R^3} 
\; \int\limits_{\R^3} 
dk_1 dk_2 w^{(2,0)}(k_1,\lambda_1;k_2,\lambda_2) a^*_{\lambda_1}(k_1)
a^*_{\lambda_2}(k_2) \hspace{1.9cm} \label{eq506}\\
W^{(0,2)}&=&\sum_{\lambda_1,\lambda_2 \in \Z_2} \; \int\limits_{\R^3} 
\, \int\limits_{\R^3} 
dk_1 dk_2 w^{(0,2)}(k_1,\lambda_1;k_2,\lambda_2) a_{\lambda_1}(k_1)
a_{\lambda_2}(k_2) \label{eq507}\\
W^{(1,1)}&=&\sum_{\lambda_1,\lambda_2 \in \Z_2} \; \int\limits_{\R^3} 
\; \int\limits_{\R^3} 
dk_1 dk_2 w^{(1,1)}(k_1,\lambda_1;k_2,\lambda_2) a^*_{\lambda_1}(k_1)
a_{\lambda_2}(k_2). \label{eq508}
\end{eqnarray}
the supscript indicates the total number of created and annihilated photons
resp. a pair of supscripts indicates the number of created and annihilated 
photons. In order to get at least a symmetric interaction we have to require
\begin{eqnarray}
w^{(1,0)}(k,\lambda)&=&(w^{(0,1)}(k,\lambda))^* \label{gl2}\\
w^{(2,0)}(k_1,\lambda_1,k_2,\lambda_2)&=&
(w^{(0,2)}(k_1,\lambda_1,k_2,\lambda_2))^*
\label{gl3}
\end{eqnarray}
As usual, we assume, that the interactions $w^{(m,n)}$, $m+n=2$ can be
factorised: 
Let $\mu$ be the measure on the Borel sets of $\R^3 \times \Z_2$, which is 
the sum of the measures with Lebesgue density $1+\frac{1}{\omega(\cdot)}$
on $\R^3$. Let $L(\Hel)$ denote the bounded operators on $\Hel$.
\begin{hypothesis} \label{hyp5}
There is a $\fG \in L^2((\R^3 \times \Z_2,\mu), L(\Hel)^3)$,
i.e.
\[\fG(k,\lambda)=\left( \begin{array}{c} \fG_1(k,\lambda) \\
\fG_2(k,\lambda) \\ \fG_3(k,\lambda) \end{array} \right) \]
consisting of bounded operators
$\fG_1(k,\lambda)$, $\fG_2(k,\lambda)$, $\fG_3(k,\lambda)$
on $\hel$ for $\mu$-almost every $(k,\lambda) \in \R^3 \times \Z_2$, 
and 
\begin{equation} \label{gl21}
\sum_{\lambda \in \Z_2}\int\limits_{\R^3} dk \|\fG(k,\lambda)\|^2 
(1+\omega(k)) < \infty,
\end{equation}
such that
\begin{eqnarray}
w^{(2,0)}(k,\lambda,k',\lambda')&=&\sum_{\iota=1}^3
\fG_{\iota}(k,\lambda) \fG_{\iota}(k',\lambda') \label{gl53}\\
w^{(1,1)}(k,\lambda,k',\lambda')&=&\sum_{\iota=1}^3
\Big(\fG_{\iota}(k,\lambda)^* \fG_{\iota}(k',\lambda') +
\fG_{\iota}(k,\lambda) \fG_{\iota}(k',\lambda')^* \Big) \hspace{0.9cm} 
\label{gl54}
\end{eqnarray}
\end{hypothesis} 
\noindent Hypothesis \ref{hyp5} is quite natural:
In \cite{BFS1} and \cite{BFS2} it can be seen how the Hamiltonian 
$H_g$ of the form specified in (\ref{eq503})-(\ref{gl3}) is related to the
standard model of quantum electrodynamics and some of it's approximations.
(\ref{gl53}) and (\ref{gl54}) are part of this type of models.
(\ref{gl21}) is still true in the usual minimal coupling model, where
\[ \fG(k,\lambda)=
\frac{\kappa(k)}{\sqrt{\omega(k)}}\, e^{-ikx} \feps_{\lambda}(k)\]
with some ultraviolet cutoff function $\kappa$ (choosing
a  Schwarz function or the characteristic function of some box for $\kappa$) 
and 
vectors $\feps_-(k), \feps_+(k) \in \R^3$, such that $\feps_-(k), \feps_+(k),
\frac{k}{|k|}$ form an oriented orthonormal basis of $\R^3$.
\begin{hypothesis} \label{hyp4}
There is a $\zeta \geq 2$, such that:
\begin{enumerate}
\item For $\iota =1,2,3$ and $(m,n)=(1,0)$
or $(0,1)$: 
\[\fG_{\iota}(\cdot,\lambda),
w^{(m,n)}(\cdot,\lambda)(\hel-b)^{-\frac{1}{2}},
\in C^{\zeta}(\R^3\backslash \{0\},L(\Hel)).\]
\item $\partial_k^{\alpha} \fG_{\iota}(\cdot,\lambda),
\partial_k^{\alpha} w^{(m,n)}(\cdot,\lambda)(\hel-b)^{-\frac{1}{2}}
 \in L^2(K,L(\Hel))$ 
for $\iota =1,2,3$, $(m,n)=(0,1)$ or $(1,0)$ and
any index $\alpha \in \N_0^3$ with $|\alpha| \leq \zeta$ and compact 
sets $K \subseteq \R^3 \backslash \{0\}$.
\end{enumerate}
\end{hypothesis}
\noindent
Ignoring electron spin, the coupling functions in minimal coupled Pauli-Fierz
models take the form
\[w^{(1,0)}(k,\lambda)=-2\fG(k,\lambda) \cdot (-i\nabla_x), \]
and in this form we have to require some smoothness of $\fG$ plus a spacial 
decay of $\fG$ or a modified coupling for Hypothesis \ref{hyp4} to be true,
for example: \\
Take $\kappa \in \cS(\R^3,\R)$ and an orthonormal basis 
$\{\feps_-(k),\feps_+(k),\frac{k}{|k|}\}$ of $\R^3$, such that each of these
three vectors is smooth on $\R^3 \backslash \{0\}$. Let 
$\mu \in \cS(\R^3, \R)$ and $\chi \in \cS(\R^3,\R^3)$, then every
component of
\[\fG(k,\lambda)=\frac{\kappa(k)}{\sqrt{\omega(k)}} \feps_{\lambda}(k)
e^{-ikx} \mu(x) \]
or
\[\fG(k,\lambda)=\frac{\kappa(k)}{\sqrt{\omega(k)}} \feps_{\lambda}(k)
e^{-ik\chi(x)}  \]
is $\infty$-often differentiable with respect to $k$ on $\R^3 \backslash \{0\}$
and the derivatives are continuous on $\R^3 \backslash \{0\}$, hence
$L^2$ on each compactum $K \subseteq \R^3 \backslash \{0\}$. \\[0.3cm]
\noindent {\bf Hypothesis }$\mathbf (\hel,\gamma)$: \\
For $|\alpha| \leq \zeta$ and for compact sets 
$K \subseteq \R^3 \backslash \{0\}$
\begin{equation}
\int\limits_K \|(\hel-b)^{\frac{\gamma}{2}} \partial_k^{\alpha} w^{(0,1)}(k,\lambda)
(\hel-b)^{-\frac{\gamma+1}{2}} \|^2 < \infty
\end{equation}
\noindent 
Before stating the next Hypothesis, we fix some notation:
For some $b < e_0=\inf \sigma(\hel)$, which is fixed for the rest of the paper 
and for $\beta, \gamma \geq 0$ we define:
\begin{eqnarray} 
\Lambda_{\beta,\gamma}^{(1)} &:=& \hspace{-2pt}\max_{m,n \in \N_0 \atop m+n=1}
\sum_{\lambda \in \Z_2} 
\int\limits_{\R^3} dk
\frac{\|(\hel-b)^{\frac{\gamma}{2}} w^{(m,n)}(k,\lambda) 
(\hel-b)^{-\frac{\gamma+1}{2}}\|^2} {\omega(k)}(1+\omega(k))^{\beta}
\nonumber \\
\widetilde{\Lambda}_{\beta,\gamma}^{(1)}&:=& \hspace{-2pt}
\max_{m,n \in \N_0 \atop m+n=1}
\sum_{\lambda \in \Z_2}
\int\limits_{\R^3} dk \|(\hel-b)^{\frac{\gamma}{2}}
w^{(m,n)}(k,\lambda) (\hel-b)^{-\frac{\gamma+1}{2}}\|^2 
(1+\omega(k))^{\beta} \nonumber\\
\Lambda^{(2)}_{\beta,\gamma} &:=&  \sum_{\lambda \in \Z_2}
\int\limits_{\R^3} dk \frac{\|(\hel-b)^{\frac{\gamma}{2}}\fG(k,\lambda)
(\hel-b)^{-\frac{\gamma}{2}}\|^2}{\omega(k)}(1+\omega(k))^{\beta},  \nonumber\\
\widetilde{\Lambda}^{(2)}_{\beta,\gamma} &:=&  \sum_{\lambda \in \Z_2}
\int\limits_{\R^3} dk \|(\hel-b)^{\frac{\gamma}{2}}\fG(k,\lambda)
(\hel-b)^{-\frac{\gamma}{2}}\|^2 (1+\omega(k))^{\beta}  \label{gl58}
\end{eqnarray}
\noindent {\bf Hypothesis }$\mathbf (\hel,\beta,\gamma)$: \\
Given $\beta,\gamma \in \N_0$, then
$\Lambda_{\beta',\gamma'}^{(1)}$, 
$\widetilde{\Lambda}_{\beta',\gamma'}^{(1)}$,
$\Lambda_{\beta',\gamma'}^{(2)}$,
$\widetilde{\Lambda}_{\beta',\gamma'}^{(2)}$
and $\tilde{\Lambda}^{(2)}_{\frac{\beta'}{2},\gamma'}$ are finite for any
$\beta',\gamma' \in \N_0$ with $\beta' \leq \beta$ and $\gamma' \leq \gamma$.
\subsubsection{Self-adjointness and Semiboundedness of {\protect
\boldmath $H_g$}}
\noindent 
$W$ is a relatively $H_0$ bounded operator, more precisely:
\begin{lemma} \label{l1.1}
Let $\Lambda^{(1)}_{0,0}, \tilde{\Lambda}^{(1)}_{0,0}, \Lambda^{(2)}_{0,0}, 
\tilde{\Lambda}^{(2)}_{0,0} < \infty$,  then 
$W^{(1)}$ is infinitesimally $H_0$ bounded and $W^{(2)}$ is relatively $H_0$ 
bounded, satisfying
\begin{eqnarray} 
\lefteqn{\|W^{(2)} (H_0-b+1)^{-1} \| \leq \label{gl57}}\\
&\leq& \Lambda^{(2)}_{0,0} +
2 \Big[(\tilde{\Lambda}^{(2)}_{0,0}+\Lambda^{(2)}_{0,0})
\Lambda^{(2)}_{0,0} \Big]^{\frac{1}{2}} +2 \Big[(\Lambda^{(2)}_{0,0})^2 +
4 \Lambda^{(2)}_{0,0} \tilde{\Lambda}^{(2)}_{0,0} + 
(\tilde{\Lambda}^{(2)}_{0,0})^2 \Big]^{\frac{1}{2}}. \nonumber
\end{eqnarray}
In particular if $g$ is small enough, such that 
\begin{equation}
g^2 \Bigg[\Lambda^{(2)}_{0,0} +
2 \Big[(\tilde{\Lambda}^{(2)}_{0,0}+\Lambda^{(2)}_{0,0})
\Lambda^{(2)}_{0,0} \Big]^{\frac{1}{2}} +2 \Big[(\Lambda^{(2)}_{0,0})^2 +
4 \Lambda^{(2)}_{0,0} \tilde{\Lambda}^{(2)}_{0,0} + 
(\tilde{\Lambda}^{(2)}_{0,0})^2 \Big]^{\frac{1}{2}}\Bigg] < 1, \label{eq701}
\end{equation}
then $H_g$ is self-adjoint on $\D(H_0)$ and bounded from below.
\end{lemma}

\begin{proof}
The estimates 
\begin{eqnarray*}
\|W^{(0,1)} (H_f+1)^{-\frac{1}{2}} (\hel-b)^{-\frac{1}{2}} \| &\leq& 
\sqrt{\Lambda_0^{(1)}} \\
\|W^{(1,0)} (H_f+1)^{-\frac{1}{2}} (\hel-b)^{-\frac{1}{2}} \| &\leq& 
\sqrt{\widetilde{\Lambda}_0^{(1)}+\Lambda_0^{(1)}} 
\end{eqnarray*} 
are special cases of equations (\ref{gla30}) and   (\ref{gla60}). Hence
\begin{equation} \label{gl24}
\|W^{(1)} \Psi \|^2 \leq 2( \widetilde{\Lambda}_0^{(1)}+2\Lambda_0^{(1)}) 
\|(\hel-b)^{\frac{1}{2}} (H_f+1)^{\frac{1}{2}} \Psi \|^2. 
\end{equation}
For any $\varepsilon >0$ 
\begin{eqnarray*}
0 &\leq& \| \varepsilon (\hel-b)(H_f+1) \Psi -\frac{1}{\varepsilon} \Psi \|^2
= \\
&=& \varepsilon^2 \|(\hel-b)(H_f+1) \Psi \|^2 + 
\frac{1}{\varepsilon^2} \|\Psi\|^2
-2 \Re \lkl \Psi, (\hel-b)(H_f+1) \Psi \rkl .
\end{eqnarray*}
$\hel-b$ and $H_f+1$ are self-adjoint positive commuting operators, so
by spectral calculus the last inequality implies:
\begin{eqnarray}
\|(\hel-b)^{\frac{1}{2}} (H_f+1)^{\frac{1}{2}} \Psi \|^2 &=&
\Re \lkl \Psi, (\hel-b)(H_f+1) \Psi \rkl \leq \nonumber \\
&\leq& \frac{\varepsilon^2}{2} \|(\hel-b)(H_f+1) \Psi\|^2 + 
\frac{1}{2\varepsilon^2} \|\Psi\|^2 \hspace{1.5cm}\label{gl25}
\end{eqnarray}
Combining (\ref{gl24}) and (\ref{gl25}), $W^{(1)}$ is 
infinitesimally $(\hel-b)(H_f+1)$ bound. 
$\D(H_0)=\D(\hel \otimes \one) \cap \D(\one \otimes H_f)=\D(\hel H_f)$, so
$(\hel-b)(H_f+1)$ is $H_0-b+1$ bounded. $\sigma(H_0) \not=0$, so there
are $a_1,a_2 \in ]0,\infty[$, such that
\[\|(\hel-b)(H_f+1) \Psi \| \leq a_1 \|(H_0-b+1) \Psi \| + a_2 \|\Psi\| \]
for any $\Psi \in \D(H_0)$, see e.g. \cite{HS} Prop. 13.2. In particular
$W^{(1)}$ is infinitesimally $H_0$-bounded.
$H_f+1 \geq 0$ and $H_0-b+1 \geq 0 $ commute, so
\begin{eqnarray*}
0 &\leq& 
\lkl (H_0-b+1)^{-\frac{1}{2}} \Psi, (H_f+1)(H_0-b+1)^{-\frac{1}{2}} \Psi \rkl 
=\\
&=& \lkl \Psi, (H_f+1) (H_0-b+1)^{-1} \Psi \rkl \leq \\
&\leq & \lkl (H_0-b+1)^{-\frac{1}{2}} \Psi, 
(H_f+1+\hel -b)(H_0-b+1)^{-\frac{1}{2}} \Psi \rkl = \\
&=& \lkl \Psi, (H_0-b+1)(H_0-b+1)^{-1} \Psi \rkl = \|\Psi\|^2,
\end{eqnarray*}
and estimate (\ref{gl57}) follows from
\begin{eqnarray*}
\|W^{(2,0)} (H_f+1)^{-1} \| &\leq & \Lambda^{(2)}_{0,0} \\
\|W^{(1,1)} (H_f+1)^{-1} \| &\leq & 2 [(\tilde{\Lambda}^{(2)}_{0,0}+\Lambda^{(2)}_{0,0})
\Lambda^{(2)}_{0,0} ]^{\frac{1}{2}} \\
\|W^{(0,2)} (H_f+1)^{-1} \| &\leq & 2 [(\Lambda^{(2)}_{0,0})^2 +
4 \Lambda^{(2)}_{0,0} \tilde{\Lambda}^{(2)}_{0,0} + 
(\tilde{\Lambda}^{(2)}_{0,0})^2 ]^{\frac{1}{2}}
\end{eqnarray*}
which are special cases of equations  (\ref{gla32}),
(\ref{gla33}) and (\ref{gla34}).
When (\ref{eq701}) is fulfilled, then $W=gW^{(1)}+g^2W^{(2)}$ is $H_0$-bounded
with bound $<1$ and Kato-Rellich Theorem implies 
self-adjointness of $H_g$ on $\D(H_0)$ and in particular, $H_g$ is bounded 
from below.
\end{proof}
\subsubsection{Properties of the ground state}
\begin{hypothesis} \label{H-2}
The interacting Hamiltonian possesses a normalised ground state 
$\gs \in \hr$,
$\|\gs\|=1$. The infimum of the spectrum $E_g:=\inf \sigma(H_g)$ is an
eigenvalue of $H_g$ with corresponding eigenvector $\gs$:
\begin{equation} \label{eq220}
H_g \gs=E_g \gs
\end{equation}
and $E_g < e_1:= \inf \sigma(\hel) \backslash \{e_0\}$. 
The projection $P^{\perp}_{\go}$ onto the orthogonal complement of the
one dimensional space spanned by $\go$ satisfies:
\begin{eqnarray} 
\|P^{\perp}_{\go} \gs \| &\leq& c_1g \label{eq221}
\end{eqnarray}
for some $c_1 < \infty$. There is some compact neighbourhood $U$ of $0$, such
that $g \mapsto E_g$ is continuous on $U$ and given $N \in \N$ and 
$0 < \tilde{r} < r < \infty$
\begin{equation} \label{gl26}
\limsup_{R \to \infty} \sup_{g \in U} \|\one_{\{|x| \geq R\}} 
H_{f,(\tilde{r},r)}^{\frac{N}{2}} \gs \| =0
\end{equation}
\end{hypothesis}
\noindent
Existence of a ground state has been proven for many variants of the 
Pauli-Fierz model; an incomplete list is \cite{AH1}, \cite{AH2},
\cite{Ge}, \cite{GLL}, \cite{Hi1}, \cite{LL}.
An existence proof for $\gs$, that gives the overlap (\ref{eq221}) 
with the vacuum $\go$ and an exponential decay $\|e^{\alpha |x|} \gs\| < \infty$
(for some $\alpha >0$) is found in \cite{BFS1} and \cite{BFS2}. In \cite{Hi2}
decay of powers of the photon-number $\|(\mathbf{N}^{\frac{k}{2}} \gs)(x)\|_{\F}$ 
with respect to the electron coordinate $x$ is proven. 
Due to the cutoffs this implies decay, when replacing
$\mathbf N$ with $H_{f,(\tilde{r},r)}$.
A stronger version of (\ref{eq221}) is needed for our problem;
the proof of Lemma \ref{lemma1} combines the relative 
bounds of Lemma \ref{la2} and \ref{la4} with some ideas from the proof 
of exponential decay in \cite{BFS1}:
\begin{lemma} \label{lemma1}
Let $0 \leq \tilde{r} < r \leq \infty$ and $\alpha, \beta, \gamma \in \N_0$, 
such that
Hypothesis \ref{H-1}, \ref{hyp5}, \ref{H-2} and $(\hel,\beta,\gamma)$ 
hold true and
$\Lambda^{(1)}_{0,\gamma}, \Lambda^{(2)}_{0,\gamma}, 
\widetilde{\Lambda}^{(2)}_{0,\gamma} < \infty$. Then
there is a $c_2=c_2(\alpha,\beta,\gamma) < \infty$, such that
\[\|(\hel-b)^{\frac{\gamma}{2}} (H_f+1)^{\frac{\beta}{2}} 
(H_{f,(\tilde{r},r)}+1)^{\frac{\alpha}{2}} (\gs-\go) \| \leq c_2 g \]
\end{lemma}
\begin{proof}
By Hypothesis \ref{H-2}
$\|P^{\perp}_{\go} \gs \| = \|\gs-\lkl \gs,\go\rkl \go \| \leq c_1 g$,
hence \linebreak $|\lkl \gs,\go \rkl | = 1-\cO(g)$ and
\begin{equation} \label{eq600}
\|\gs-\go\| \leq \|\gs-\lkl \gs,\go \rkl \go \| + \|\go\| |1-\lkl \gs,\go
\rkl |= \cO(g).
\end{equation}
Let $e_0$ be the ground state energy of $\hel$ and 
$e_1=\inf\sigma(\hel)\backslash \{e_0\}$ 
the energy of the first excited state, if there are more bound electron states,
resp. the ionisation threshold of $\hel$, if there is just one bound electron
state and choose $e' \in]e_0,e_1[$, such that 
$e' >E_g$. From $H_f \geq 0$ and $e' <e_1$ we conclude
\[\one_{]-\infty,e'[}(H_0)=\one_{\{e_0\}}(\hel) \one_{]-\infty,e'[}(H_0)=
\one_{[0,|e_0-e'|[}(H_f)\one_{\{e_0\}}(\hel) \one_{]-\infty,e'[}(H_0),\]
hence 
\begin{eqnarray*}
\lefteqn{(\hel-b)^{\frac{\gamma}{2}}(H_f+1)^{\frac{\alpha+\beta}{2}} 
\one_{]-\infty,e'[}(H_0)=}\\
&=&(\hel-b)^{\frac{\gamma}{2}}\one_{\{e_0\}}(\hel) (H_f+1)^{\frac{\alpha+\beta}{2}} 
\one_{[0,|e_0-e'|[}(H_f) \one_{]-\infty,e'[}(H_0) 
\end{eqnarray*}
is bounded. $H_f$, $H_{f,(\tilde{r},r)}$, $\hel$ and $H_0$ commute, so from 
$H_{f,(\tilde{r},r)} \leq H_f$ we conclude
\begin{eqnarray*}
0 &\leq& \|(\hel-b)^{\frac{\gamma}{2}} (H_f+1)^{\frac{\beta}{2}} 
(H_{f,(\tilde{r},r)}+1)^{\frac{\alpha}{2}} \one_{]-\infty,e'[}(H_0) \Psi \|^2=\\
&=& \lkl \Psi, (\hel-b)^{\gamma} (H_f+1)^{\beta} (H_{f,(\tilde{r},r)}+1)^{\alpha} 
\one_{]-\infty,e'[}(H_0)\Psi \rkl \leq \\
&\leq& \lkl \Psi, (\hel-b)^{\gamma} (H_f+1)^{\alpha+\beta} \one_{]-\infty,e'[}(H_0)
\Psi \rkl = \\
&=& \|(\hel-b)^{\frac{\gamma}{2}} (H_f+1)^{\frac{\alpha+\beta}{2}} 
\one_{]-\infty,e'[}(H_0) \Psi \|^2,
\end{eqnarray*}
so by (\ref{eq600}) it is enough to prove
\[\|(\hel-b)^{\frac{\gamma}{2}} (H_f+1)^{\frac{\beta}{2}} 
(H_{f,(\tilde{r},r)}+1)^{\frac{\alpha}{2}} \one_{[e',\infty[}(H_0)(\gs-\go) \|
\leq \cO(g). \]
But as $\go=\varphi_0 \otimes \Omega=\one_{\{e_0\}}(H_0) \go$ and
$e_0<e'$, we get
\[ \one_{[e',\infty[}(H_0)(\gs-\go)= \one_{[e',\infty[}(H_0)\gs \]
By $e' > E_g$ we can choose $\chi \in C_0^{\infty}(]-\infty,e'[)$, such that
$\chi(E_g)=1$, hence $\chi(H_g) \gs=\gs$ and 
$\one_{[e',\infty[}(H_0) \chi(H_0)=0$. Choose an almost analytic extension
$\tilde{\chi}$ of $\chi$ in some compact set 
$\cM \subseteq ]-\infty,e'[+i\R$, such that for $z=x+iy$
\[\frac{\partial \tilde{\chi}}{\partial \ol{z}}:= \frac{1}{2} \Big(
\frac{\partial \tilde{\chi}}{\partial x}  
+i \frac{\partial \tilde{\chi}}{\partial y} \Big) \]
satisfies
\begin{equation} \label{eq580}
\Big| \frac{\partial \tilde{\chi}} {\partial \ol{z}} \Big| \leq
\cO(|\Im z|^2),
\end{equation}
see e.g. \cite {Davis} chapter 2.2 for the explicit construction.
Introducing the complex measure
$d\mu(z):=\frac{1}{\pi} \frac{\partial \tilde{\chi}}{\partial \ol{z}} dx dy$
spectral calculus implies
\begin{eqnarray}
\lefteqn{\| (\hel-b)^{\frac{\gamma}{2}}  
(H_f+1)^{\frac{\beta}{2}}(H_{f,(\tilde{r},r)}+1)^{\frac{\alpha}{2}} 
\one_{[e',\infty[}(H_0)\gs\| = \label{eq570} }\\
&=&\|(\hel-b)^{\frac{\gamma}{2}} 
(H_f+1)^{\frac{\beta}{2}}(H_{f,(\tilde{r},r)}+1)^{\frac{\alpha}{2}} \one_{[e',\infty[}(H_0)
(\chi(H_g)-\chi(H_0)) \gs \| \nonumber \\
&=&\Bigg\|(\hel-b)^{\frac{\gamma}{2}} (H_{f,(\tilde{r},r)}+1)^{\frac{\alpha}{2}} 
\one_{[e',\infty[}(H_0) (H_f+1)^{\frac{\beta}{2}}\nonumber \\[-0.3cm]
&& \hspace{2cm}
\int\limits_{\cM} d\mu(z) [(H_g-z)^{-1}-(H_0-z)^{-1}] \gs\Bigg\|  \nonumber\\
&=&\Bigg \|\one_{[e',\infty[}(H_0) (\hel-b)^{\frac{\gamma}{2}}  
(H_f+1)^{\frac{\beta}{2}} (H_{f,(\tilde{r},r)}+1)^{\frac{\alpha}{2}} 
\nonumber \\[-0.3cm]
&& \hspace{2cm} \int\limits_{\cM}  d\mu(z) 
(H_0-z)^{-1}W(H_g-z)^{-1} \gs\Bigg \|  \nonumber
\end{eqnarray}
The eigenvalue equation $H_g \gs =E_g \gs$ implies $\gs=(H_g-E_g+1)^l \gs$
for any $l \in \Z$. 
Choose $\eta \in \N$, such that
$\eta \geq \frac{\alpha+\beta+\gamma}{2}+2$, then
\[(H_{f,(\tilde{r},r)}+1)^{\frac{\alpha}{2}}  (H_f+1)^{\frac{\beta}{2}+1} 
(\hel-b)^{\frac{\gamma}{2}+1} (H_g-E_g+1)^{-\eta}\] 
is bounded.
So commuting $(H_f+1)^{\frac{\beta}{2}}$ and 
$(H_0-z)^{-1}$,
\begin{eqnarray}
\lefteqn{\Big\| \int\limits_{\cM} d\mu(z) (\hel-b)^{\frac{\gamma}{2}}
(H_0-z)^{-1} (H_f+1)^{\frac{\beta}{2}} (H_{f,(\tilde{r},r)}+1)^{\frac{\alpha}{2}} 
W(H_g-z)^{-1} \gs \Big\|=}\nonumber\\[-0.3cm]
&=&\Bigg\| \int\limits_{\cM} d\mu(z) (\hel-b)^{\frac{\gamma}{2}}
(H_0-z)^{-1} 
(H_f+1)^{\frac{\beta}{2}}(H_{f,(\tilde{r},r)}+1)^{\frac{\alpha}{2}} W \nonumber\\[-0.8cm] 
&& \hspace{5.5cm}
(H_g-E_g+1)^{-\eta}(H_g-z)^{-1} \gs \Bigg\| \nonumber\\
&\leq& \|(\hel-b)^{\frac{\gamma}{2}}  
(H_f+1)^{\frac{\beta}{2}}(H_{f,(\tilde{r},r)}+1)^{\frac{\alpha}{2}} W 
(H_{f,(\tilde{r},r)}+1)^{-\frac{\alpha}{2}} (H_f+1)^{-\frac{\beta}{2}-1}
\nonumber\\
&& (\hel-b)^{-\frac{\gamma}{2}-1}\|
\int\limits_{\cM} d\mu(z) \|(H_0-z)^{-1} \| | (E_g-z)^{-1}|\nonumber\\
&& \|(H_{f,(\tilde{r},r)}+1)^{\frac{\alpha}{2}} (H_f+1)^{\frac{\beta}{2}+1} 
(\hel-b)^{\frac{\gamma}{2}+1} (H_g-E_g+1)^{-\eta} \| .
\label{gl5}
\end{eqnarray}
The integrand is bounded by $|\Im z|^{-2}$, so by compactness of $\cM$ 
and the bound (\ref{eq580}), this integral is finite, 
and due to Lemma \ref{la2} and \ref{la4}
\[
\|(\hel-b)^{\frac{\gamma}{2}}  (H_f+1)^{\frac{\beta}{2}} 
(H_{f,(\tilde{r},r)}+1)^{\frac{\alpha}{2}} 
W^{(j)}  (H_{f,(\tilde{r},r)}+1)^{-\frac{\alpha}{2}} (H_f+1)^{-\beta-1} 
(\hel-b)^{-\frac{\gamma}{2}-1}\| \]
remains bounded for $j=1,2$. If we now remember $W=gW^{(1)}+g^2W^{(2)}$, we 
see, that the right hand side of (\ref{gl5}) is $\cO(g)$.
\end{proof}
\begin{corollary} \label{co1.3}
Let $U \subseteq \R$ be compact, such that (\ref{eq701}) is true 
for all $g \in U$. Let the assumptions of Lemma \ref{lemma1}
be satisfied, then there is a $c_3=c_3(\alpha,\beta,\gamma, U) < \infty$, 
such that
\begin{equation}
\label{gl11}
\sup_{g \in U} \|(\hel-b)^{\frac{\gamma}{2}} 
(H_f+1)^{\frac{\beta}{2}} (H_{f,(\tilde{r},r)}+1)^{\frac{\alpha}{2}} \gs \| \leq c_3
\end{equation}
\end{corollary}
\begin{proof}
Choosing $e'$ as in the proof of Lemma \ref{lemma1},
\[(\hel-b)^{\frac{\gamma}{2}} 
(H_f+1)^{\frac{\beta}{2}} (H_{f,(\tilde{r},r)}+1)^{\frac{\alpha}{2}} 
\one_{]-\infty,e'[}(H_0) \]
is a bounded operator. Due to the normalisation condition $\|\gs\|=1$
\begin{eqnarray*}
\lefteqn{\sup_{g \in U} \|(\hel-b)^{\frac{\gamma}{2}} 
(H_f+1)^{\frac{\beta}{2}}  (H_{f,(\tilde{r},r)}+1)^{\frac{\alpha}{2}} 
\one_{]-\infty,e'[}(H_0) \gs \| \leq }\\
&\leq & \| (\hel-b)^{\frac{\gamma}{2}} 
(H_f+1)^{\frac{\beta}{2}} (H_{f,(\tilde{r},r)}+1)^{\frac{\alpha}{2}} 
\one_{]-\infty,e'[}(H_0) \| 
\end{eqnarray*}
gives a uniform bound. Resolvent equation implies
\begin{eqnarray*}
\lefteqn{(H_g-E_g+1)^{-1}-(H_h-E_h+1)^{-1}=(H_h-E_h+1)^{-1}}\\
&=&\Big[(h-g)W^{(1)}+(h^2-g^2)W^{(2)}+E_g-E_h \Big](H_g-E_g+1)^{-1},
\end{eqnarray*}
so due to the relative bounds on the interaction and continuity of the
ground state energies, $g \mapsto (H_g-E_g+1)^{-1}$ is continuous.
Therefore the $g$-dependent terms in (\ref{gl5}) can be estimated 
uniform on $U$.
\end{proof}
\subsection{Preparation of initial states}
\begin{hypothesis} \label{H-3}
We start with a photon cloud 
$A=a^*(f_1) \cdots a^*(f_N)$ with smooth momentum distributions
$f_1,...,f_N \in C_0^\infty(\R^3 \backslash \{0\})$ of compact support
away from zero momentum. As in (\ref{gl1.3}), we then use the incoming 
scattering state
\[ A(t) \gs = e^{-itH_g} e^{itH_0} A e^{-itH_0} e^{itH_g} \gs =
e^{-itH_g} \prod_{j=1}^N a_{\lambda_j}^*(e^{-it\omega}f_j) e^{itH_g} \gs \]
in the limit $t \to \infty$ as initial state.
\end{hypothesis}

%
\section{Asymptotic expansions} \label{sec2}
\setcounter{equation}{0}
In this section we develop asymptotic expressions for the full
interacting dynamics applied to photon clouds plus ground state. For
this purpose, we define the free Heisenberg time evolution
\begin{equation} \label{gl3.1}
Z_t:=e^{-itH_0} Z e^{itH_0} 
\end{equation}
on the domain $\D(Z_t):=\{\Psi \in \hr: e^{itH_0} \Psi \in
\D(Z)\}$ of an operator $Z$ in $\hr$. Using this free time evolution,
we find an asymptotic expansion of $e^{-i\tau (H_g-E_g)} A(t) \gs$ for 
intermediate times $g^{-\alpha} < \tau < g^{-\beta}$ as $g \searrow 0$ 
with $0 < \alpha < \beta < 1$. 
\subsection{Rewriting the time evolution} \label{sec2.1}
As a part of this program, we have to supply two
kind of technical lemma. The first kind, allows us to rewrite 
$e^{-i\tau(H_g-E_g)} A(t) \gs$ in
terms of $A_{\tau} \gs$ plus an integral, where several commutators of the
free Heisenberg time evolution of the interaction $W$ and 
the photon cloud $A$ come into play. 
\begin{theorem} \label{Th1.1}
Let $\Lambda^{(1)}_{0,0}, \tilde{\Lambda}^{(1)}_{0,0}, \Lambda^{(2)}_{0,0}, 
\tilde{\Lambda}^{(2)}_{0,0} < \infty$,
$n \in \N$ and let (\ref{eq701}) be satisfied.
Let $Z \in L(\D(H_0^n),\D(H_0))$ be a bounded operator from
$\D(H_0^n)$ into $\D(H_0)$. Given $\tau \in \R$ and
$\Psi \in \Ran (H_g-i)^{-n}$ the map
\begin{equation} \label{eq120}
\begin{array}{rcl} h_{\tau,Z}: \R & \longrightarrow & \hr \\
s & \longmapsto & e^{-isH_g}e^{isH_0} Z_{\tau} e^{-isH_0} e^{isH_g} \Psi
\end{array}
\end{equation}
is differentiable with derivative
\begin{equation} \label{eq121}
h_{\tau,Z}'(s)=
-ie^{-isH_g} e^{isH_0} [W_s, Z_{\tau}] e^{-isH_0}e^{isH_g} \Psi.
\end{equation}
\end{theorem}
\begin{proof}
{From} the general definitions of sums and products of operators, we
conclude $\D(H_g^l) \subseteq \D(H_0^l)$ for all $l \in \N$. The
unitary groups $(e^{-isH_g})_{s \in \R}$ and $(e^{-isH_0})_{s \in \R}$
leave $\D(H_g^l)$ respectively $\D(H_0^l)$ invariant, hence we get
$e^{-isH_0} e^{isH_g} \Psi \in \D(H_0^n)$ and 
$Z_{\tau} e^{-isH_0} e^{isH_g} \Psi \in \D(H_0)$. {From} Lemma \ref{l1.1} 
we know $\D(H_g)=\D(H_0)$, so this subspace is invariant
under $e^{\pm isH_g}$ and $e^{\pm isH_0}$ and the function $h_{\tau,Z}$ is
well defined. We look at the restrictions of the operators $e^{\pm isH_0}$ and
$e^{\pm isH_g}$ to $\D(H_0)$ as bounded operators on $\D(H_0)$
and apply results on one parameter unitary groups, see \cite{Ru} 13.35
to get on $\D(H_0)=\D(H_g)$:
\[\frac{d}{ds}(e^{\pm isH_0})=\pm iH_0 e^{\pm isH_0}\]
\[\frac{d}{ds}(e^{\pm isH_g})=\pm iH_g e^{\pm isH_g}.\]
Now the chain rule implies
\[\frac{d}{ds}(e^{\mp is H_g} e^{\pm isH_0})=
\mp ie^{\mp isH_g} (H_g-H_0) e^{\pm isH_0}\]
\[\frac{d}{ds}(e^{\pm is H_0} e^{\mp isH_g})=
\pm ie^{\pm isH_0} (H_0-H_g) e^{\mp isH_g}\]
The assumption $Z \in L(\D(H_0^n),\D(H_0))$ implies
$Z_{\tau} \in L(\D(H_0^n),\D(H_0))$, hence for each 
$\Psi \in \Ran (H_g-i)^{-n}$,
the map
\[\begin{array}{rclcl}
h_{\tau,Z}:\R & \hspace{-1.5pt}\rightarrow \hspace{-1.5pt} & 
L(\D(H_0)) \times \hr & \rightarrow & \hr\\
s &\hspace{-1.5pt} \mapsto \hspace{-1.5pt} & 
(e^{-isH_g} e^{isH_0}, Z_{\tau} e^{-isH_0}e^{isH_g} \Psi) &
\mapsto & e^{-isH_g}e^{isH_0} Z_{\tau} e^{-isH_0} e^{isH_g} \Psi
\end{array} \]
is differentiable, see \cite{Di} 8.1.4, with derivative
\begin{eqnarray*} 
h_{\tau,Z}'(s)&=&
-i e^{-isH_g} (H_g-H_0) e^{isH_0} Z_{\tau} e^{-isH_0} e^{isH_g} \Psi+\\
&&+ i e^{-isH_g} e^{isH_0} Z_{\tau} e^{-isH_0} (H_g-H_0) e^{isH_g} \Psi\\
&=& -ie^{-isH_g} e^{isH_0} [(H_g-H_0)_s, Z_{\tau}] e^{-isH_0}e^{isH_g} \Psi=\\
&=& -ie^{-isH_g} e^{isH_0} [W_s, Z_{\tau}] e^{-isH_0}e^{isH_g} \Psi.\\[-1.4cm]  
\end{eqnarray*}
\end{proof}
\begin{corollary} \label{co1.1}
Under the assumptions of Theorem \ref{Th1.1} the time evolution is
\begin{eqnarray*}
e^{-i\tau H_g }Z(t)e^{i\tau H_g} \Psi&=&
Z_{\tau}\Psi - i \int\limits_0^{t+\tau} ds \,
e^{-isH_g}e^{isH_0}[W_s,Z_{\tau}]e^{-isH_0} e^{isH_g} \Psi \\
e^{-i\tau H_g} e^{i\tau H_0} Z e^{-i\tau H_0} e^{i\tau H_g} \Psi &=&
Z \Psi 
- i \int\limits_0^{\tau} ds \, e^{-isH_g} e^{isH_0} [W_s,Z] e^{-isH_0} e^{isH_g}
\Psi
\end{eqnarray*}
\end{corollary}
\begin{proof}
Using Theorem \ref{Th1.1}, the differentiability of $h_{\tau,Z}$ implies
\begin{eqnarray} 
e^{-i\tau H_g} Z(t) e^{i\tau H_g} \Psi  &=& 
e^{-i(t+\tau)H_g} e^{i(t+\tau)H_0} Z_{\tau}
e^{-i(t+\tau)H_0}e^{i(t+\tau)H_g} \Psi= \label{eq122} \\
&=&Z_{\tau} \Psi + h_{\tau,Z}(s)|_{s=0}^{s=t+\tau}=
Z_{\tau} \Psi + \int\limits_0^{t+\tau} ds h_{\tau,Z}'(s)=
\nonumber\\ 
&=&Z_{\tau} \Psi -i \int\limits_0^{t+\tau} ds 
e^{-isH_g}e^{isH_0}[W_s,Z_{\tau}] e^{-isH_0} e^{isH_g} \Psi\nonumber
\end{eqnarray}
and in the same way
\begin{eqnarray*}
e^{-i\tau H_g} e^{i\tau H_0} Z e^{-i\tau H_0} e^{i\tau H_g} \Psi &=&
Z \Psi + h_{0,Z}(s)|_{s=0}^{s=\tau}=\\
&=&Z \Psi
- i \int\limits_0^{\tau} ds \, e^{-isH_g} e^{isH_0} [W_s,Z] e^{-isH_0} e^{isH_g}
\Psi.
\nonumber \\[-1cm]  \nonumber 
\end{eqnarray*}
\end{proof}
\vspace*{0.3cm}
\subsection{Commutator estimates} \label{sec2.2}
The second kind of lemma, which we are going to prove now, establishes
some control on the time decay of the commutators in Corollary
\ref{co1.1}. These results are needed for an error bound of the
asymptotic expansions.
\begin{lemma} \label{l2.1}
Suppose Hypothesis \ref{H-1}, \ref{hyp4}, \ref{H-3} and
$(\hel,\gamma)$ hold true and that 
$\Lambda^{(1)}_{0,\gamma}, \Lambda^{(1)}_{\beta,\gamma},
\tilde{\Lambda}^{(1)}_{\beta,\gamma} < \infty$.
Let
\begin{eqnarray}
\tilde{r} &<& \inf \{\omega(k): k \in \supp f_j, j=1,...,N\} \nonumber\\
r &>& \sup\{\omega(k): k \in \supp f_j, j=1,...,N\} \label{gl201}
\end{eqnarray}
and
$\lambda_1,...,\lambda_N \in \Z_2$, 
then for $A=a^*_{\lambda_1}(f_1) \cdots a^*_{\lambda_N}(f_N)$ and all 
$s \in \R$
\[(H_f+1)^{\frac{\beta}{2}} (\hel-b)^{\frac{\gamma}{2}} 
[W^{(1)},A_{s}](\hel-b)^{-\frac{\gamma+1}{2}} 
(H_f+1)^{-\frac{\beta+1}{2}} (H_{f,(\tilde{r},r)}+1)^{-\frac{N}{2}} \]
defines a bounded operator on $\hr$ and there is 
$c_4=c_4(\beta,\gamma) < \infty$, which can be chosen independent of $s$, 
such that
\begin{eqnarray} 
&& \hspace{-0.5cm}\| (H_f+1)^{\frac{\beta}{2}} (\hel-b)^{\frac{\gamma}{2}} 
[W^{(1)},A_{s}](\hel-b)^{-\frac{\gamma+1}{2}} 
(H_f+1)^{-\frac{\beta+1}{2}} (H_{f,(\tilde{r},r)}+1)^{-\frac{N}{2}} \| \nonumber \\
&&  \hspace{1.5cm} \leq c_4(1+|s|)^{-\zeta}. \label{eq123} 
\end{eqnarray}
\end{lemma}
\begin{proof}
We apply the pull-through formula to obtain the free time evolution of
$A$:
\begin{equation} \label{eq34}
A_{s}=a^*_{\lambda_1}(e^{-is\omega} f_1) \cdots 
a^*_{\lambda_N}(e^{-is\omega}f_N), 
\end{equation}
then for 
$\Phi \in \Ran (\hel-b)^{-\frac{\gamma+1}{2}}  
(H_f+1)^{-\frac{\beta+1}{2}}(H_{f,(\tilde{r},r)}+1)^{-\frac{N}{2}}$ all terms 
\begin{eqnarray*}  
&&\hspace{-1cm} (H_f+1)^{\frac{\beta}{2}} (\hel-b)^{\frac{\gamma}{2}}
W^{(1)} A_{s} \Phi= \\
&=& \Big( (H_f+1)^{\frac{\beta}{2}} (\hel-b)^{\frac{\gamma}{2}}
W^{(1)} (\hel-b)^{-\frac{\gamma+1}{2}}(H_f+1)^{-\frac{\beta+1}{2}}\Big) 
\\[-0.3cm]
&& \Big((H_f+1)^{\frac{\beta+1}{2}} 
\prod_{j=1}^N a^*_{\lambda_j}(e^{-is\omega} f_j)
(H_f+1)^{-\frac{\beta+1}{2}}(H_{f,(\tilde{r},r)}+1)^{-\frac{N}{2}} \Big)\\[-0.2cm] 
&& (\hel-b)^{\frac{\gamma+1}{2}}(H_f+1)^{\frac{\beta+1}{2}}
(H_{f,(\tilde{r},r)}+1)^{\frac{N}{2}} \Phi \\[0.3cm] 
&& \hspace{-1cm}
(H_f+1)^{\frac{\beta}{2}} (\hel-b)^{\frac{\gamma}{2}}A_{s} W^{(1)} \Phi= 
\\[-0.2cm]
&=&\Big(
(H_f+1)^{\frac{\beta}{2}} \prod_{j=1}^N a^*_{\lambda_j}(e^{-is\omega} f_j)
(H_f+1)^{-\frac{\beta}{2}}(H_{f,(\tilde{r},r)}+1)^{-\frac{N}{2}}\Big) \\[-0.1cm]
&& \Big((H_{f,(\tilde{r},r)}+1)^{\frac{N}{2}}
(H_f+1)^{\frac{\beta}{2}} (\hel-b)^{\frac{\gamma}{2}}
W^{(1)} (\hel-b)^{-\frac{\gamma+1}{2}}(H_f+1)^{-\frac{\beta+1}{2}}\\
&& (H_{f,(\tilde{r},r)}+1)^{-\frac{N}{2}}\Big)
(\hel-b)^{\frac{\gamma+1}{2}}(H_f+1)^{\frac{\beta+1}{2}}
(H_{f,(\tilde{r},r)}+1)^{\frac{N}{2}} \Phi 
\end{eqnarray*} 
are well defined by Corollary \ref{coa2} and Lemma \ref{la2}. 
So the commutator is written as
\begin{eqnarray} 
&& \hspace{-1cm}
(H_f+1)^{\frac{\beta}{2}} (\hel-b)^{\frac{\gamma}{2}} 
[W^{(1)},A_{s}](\hel-b)^{-\frac{\gamma+1}{2}} 
(H_f+1)^{-\frac{\beta+1}{2}} (H_{f,(\tilde{r},r)}+1)^{-\frac{N}{2}} \Psi\nonumber  \\
&=&\sum_{\lambda \in \Z_2} \;\int\limits_{\R^3} dk \int\limits_{\R^3} dk_1
\cdots \int\limits_{\R^3} dk_N  e^{-is\omega(k_1)} f_1(k_1) \cdots
e^{-is\omega(k_N)} f_N(k_N) \nonumber \\
&& \hspace{0.3cm} \Big\{ (\hel-b)^{\frac{\gamma}{2}}
 w^{(1,0)}(k,\lambda) (\hel-b)^{-\frac{\gamma+1}{2}} 
\nonumber\\[-0.3cm]
&& \hspace{0.6cm} (H_f+1)^{\frac{\beta}{2}}
\Big[ a^*_{\lambda}(k), \prod_{j=1}^N a^*_{\lambda_j}(k_j) \Big]
(H_f+1)^{-\frac{\beta+1}{2}}+ \nonumber \\
&&\hspace{0.6cm} +(\hel-b)^{\frac{\gamma}{2}} w^{(0,1)}(k,\lambda) 
(\hel-b)^{-\frac{\gamma+1}{2}} \nonumber\\
&&\hspace{0.6cm} (H_f+1)^{\frac{\beta}{2}}
\Big[ a_{\lambda}(k), \prod_{j=1}^N a^*_{\lambda_j}(k_j) \Big]
(H_f+1)^{-\frac{\beta+1}{2}}
\Big\} (H_{f,(\tilde{r},r)}+1)^{-\frac{N}{2}}\Psi \label{eq35}
\end{eqnarray}
Inserting the two commutators
\begin{eqnarray*}
\big[ a^*_{\lambda}(k), \prod_{j=1}^N a^*_{\lambda_j}(k_j) \big] &=& 0 
\label{eq36} \\
\big[ a_{\lambda}(k), \prod_{j=1}^N a^*_{\lambda_j}(k_j) \big] &=&
\sum_{j=1}^N \delta(k-k_j) \delta_{\lambda,\lambda_j}
\prod_{l=1 \atop l \not = j}^N a^*_{\lambda_l }(k_l)
\label{eq37}
\end{eqnarray*}
into (\ref{eq35}), we get
\begin{eqnarray}  
&& \hspace{-1cm} (H_f+1)^{\frac{\beta}{2}} (\hel-b)^{\frac{\gamma}{2}} 
[W^{(1)},A_{s}](\hel-b)^{-\frac{\gamma+1}{2}} 
(H_f+1)^{-\frac{\beta+1}{2}} (H_{f,(\tilde{r},r)}+1)^{-\frac{N}{2}} \Psi \nonumber \\
&=&
\sum_{j=1}^N \int\limits_{\R^3} dk (\hel-b)^{\frac{\gamma}{2}}
w^{(0,1)}(k,\lambda_j) (\hel-b)^{-\frac{\gamma+1}{2}} 
e^{-is\omega(k)} f_j(k) \nonumber\\[-0.6cm]
&& \hspace{1.2cm}(H_f+1)^{\frac{\beta}{2}}
\prod_{l=1 \atop l \not = j}^N
a^*_{\lambda_l}(e^{-is\omega}f_l) (H_f+1)^{-\frac{\beta+1}{2}} 
(H_{f,(\tilde{r},r)}+1)^{-\frac{N}{2}} \Psi \label{eq38}
\end{eqnarray}
{From} Corollary \ref{coa2} we conclude, that 
\[\Psi_{j,s}:=(H_f+1)^{\frac{\beta}{2}} \prod_{l=1 \atop l \not=j}^N
a_{\lambda_l}^*(e^{-is\omega}f_l) (H_f+1)^{-\frac{\beta+1}{2}}
(H_{f,(\tilde{r},r)}+1)^{-\frac{N}{2}} \Psi \]
is a well defined element of $\F$ with
$\displaystyle \sup_{s \in \R \atop j=1,...,N} \|\Psi_{j,s}\| < 
c(\|f_1\|_{\omega}, ... ,\|f_N\|_{\omega}) \|\Psi\|$, with a finite constant
$c(\|f_1\|_{\omega}, ... ,\|f_N\|_{\omega})$ depending only on 
the wighted $L^2$ norms given by
\[\|f_j\|_{\omega}^2:= \int\limits_{\R^3} |f_j(k)|^2 (1+\frac{1}{\omega(k)}) 
dk, \] 
which are finite for $f_j \in C_0^{\infty}(\R^3 \backslash \{0\})$.
Due to Hypothesis $(\hel,\gamma)$
\[T(k,\lambda):=(\hel-b)^{\frac{\gamma}{2}} w^{(0,1)}(k,\lambda) 
(\hel-b)^{-\frac{\gamma+1}{2}}\]
and all partial $k$-derivatives of order $\leq \zeta$ are square integrable
on each compactum $K \subseteq \R^3 \backslash \{0\}.$
Then the commutator takes the form
\begin{eqnarray}
&& \hspace{-1cm}(H_f+1)^{\frac{\beta}{2}} (\hel-b)^{\frac{\gamma}{2}} 
[W^{(1)},A_{s}](\hel-b)^{-\frac{\gamma+1}{2}} 
(H_f+1)^{-\frac{\beta+1}{2}} (H_{f,(\tilde{r},r)}+1)^{-\frac{N}{2}} \Psi \nonumber \\
&=& \sum_{j=1}^N \int\limits_{\R^3} dk e^{-is\omega} f_j(k) T(k,\lambda_j)
\Psi_{j,s} \label{eq750}
\end{eqnarray}
The support of $f_j$ is located away from the origin.
So $\nabla_k \omega(k)= \frac{k}{|k|}$ and \linebreak 
$\frac{i}{s} \frac{k}{|k|} \cdot \nabla_k e^{-is\omega(k)}= e^{-is\omega(k)}$
on $\supp f_j$. For $\Phi \in  \hr$ by $\zeta$ times  partial integration,
\begin{eqnarray} \lefteqn{
\Big| \Big \lkl \Phi, \sum_{j=1}^ N \int\limits_{\R^3} dk 
e^{-is\omega(k)} T(k,\lambda_j) f_j(k) \Psi_{j,s} \Big \rkl \Big|=\nonumber} 
\\
&=& \Big| \sum_{j=1}^ N \int\limits_{\R^3} dk \Big \lkl \Phi,
\Big( \Big[\frac{i}{s} \frac{k}{|k|} \cdot \nabla_k\Big]^{\zeta}
e^{-is\omega(k)} \Big) T(k,\lambda_j) f_j(k) \Psi_{j,s} \Big \rkl \Big| = 
\nonumber\\
&=& \frac{1}{s^{\zeta}} \Big| \sum_{j=1}^N \int\limits_{\R^3} dk 
\Big \lkl \Phi,
e^{-is\omega(k)} \Big[\nabla_k \frac{k}{|k|} \Big]^{\zeta} 
\big(T(k,\lambda_j) f_j(k) \big) 
\Psi_{j,s}  \Big \rkl \Big | \leq \nonumber \\
&\leq& \frac{1}{s^{\zeta}} \|\Phi \| \sum_{j=1}^N \|\Psi_{j,s}\| 
\int\limits_{\supp f_j} dk \|\Big[\nabla_k \frac{k}{|k|} \Big]^{\zeta} 
(T(k,\lambda_j) f_j(k))\| \label{eq521}
\end{eqnarray}
$\Big[\nabla_k \frac{k}{|k|} \Big]^{\zeta} (T(k,\lambda) f_j(k))$
is a sum of
$(\hel-b)^{\frac{\gamma}{2}} \nabla_k^{\alpha_1} w^{(0,1)}(k,\lambda) 
(\hel-b)^{-\frac{\gamma+1}{2}} $
multiplied with some derivatives 
$(\nabla_k^{\alpha_2} \frac{k}{|k|})(\nabla_k^{\alpha_3}f_j)$ for
$|\alpha_1|,|\alpha_2|,|\alpha_3| \leq \zeta$. 
So all these terms are integrable on the support of $f_j$ and 
\[ \Big| \Big \lkl \Phi, \sum_{j=1}^N \int\limits_{\R^3} dk 
e^{-is\omega(k)} T(k,\lambda_j) f_j(k)
\Psi_{j,s} \Big \rkl \Big| \leq c_4(1+|s|)^{-\zeta} \|\Psi\| \|\Phi\| \]
for some $c_4$ depending on $\beta, \gamma$,
$\|f_1\|_{\omega},...,\|f_N\|_{\omega}$,
$\|(\nabla_k^{\alpha_2} \frac{k}{|k|})(\nabla_k^{\alpha_3}f_j)\|_{L^2(\supp f_j)}$
and 
$\|(\hel-b)^{\frac{\gamma}{2}} \nabla_k^{\alpha_1} w^{(0,1)}(\cdot,\lambda) 
(\hel-b)^{-\frac{\gamma+1}{2}}\|_{L^2(\supp f_j)}$
for
$|\alpha_1|,|\alpha_2|,|\alpha_3| \leq \zeta$, $j=1,...,N$
but chosen independent of $s \in \R$. 
\end{proof}
\begin{remark}
\end{remark} \vspace*{-0.2cm} 
\noindent In the proof we have seen, that the requirement 
$f_j \in C_0^{\infty}(\R^3 \backslash \{0\})$ could be relaxed in several 
aspects:
\begin{itemize}
\item Work with $f_j \in C_0^{\zeta}(\R^3 \backslash \{0\})$.
\item If $\omega$ is smooth on $\R^3$, e.g. by introducing a photon rest mass
and working with
$\omega(k)=\sqrt{k^2 +m^2}$, then we could allow $f_j|_U \not =0$ on every
neighbourhood $U$ of $0$.
\item If we want to get rid of the compact support of $f_j$, we have to impose
suitable integration conditions on $T(k,\lambda)$, $f_j$, $\omega(k)$ and its
derivatives plus some extra boundary conditions at $\infty$, 
to get a finite bound in (\ref{eq521}).
\end{itemize}
All those points would make life much more complicated, which we want to avoid.
\begin{lemma} \label{l1.2}
Suppose Hypothesis \ref{H-1}, \ref{hyp5}, \ref{hyp4}, \ref{H-3} 
hold true and $\Lambda^{(2)}_{0,0}, \tilde{\Lambda}^{(2)}_{0,0}  < \infty$.
Let $f_1,...,f_N \in C_0^{\infty}(\R^3 \backslash \{0\})$,
$\lambda_1,...,\lambda_N \in \Z_2$ and choose $\tilde{r},r$ as in (\ref{gl201}),
then for $A=a^*_{\lambda_1}(f_1) \cdots a^*_{\lambda_N}(f_N) $ 
\[[W^{(2)},A_{s}] (H_f+1)^{-1} (H_{f,(\tilde{r},r)}+1)^{-\frac{N}{2}} \]
defines a bounded operator on $\hr$ and there is a constant
$c_5 < \infty$, which can be chosen independent of $s \in \R$, such that
\begin{equation} 
\| [W^{(2)},A_{s}] (H_f+1)^{-1} (H_{f,(\tilde{r},r)}+1)^{-\frac{N}{2}} \| 
\leq c_5 (1+|s|)^{-\zeta} \label{eq751} 
\end{equation}
\end{lemma}
\begin{proof}
Separating electron and photon part for any $\Phi \in \Hel \otimes \F$, 
the commutator can be written as
\begin{eqnarray}
&& \hspace{-1.2cm} 
[W^{(2)},A_{s}]
(H_f+1)^{-1} (H_{f,(\tilde{r},r)}+1)^{-\frac{N}{2}}  \Phi \label{eq12} \\
&=& \sum_{\lambda,\lambda' \in \Z_2} \int\limits_{\R^3}
\int\limits_{\R^3} dk dk' \Bigg\{ 
w^{(2,0)}(k,\lambda;k',\lambda')  \Big[a_{\lambda}^*(k)a_{\lambda'}^*(k'), 
\prod_{j=1}^N a_{\lambda_j}^{*}(e^{-is\omega} f_j)\Big] 
\nonumber \\
&& \hspace{3.3cm} + 
w^{(1,1)}(k,\lambda;k',\lambda')  \Big[a_{\lambda}^*(k)a_{\lambda'}(k'),
\prod_{j=1}^N a_{\lambda_j}^{*}(e^{-is\omega} f_j)\Big] 
\nonumber \\
&& \hspace{3.3cm} + 
w^{(0,2)} (k,\lambda;k',\lambda')  \Big[ a_{\lambda}(k)a_{\lambda'}(k'),
\prod_{j=1}^N a_{\lambda_j}^{*}(e^{-is\omega} f_j)\Big] \Bigg\}
\nonumber\\
&&\hspace{4cm}(H_f+1)^{-1} (H_{f,(\tilde{r},r)}+1)^{-\frac{N}{2}} \Phi. \nonumber
\end{eqnarray}
The commutator relations
\begin{eqnarray}
&&\Big[a_{\lambda}^*(k)a_{\lambda'}^*(k'), 
\prod_{j=1}^N a_{\lambda_j}^{*}(e^{-is\omega} f_j)\Big]=0 \label{eq13} \\
&& \Big[a_{\lambda}^*(k)a_{\lambda'}(k'), 
\prod_{j=1}^N a_{\lambda_j}^*(e^{-is\omega} f_j)\Big]= \label{eq14} \\[-0.3cm]
&=& a_{\lambda}^*(k) \sum_{j=1}^N \delta_{\lambda' \lambda_j}
e^{-is\omega(k')} f_j(k') 
\prod_{l=1 \atop l \not = j}^N a_{\lambda_l}^*(e^{-is\omega} f_l)
\nonumber
\end{eqnarray}
\begin{eqnarray}
\lefteqn{ \Big[a_{\lambda}(k)a_{\lambda'}(k'), 
\prod_{j=1}^N a_{\lambda_j}^{*}(e^{-is\omega} f_j)\Big]= \label{eq15}} \\
&=& a_{\lambda}(k) \sum_{j=1}^N \delta_{\lambda' \lambda_j}
e^{-is\omega(k')} f_j(k') 
\prod_{l=1 \atop l \not = j}^N a_{\lambda_l}^*(e^{-is\omega} f_l) +
\nonumber \\
&& +a_{\lambda'}(k') \sum_{j=1}^N \delta_{\lambda \lambda_j}
e^{-is\omega(k)} f_j(k) 
\prod_{l=1 \atop l \not = j}^N a_{\lambda_l}^*(e^{-is\omega} f_l) 
\nonumber\\
&& -\sum_{j=1}^N \delta_{\lambda \lambda_j} e^{-is\omega(k)} f_j(k) 
\sum_{l=1 \atop l \not =j}^N \delta_{\lambda_l,\lambda'} e^{-is\omega(k')}
f_l(k')
\prod_{m=1 \atop m \not = j,l}^N a_{\lambda_m}^*(e^{-is\omega} f_m) 
\nonumber
\end{eqnarray}
simplify (\ref{eq12}). If $\fG(k,\lambda)^{\#}$ denotes either 
$\fG(k,\lambda)$ or its adjoint $\fG(k,\lambda)^*$, then 
$\int\limits_{\R^3}dk \fG_{\iota}(k,\lambda)^{\#} e^{-is\omega(k)} f_j(k)$
defines a bounded operator on $\Hel$, so after inserting
$\one=(H_f+1)^{-\frac{1}{2}} (H_f+1)^{\frac{1}{2}}$ and commutation
\begin{eqnarray}
&& \hspace{-0.7cm}
[W^{(2)},A_{s}] (H_f+1)^{-1} (H_{f,(\tilde{r},r)}+1)^{-\frac{N}{2}} \Phi
\label{eq16} \\
&\hspace{-0.3cm}=& \sum_{j=1}^N \sum_{\iota=1}^3 
\sum_{\lambda \in \Z_2} \int\limits_{\R^3} dk 
\fG_{\iota}(k,\lambda)^* a^*_{\lambda}(k) (H_f+1)^{-\frac{1}{2}} \hspace{-2pt}
\int\limits_{\R^3}  dk' 
\fG_{\iota}(k',\lambda_j) e^{-is\omega(k')} f_j(k') 
\nonumber\\[-0.4cm]
&& \hspace{2.8cm} (H_f+1)^{\frac{1}{2}}
\prod_{l=1 \atop l \not = j}^N a_{\lambda_l}^*(e^{-it\omega} f_l) 
(H_f+1)^{-1} (H_{f,(\tilde{r},r)}+1)^{-\frac{N}{2}} \Phi
\nonumber \\[-0.2cm]
&&\hspace{-0.3cm}+ \sum_{j=1}^N \sum_{\iota=1}^3 
\sum_{\lambda \in \Z_2} \int\limits_{\R^3} dk 
\fG_{\iota}(k,\lambda) a^*_{\lambda}(k) (H_f+1)^{-\frac{1}{2}}\hspace{-2pt}
\int\limits_{\R^3} \hspace{-2pt} dk' 
\fG_{\iota}(k',\lambda_j)^* e^{-is\omega(k')} f_j(k')
\nonumber\\[-0.4cm]
&& \hspace{3.2cm} (H_f+1)^{\frac{1}{2}}
\prod_{l=1 \atop l \not = j}^N a_{\lambda_l}^*(e^{-is\omega} f_l) 
(H_f+1)^{-1} (H_{f,(\tilde{r},r)}+1)^{-\frac{N}{2}} \Phi
\nonumber \\[-0.2cm]
&&\hspace{-0.3cm}+ \sum_{j=1}^N \sum_{\iota=1}^3 
\sum_{\lambda \in \Z_2} 
\int\limits_{\R^3} \hspace{-2pt} dk' 
\fG_{\iota}(k',\lambda_j)^* e^{-is\omega(k')} f_j(k') \hspace{-3pt}
\int\limits_{\R^3} \hspace{-2pt} dk \fG_{\iota}(k,\lambda)^*
a_{\lambda}(k) (H_f+1)^{-\frac{1}{2}}
\nonumber\\[-0.4cm]
&& \hspace{3.2cm} (H_f+1)^{\frac{1}{2}}
\prod_{l=1 \atop l \not = j}^N a_{\lambda_l}^*(e^{-is\omega} f_l) 
(H_f+1)^{-1} (H_{f,(\tilde{r},r)}+1)^{-\frac{N}{2}} \Phi
\nonumber \\
&&\hspace{-0.3cm}+ \sum_{j=1}^N \sum_{\iota=1}^3 
\sum_{\lambda \in \Z_2} \int\limits_{\R^3} dk' 
\fG_{\iota}(k',\lambda)^*
a_{\lambda}(k') (H_f+1)^{-\frac{1}{2}} \hspace{-2pt}
\int\limits_{\R^3} \hspace{-2pt} dk
\fG_{\iota}(k,\lambda_j)^* e^{-is\omega(k)} f_j(k) 
\nonumber\\[-0.2cm]
&& \hspace{2.4cm} (H_f+1)^{\frac{1}{2}}
\prod_{l=1 \atop l \not = j}^N a_{\lambda_l}^*(e^{-is\omega} f_l) 
(H_f+1)^{-1} (H_{f,(\tilde{r},r)}+1)^{-\frac{N}{2}} \Phi
\nonumber \\
&&- \sum_{j=1}^N \sum_{\iota=1}^3 
\sum_{l=1 \atop l \not=j}^N 
\int\limits_{\R^3} dk' \fG_{\iota}(k',\lambda_l)^* 
e^{-is\omega(k')}f_l(k')
\int\limits_{\R^3} \hspace{-2pt} dk
\fG_{\iota}(k,\lambda_j)^* e^{-is\omega(k)} f_j(k) 
 \nonumber\\[-0.2cm]
&& \hspace{2.2cm}  
\prod_{m=1 \atop m \not = j,l}^N a_{\lambda_m}^*(e^{-is\omega} f_m) 
(H_f+1)^{-1} (H_{f,(\tilde{r},r)}+1)^{-\frac{N}{2}}
\Phi.
\nonumber 
\end{eqnarray}
Now as in the proof of Lemma \ref{l2.1} by non stationary phase method each 
expression of the form 
$\int\limits_{\R^3}dk \fG_{\iota}(k,\lambda)^{\#} e^{-is\omega(k)} f_j(k)$
defines a bounded operator on $\Hel$, which has norm of order
$\cO(1+|s|)^{-\zeta}$. The estimates 
\begin{eqnarray}
\|\sum_{\lambda \in \Z_2} \int\limits_{\R^3} dk \fG_{\iota}(k,\lambda)^{\#} 
a_{\lambda}(k) \Psi \|
&\leq&  \Big( \sum_{\lambda \in \Z_2} \int\limits_{\R^3} dk 
\frac{\|\fG_{\iota}(k,\lambda) \|^2}{\omega(k)} \Big)^{\frac{1}{2}}
\|H_f^{\frac{1}{2}} \Psi \|^2 \leq \nonumber\\
&\leq& \sqrt{\Lambda^{(2)}_{0,0}} \|H_f^{\frac{1}{2}} \Psi \| \label{eq541}
\end{eqnarray}
\begin{eqnarray}
\lefteqn{
\|\sum_{\lambda \in \Z_2} \int\limits_{\R^3} dk \fG_{\iota}(k,\lambda)^{\#} 
a^*_{\lambda}(k) \Psi \|^2 =
\sum_{\lambda \in \Z_2} \int\limits_{\R^3} dk 
\| \fG_{\iota}(k,\lambda) \Psi \|^2 +\label{eq542}}\\
&=& \sum_{\lambda,\lambda' \in \Z_2} \int\limits_{\R^3} dk \int\limits_{\R^3}
dk' \Big\lkl a_{\lambda'}(k') \fG_{\iota}(k,\lambda)^{\#} \Psi, 
a_{\lambda}(k) 
\fG_{\iota}(k',\lambda')^{\#} \Psi \Big \rkl \nonumber \\
&\leq& 
\sum_{\lambda,\lambda' \in \Z_2} \int\limits_{\R^3} dk \int\limits_{\R^3}
dk' \|\fG_{\iota}(k,\lambda)\|\|a_{\lambda'}(k') \Psi \|
\|\fG_{\iota}(k',\lambda')\|\|a_{\lambda}(k) \Psi \|+\nonumber\\
&&+ \sum_{\lambda \in \Z_2} \int\limits_{\R^3} dk 
\| \fG_{\iota}(k,\lambda) \Psi \|^2 
\leq  \Lambda^{(2)}_{0,0} (\|H_f^{\frac{1}{2}} \Psi \|^2 + \|\Psi\|^2)=\nonumber\\
&=&\Lambda^{(2)}_{0,0}\|(H_f+1)^{\frac{1}{2}} \Psi\|^ 2 \nonumber
\end{eqnarray}
prove, that 
$\displaystyle \sum_{\lambda \in \Z_2} \int\limits_{\R^3} dk 
\fG(k,\lambda)^{\#} a_{\lambda}^{\#}(k) (H_f+1)^{-\frac{1}{2}}$ 
define bounded operators.
So Corollary \ref{coa2}, which implies 
\[ \displaystyle \sup_{s \in \R} 
\Big\|(H_f+1)^{\frac{1}{2}} \prod_{l=1 \atop l \not = j}^N 
a_{\lambda_l}^*(e^{-is\omega} f_l) (H_f+1)^{-1} (H_{f,(\tilde{r},r)}+1)^{-\frac{N}{2}} 
\Phi \Big\| < \infty \]
finishes the proof.
\end{proof}
\begin{corollary} \label{co2.5}
Under the hypothesis of Lemma \ref{l2.1} and \ref{l1.2} and with the 
constants $c_4,c_5$ from there, for any $(s,t) \in \R^2$
\begin{eqnarray} 
&&\hspace{-1cm}\| (H_f+1)^{\frac{\beta}{2}} (\hel-b)^{\frac{\gamma}{2}} 
[W^{(1)}_t,A_{s}](\hel-b)^{-\frac{\gamma+1}{2}} 
(H_f+1)^{-\frac{\beta+1}{2}} (H_{f,(\tilde{r},r)}+1)^{-\frac{N}{2}} \|
\nonumber \\
&& \hspace{1cm}\leq c_4(1+|t-s|)^{-\zeta} \label{gl4} 
\end{eqnarray}
\begin{equation}
\| [W^{(2)}_t,A_{s}] (H_f+1)^{-1} (H_{f,(\tilde{r},r)}+1)^{-\frac{N}{2}} \| 
\leq c_5 (1+|t-s|)^{-\zeta}
\end{equation}
\end{corollary}
\begin{proof}
$e^{itH_0} [W^{(j)}_t,A_s]e^{-itH_0}=[W^{(j)}_t,A_s]_{-t}=[W^{(j)},A_{s-t}]$ 
and $H_0$ commutes with 
$\hel$, $H_f$ and $H_{f,(\tilde{r},r)}$, so due to unitary of $e^{\pm itH_0}$
\begin{eqnarray*}
\lefteqn{\| [W^{(2)}_t,A_{s}] (H_f+1)^{-1} (H_{f,(\tilde{r},r)}+1)^{-\frac{N}{2}} \|=} \\
&=& \| e^{itH_0} [W^{(2)}_t,A_{s}] (H_f+1)^{-1} (H_{f,(\tilde{r},r)}+1)^{-\frac{N}{2}} 
e^{-itH_0} e^{itH_0}\| \leq \\
&\leq& \| [W^{(2)},A_{s-t}] (H_f+1)^{-1} (H_{f,(\tilde{r},r)}+1)^{-\frac{N}{2}} \| \leq
c_5 (1+|t-s|)^{-\zeta},
\end{eqnarray*}
the estimate for the $W^{(1)}$ commutator is proven the same way.
\end{proof}
\begin{corollary} \label{l2.2}
Under the hypothesis \ref{H-1}, \ref{hyp5}, \ref{hyp4}, \ref{H-3}, 
$(\hel,1,1)$ and $(\hel,1)$ there is
$c_6 < \infty$, which can be chosen independent of $(s,t) \in \R^2$, such that
\begin{eqnarray*} 
\|[W,[W^{(1)}_{t},A_{s}]] (\hel-b)^{-1} (H_{f,(\tilde{r},r)}+1)^{-\frac{N}{2}} 
(H_f+1)^{-\frac{3}{2}}\| \hspace{-6pt}
&\leq& \hspace{-8pt} c_6(1+|t-s|)^{-\zeta} \label{eq50} \\
\|[W^{(1)},[W^{(1)}_{t},A_{s}]] (\hel-b)^{-1} (H_{f,(\tilde{r},r)}+1)^{-\frac{N}{2}} 
(H_f+1)^{-1}\| \hspace{-6pt}
&\leq& \hspace{-8pt}c_6(1+|t-s|)^{-\zeta} \label{eq705}\\
\|[W^{(2)},[W^{(1)}_{t},A_{s}]] (\hel-b)^{-\frac{1}{2}}
(H_{f,(\tilde{r},r)}+1)^{-\frac{N}{2}} (H_f+1)^{-\frac{3}{2}}\| \hspace{-6pt}
&\leq& \hspace{-8pt} c_6(1+|t-s|)^{-\zeta} \label{eq706} 
\end{eqnarray*}
\end{corollary}
\begin{proof}
Inserting identity in form of positive and negative powers of $H_f+1$,
$H_{f,(\tilde{r},r)}+1$ and $\hel-b$
\begin{eqnarray*}
\lefteqn{
[W^{(1)},[W^{(1)}_{t},A_{s}]] (\hel-b)^{-1} (H_{f,(\tilde{r},r)}+1)^{-\frac{N}{2}} 
(H_f+1)^{-1}=}\\
&=& \Big( W^{(1)} (\hel-b)^{-\frac{1}{2}} (H_f+1)^{-\frac{1}{2}} \Big)
\Big( (\hel-b)^{\frac{1}{2}} (H_f+1)^{\frac{1}{2}} [W^{(1)}_{t},A_{s}] \\
&& \hspace{0.5cm} (\hel-b)^{-1} (H_{f,(\tilde{r},r)}+1)^{-\frac{N}{2}} (H_f+1)^{-1} \Big)-
\\
&&- \Big( [W^{(1)}_{t},A_{s}] (\hel-b)^{-\frac{1}{2}} 
(H_{f,(\tilde{r},r)}+1)^{-\frac{N}{2}} (H_f+1)^{-\frac{1}{2}} \Big) 
\Big( (\hel-b)^{\frac{1}{2}} \\
&& \hspace{0.4cm} 
(H_{f,(\tilde{r},r)}+1)^{\frac{N}{2}} 
(H_f+1)^{\frac{1}{2}}
W^{(1)} (\hel-b)^{-1} (H_{f,(\tilde{r},r)}+1)^{-\frac{N}{2}} (H_f+1)^{-1} \Big)
\end{eqnarray*}
so the commutator terms are estimated by Corollary \ref{co2.5} and the
$W^{(1)}$ terms by Lemma \ref{la2}. Using Lemma \ref{la4} for the
$W^{(2)}$ terms of the following equation
\begin{eqnarray*}
\lefteqn{
[W^{(2)},[W^{(1)}_{t},A_{s}]] (\hel-b)^{-1} (H_{f,(\tilde{r},r)}+1)^{-\frac{N}{2}} 
(H_f+1)^{-1}=}\\
&=& \Big( W^{(2)} (H_f+1)^{-1} \Big)
\Big( (H_f+1) [W^{(1)}_{t},A_{s}] \\
&& \hspace{0.5cm} (\hel-b)^{-\frac{1}{2}} (H_{f,(\tilde{r},r)}+1)^{-\frac{N}{2}} 
(H_f+1)^{-\frac{3}{2}} \Big)-
\\
&&- \Big( [W^{(1)}_{t},A_{s}] (\hel-b)^{-\frac{1}{2}} 
(H_{f,(\tilde{r},r)}+1)^{-\frac{N}{2}} (H_f+1)^{-\frac{1}{2}} \Big) 
\Big( (\hel-b)^{\frac{1}{2}} \\
&& \hspace{0.4cm} 
(H_{f,(\tilde{r},r)}+1)^{\frac{N}{2}} (H_f+1)^{\frac{1}{2}}
W^{(2)} (\hel-b)^{-1} (H_{f,(\tilde{r},r)}+1)^{-\frac{N}{2}} (H_f+1)^{-1} \Big)
\end{eqnarray*}
the claim follows
\end{proof}
\subsection{Existence of incoming scattering states} \label{sec2.3}
\begin{lemma}
Suppose Hypothesis \ref{H-1}, \ref{hyp5}, \ref{hyp4} and \ref{H-2} and
let $\displaystyle A=\prod_{j=1}^N a^*_{\lambda_j}(f_j)$ 
be as in Hypothesis \ref{H-3}, then
\[A(\infty) \gs = \lim_{t \to \infty} A(t) \gs \]
exists.
\end{lemma}
\begin{proof}
By Corollary \ref{co1.1} for $t,s \in \R$, $s \leq t$:
\[ \|A(t) \gs - A(s) \gs \| =
\Big\| \int\limits_s^t dq e^{-iq(H_g-E_g)}[W,A_{-q}] \gs \Big\| \leq 
\int\limits_s^t \|[W,A_{-q}] \gs \| dq\]
Choose $\tilde{r},r$ as in (\ref{gl201}).
An application of the commutator estimates from Corollary \ref{l2.2} and
Corollary \ref{co1.3} imply:
\begin{eqnarray*} 
\int\limits_s^t \|[W,A_{-q}] \gs \| dq \hspace{-4pt} &\leq& \hspace{-4pt}
\|(\hel-b)^{\frac{1}{2}} (H_f+1) (H_{f,(\tilde{r},r)}+1)^{\frac{N}{2}} \gs \|  \\
&& \hspace{-4pt}\int\limits_s^t \|[W,A_{-q}] (\hel-b)^{-\frac{1}{2}}(H_f+1)^{-1} 
(H_{f,(\tilde{r},r)}+1)^{-\frac{N}{2}} \| dq \\
&\leq& \hspace{-4pt}c_3(N,2,1) c_6 \int\limits_s^t (1+|q|)^{-\zeta} dq 
\stackrel{s,t \to \infty}{\longrightarrow} 0.
\end{eqnarray*}
So $A(t) \gs $ is Cauchy and $A(\infty) \gs=\displaystyle \lim_{t \to \infty}
A(t) \gs$ exists.
\end{proof}
\subsection{An asymptotic expansion, that is correct in second order}
Lemma \ref{l2.9} and Theorem \ref{co2.1} are closely related. Lemma \ref{l2.9}
establishes an upper bound on the the projection onto $\Ran F_R$ of the 
first term in the expansion (\ref{eq215}) in the case $R \to \infty$.
Thus it justifies, why we can neglect the structure of the 
$\co(g)$ terms appearing in Theorem \ref{co2.1} for the leading order
term of the ionisation probability.
\begin{lemma} \label{l2.9}
Suppose Hypothesis \ref{H-2} and let $U$ be as in Hypothesis \ref{H-2} and 
$\tau(g) \nearrow \infty$ as $g \searrow 0$, then
\[\limsup_{R \to \infty} \sup_{g \in U} \|F_R A_{\tau(g)} \gs \|=0 \]
\end{lemma}
\begin{proof}
$\hel=\hel \otimes \one_{\F}$ commutes with $A=\one_{\Hel} \otimes A$, so
$A_{\tau}=\one \otimes e^{-i\tau H_f}Ae^{i\tau H_f}$ 
and pull through formula implies
\[e^{-i\tau H_f}Ae^{i\tau H_f}=a^*_{\lambda_1}(e^{-i\tau\omega}f_1) \cdots 
a^*_{\lambda_N}(e^{-i\tau \omega} f_N).\]
In particular $A_{\tau}$ and $F_R$ commute. Choose the regularization parameters
$0 < \tilde{r} < r < \infty$ such that
\begin{eqnarray*}
\tilde{r} & < & \inf \{\omega(k): k \in \supp (f_j), j=1,...,N\} \\
r & > & \sup \{\omega(k): k \in \supp (f_j), j=1,...,N\} 
\end{eqnarray*}
then
\[\|F_R A_{\tau(g)} \gs\| \leq \|A_{\tau(g)}(H_{f,(\tilde{r},r)}+1)^{-\frac{N}{2}}\| 
\, \|(H_{f,(\tilde{r},r)}+1)^{\frac{N}{2}} F_R \gs \|.\] 
Corollary \ref{coa2} implies
$\displaystyle \sup_{\tau \in \R} \|A_{\tau}(H_{f,(\tilde{r},r)}+1)^{-\frac{N}{2}}\|
\leq C\|f_1\|_{\omega} \cdots \|f_N\|_{\omega}< \infty$.
Let $U$ be the compact neighbourhood of $0$ from Hypothesis \ref{H-2}, then
\[\limsup_{n \to \infty} \sup_{g \in U} \|F_R A_{\tau(g)} \gs\| \leq
C\|f_1\|_{\omega} \cdots \|f_N\|_{\omega} \limsup_{R \to \infty} \sup_{g \in U}
\|F_R H_{f,(\tilde{r},r)} \gs\|=0. \]
\end{proof}
\begin{theorem} \label{co2.1}
Suppose Hypothesis \ref{H-1}, \ref{hyp5}, \ref{hyp4}, \ref{H-2}, \ref{H-3},
$(\hel,1)$ and $(\hel,1,1)$, let
$H_g$ be self-adjoint and let 
$0  < \alpha < \beta <  1 $ and $g^{-\alpha} < \tau= \tau(g) < g^{-\beta}$
as $g \searrow 0$, then
there is some $\cR(\tau(g),t) \in \hr$, such that 
\[\sup_{t \geq g^{-1}} \|\cR(\tau(g),t)\| \leq \co(g) \] and
\begin{equation} \label{eq215}
e^{-i\tau(H_g-E_g)}A(t) \gs =A_{\tau} \gs -
ig \int\limits_{-\infty}^{\infty} e^{-i(\tau-r)(H_0-E_0)}
[W^{(1)},A_r] \go dr + \cR(\tau,t)
\end{equation}
\end{theorem}
\begin{proof}
Choose $\beta, \gamma$, such that $\alpha < \beta < \gamma < 1$. The time 
evolution in Corollary \ref{co1.1} gives
\begin{eqnarray*}
\lefteqn{ e^{-i\tau(H_g-E_g)} A(t) \gs=
A_{\tau} \gs -i \int\limits_0^{\tau+t} ds e^{-isH_g} e^{isH_0} [W_s,A_{\tau}]
e^{-isH_0} e^{isH_g} \gs=} \\[-0.5cm]
&=& A_{\tau} \gs -i \int\limits_0^{\tau+t} ds e^{-is(H_g-E_g)} 
[W,A_{\tau-s}] \gs= \\
&=& A_{\tau} \gs -ig \int\limits_0^{g^{-\gamma}} ds e^{-isH_g} e^{isH_0} 
[W^{(1)}_s,A_{\tau}] e^{-isH_0} e^{isH_g} \gs + \co(g),
\end{eqnarray*}
because $t \geq g^{-1}$ ensures $t +\tau \geq g^{-\gamma}$, so due to 
Lemma \ref{coa1}, the 
commutator estimates in Lemma \ref{l2.1} 
(with $c_4=c_4(0,0)$) and Corollary \ref{co1.3}, the following estimate
\begin{eqnarray}
\lefteqn{ \Big\|g\int \limits_{g^{-\gamma}}^{t+\tau} e^{-is(H_g-E_g)}
[W^{(1)},A_{\tau-s}] \gs ds \Big\| \leq
g \int \limits_{g^{-\gamma}}^{\infty} \|[W^{(1)},A_{\tau-s}] \gs \|ds
\leq} \nonumber \\
&\leq& g c_4
\|(\hel-b)^{\frac{1}{2}} (H_f+1)^{\frac{1}{2}} (H_{f,(\tilde{r},r)}+1)^{\frac{N}{2}} \gs\|
 \int \limits_{g^{-\gamma}}^{\infty} (1+|\tau-s|)^{-\zeta} ds \leq \nonumber \\
&\leq& g c_3(N,1,1) c_4  \int \limits_{g^{-\gamma}}^{\infty} (1+s-\tau)^{-\zeta} 
ds =
g c_3 c_4  \int \limits_{g^{-\gamma}-\tau}^{\infty} (1+r)^{-\zeta} dr \leq 
\nonumber \\
&\leq& g c_3 c_4  \int \limits_{g^{-\gamma}-g^{-\beta}}^{\infty} r^{-\zeta} 
dr =
\frac{g c_3 c_4}{\zeta-1} g^{-\gamma(1-\zeta)}(1-g^{\gamma-\beta})^{1-\zeta}=
\co(g).
\label{gl14}
\end{eqnarray}
is uniform in $t \geq g^{-1}.$ 
The estimate for the $W^{(2)}$ commutator in Lemma \ref{l1.2} 
implies
\begin{eqnarray*}
\lefteqn{
\Big\|g^2 \int\limits_0^{t+\tau} e^{-is(H_g-E_g)}[W^{(2)},A_{\tau-s}] \gs ds
\Big\| 
\leq g^2 \int\limits_0^{t+\tau} \|[W^{(2)},A_{\tau-s}] \gs \| ds \leq }\\
&\leq & g^2 c_3(N,2,0) c_5 \big[\int\limits_{\tau-1}^{\tau+1} ds + 
\int\limits_{\tau+1}^{\infty} (s-\tau)^{-\zeta} ds+
\int\limits_{-\infty}^{\tau-1} (\tau-s)^{-\zeta} ds\big]=\\
&=&2 g^2 c_3 c_5(1+\frac{1}{\zeta-1})
\end{eqnarray*}
Corollary \ref{co1.1} applied again yields
\begin{eqnarray*}
\lefteqn{ g \int\limits_0^{g^{-\gamma}} e^{-isH_g} e^{isH_0} 
[W^{(1)}_s,A_{\tau}] e^{-isH_0} e^{isH_g} \gs ds =
g \int\limits_0^{g^{-\gamma}} [W^{(1)}_s,A_{\tau}] \gs ds -}\\
&&-ig  \int\limits_0^{g^{-\gamma}} ds \int\limits_0^s dq e^{-iqH_g} e^{iqH_0} 
[W_q,[W^{(1)}_s,A_{\tau}]] e^{-iqH_0} e^{iqH_g} \gs  =\\
&=& g \int\limits_0^{g^{-\gamma}} [W^{(1)}_s,A_{\tau}] \gs ds -
ig  \int\limits_0^{g^{-\gamma}} ds \int\limits_0^s dq e^{-iq(H_g-E_g)}
[W,[W^{(1)}_{s-q},A_{\tau-q}]] \gs 
\end{eqnarray*}
and due to Corollary \ref{l2.2} and the choice $\tau < g^{-\gamma}$ with 
$\gamma < 1$
\begin{eqnarray*}
\lefteqn{
g^2 \Big\| \int\limits_0^{g^{-\gamma}} ds \int\limits_0^s dq e^{-iq(H_g-E_g)}
[W^{(1)},[W^{(1)}_{s-q},A_{\tau-q}]] \gs \Big\| \leq }\\
&\leq& 
g^2 c_3(N,2,2) c_6 \int\limits_0^{g^{-\gamma}} ds \int\limits_0^s dq 
(1+|\tau-s|)^{-\zeta} \leq\\
&\leq &
g^2 c_3 c_6 \int\limits_0^{g^{-\gamma}} s (1+|\tau-s|)^{-2} ds = \\
&=& g^2 c_3 c_6 \big[ 2\tau -1 +\frac{1-\tau}{1+g^{-\gamma}-\tau}-
\log\frac{1+g^{-\gamma}-\tau}{1+\tau} \big] \leq \cO(g^2 \tau) =\co(g),
\end{eqnarray*}
and the $[W^{(2)},[W^{(1)}_{s-q},A_{\tau-q}]]$ term is estimated similar
and gives an $\co(g^2)$-term, hence 
\begin{equation} \label{gl12}
\Big\| e^{-i\tau(H_g-E_g)}A(t) \gs -A_{\tau} \gs 
+ig \int\limits_0^{g^{-\gamma}} ds
[W^{(1)}_s,A_{\tau}] \gs \Big\| \leq  \co(g). 
\end{equation}
According to Corollary \ref{co2.5} and Lemma \ref{lemma1}
\begin{eqnarray*}
\|[W^{(1)}_s,A_{\tau}] (\gs-\go) \| \hspace{-5pt} &\leq& \hspace{-5pt}
\|[W^{(1)}_s,A_{\tau}] (\hel-b)^{-\frac{1}{2}} (H_f+1)^{-\frac{1}{2}}
(H_{f,(\tilde{r},r)}+1)^{-\frac{N}{2}} \| \\
&& \| (\hel-b)^{\frac{1}{2}} (H_f+1)^{\frac{1}{2}} 
(H_{f,(\tilde{r},r)}+1)^{\frac{N}{2}} 
(\gs-\go) \| \\
&\leq&  g c_2 c_4 (1+|\tau-s|)^{-\zeta},
\end{eqnarray*}
and even in the worst case $\zeta =2$
\begin{eqnarray*}
\lefteqn{ 
g \Big\| \int\limits_0^{g^{-\gamma}} ds [W^{(1)}_s,A_{\tau}] (\gs-\go) \Big\| 
\leq g^2 c_2 c_4 \int\limits_0^{g^{-\gamma}}(1+|\tau-s|)^{-2} ds = }\\
&=& g^2 c_2 c_4 \Big[2-\frac{1}{1+\tau}-\frac{1}{1+g^{-\gamma}-\tau}\Big] = 
\cO(g^2)
\end{eqnarray*}
is only a term of lower order, so (\ref{gl12}) becomes
\begin{equation} \label{gl13}
\Big\| e^{-i\tau(H_g-E_g)}A(t) \gs -A_{\tau} \gs 
+ig \int\limits_0^{g^{-\gamma}} ds
[W^{(1)}_s,A_{\tau}] \go \Big\| \leq  \co(g). 
\end{equation}
In the domain of the commutator $[W^{(1)}_s,A_{\tau}]=
e^{-isH_0}[W^{(1)},A_{\tau-s}] e^{isH_0}$, so
\begin{eqnarray*} 
\lefteqn{ g \int\limits_0^{g^{-\gamma}} ds [W^{(1)}_s,A_{\tau}] \go =
g \int\limits_0^{g^{-\gamma}} ds e^{-is(H_0-E_0)} [W^{(1)},A_{\tau-s}] \go= }\\
&=& g \int\limits_{\tau-g^{-\gamma}}^{\tau} dr \,
e^{-i(\tau-r)(H_0-E_0)} [W^{(1)},A_r] \go
\end{eqnarray*}
Due to the choice of $\gamma$ and the assumption on $\tau$, 
from which we concluded $\tau < g^{-\beta}$ with 
$\beta < \gamma$,
we get $g^{-\gamma} - \tau \geq g^{-\gamma}(1-g^{\gamma -\beta})=
\cO(g^{-\gamma})$
so in analogy to (\ref{gl14})
\[g \Big\| \int\limits_{-\infty}^{\tau-g^{-\gamma}} dr \, e^{-i(\tau-r)(H_0-E_0)} 
[W^{(1)},A_r] \go \Big \| \leq \co(g) \]
and the bound $g^{-\alpha} < \tau$ and the analogon of (\ref{gl14}) implies
\[g \Big \| \int\limits_{\tau}^{\infty} dr \, e^{-i(\tau-r)(H_0-E_0)} 
[W^{(1)},A_r] \go \Big\| \leq \co(g) \]
Plugging this estimates into (\ref{gl13}), we get:
\[ \Big\| e^{-i\tau(H_g-E_g)}A(t) \gs -A_{\tau} \gs +
ig \int\limits_{-\infty}^{\infty} e^{-i(\tau-r)(H_0-E_0)}
[W^{(1)},A_r] \go dr \Big\| \leq  \co(g),  \]
uniform in $t \geq g^{-1}$.
\end{proof}
\section{Formulas for the ionisation probability in leading orders}
\label{sec3}
\setcounter{equation}{0}
The goal of this section is a derivation of Einstein's description of
the photoelectric effect out of our quantum electrodynamical model. In
a few steps, we will see, in which aspects this simple model is adequate.
\subsection{Ionisation probability vanishes in zeroth order} \label{s3.1} 
\begin{theorem} \label{thm3.1}
Suppose Hypothesis \ref{H-1}, \ref{hyp5}, \ref{hyp4}, \ref{H-2}, \ref{H-3}
and $(\hel,1,1)$ 
and let $H_g$ be self-adjoint.
Let $\tau(g) \nearrow \infty$ when $g \searrow 0$, then the
ionisation probability vanishes in zeroth order:
\begin{equation} \label{eq225}
Q^{(0)}(A)=\lim_{R \nearrow \infty} \, \lim_{g \searrow 0} \, 
\lim_{t \nearrow \infty}
\|F_R e^{-i\tau(g) H_g} A(t) \gs \|^2 =0
\end{equation}
\end{theorem}
\begin{proof}
Due to corollary \ref{co1.1}
\begin{eqnarray*}
e^{-i\tau (H_g-E_g)} A(t) \gs &=&
A_{\tau} \gs - ig \int\limits_0^{t + \tau} ds e^{-is(H_g-E_g)} [W^{(1)},A_{\tau-s}] 
\gs \\
&&- ig^2 \int\limits_0^{t + \tau} ds e^{-is(H_g-E_g)} [W^{(2)},A_{\tau-s}] \gs
\end{eqnarray*}
and the commutator estimates imply
\begin{eqnarray}
\lefteqn{
\Bigg\|\int\limits_0^{t + \tau} ds e^{-is(H_g-E_g)} [W^{(1)},A_{\tau-s}] \gs
\Bigg\|=
\Bigg\|\int\limits_{-t}^{\tau} ds e^{-i(\tau-r)(H_g-E_g)} [W^{(1)},A_r] \gs
\Bigg\| \nonumber }\\
&\leq& \int\limits_{-\infty}^{\infty} \|[W^{(1)},A_r] \gs \| dr \leq
c_3(N,1,1) c_4(0,0) \int\limits_{-\infty}^{\infty} (1+r)^{-\zeta} dr = \cO(1)
\hspace{1cm} \label{gl51}
\end{eqnarray}
and a similar bound for the $W^{(2)}$ commutator. In
\begin{eqnarray*}
\|F_R e^{-i\tau(g) (H_g-E_g)} A(t) \gs \|^2 &\leq&
3 \|F_R A_{\tau(g)} \gs \|^2 + \\
&&+ 3g^2 \Bigg\|\int\limits_0^{t + \tau(g)} ds e^{-is(H_g-E_g)} [W^{(1)},A_{\tau-s}] 
\gs \Bigg\|^2 \\
&&+3g^4 \Bigg\|\int\limits_0^{t + \tau(g)} ds e^{-is(H_g-E_g)} [W^{(4)},A_{\tau-s}] 
\gs \Bigg\|^2 
\end{eqnarray*}
the first term does not depend on $t$, so it vanishes in the limit 
$\displaystyle \lim_{R \to \infty} \lim_{g \searrow 0}$ 
according to Lemma \ref{l2.9}, the last two integrals are $\cO(1)$
uniform in $t$ like in (\ref{gl51}), hence they vanish in 
$\displaystyle \lim_{g \searrow 0} \lim_{t \to \infty}$
and (\ref{eq225}) is proven.
\end{proof}
\subsection{A formula for the ionisation probability in second order} 
\label{sec3.2}
\noindent In the proof of the last theorem, we have divided 
$F_R e^{-i\tau(g)(H_g-E_g)} \gs$ into the term $F_R A_{\tau(g)} \gs$, which
vanishes in the limit $R \to \infty, g \to 0$. The other term, which is
according to Corollary \ref{co1.1} just
$F_R \int\limits_0^{t+\tau(g)} e^{-is(H_g-E_g)}[W,A_{\tau(g)-s}] \gs$
contains some explicit prefactor $g$. As 
it is mentioned in the introduction, we are now going to investigate the
second order term:
\begin{theorem} \label{thm3.2}
Suppose Hypothesis \ref{H-1}, \ref{hyp5}, \ref{hyp4}, \ref{H-2}, \ref{H-3},
$(\hel,1)$ and $(\hel,1,1)$, let $H_g$ be self-adjoint.
Let $0 < \alpha < \beta < 1$ and $g^{-\alpha} < \tau(g) < g^{-\beta}$ for 
$g \searrow 0$
and set
\begin{equation} \label{gl18}
\Psi:=\Psi(A)=\lim_{t \to \infty}\int\limits_{-t}^t ds [W^{(1)}_{-s},A] \go,
\end{equation}
then the
second order of the ionisation probability is
\begin{eqnarray*}
Q^{(2)}(A)&=&
\lim_{R \nearrow \infty} \, \lim_{g \searrow 0} \, \lim_{t \to \infty} 
\Big\| F_R \hspace{-0.2cm}\int\limits_0^{t+\tau(g)} \hspace{-0.2cm}
ds \, e^{-is(H_g-E_g)} 
[W^{(1)}+gW^{(2)}, A_{\tau(g)-s}] \gs \Big \|^2 \\
&=& \|\one_{ac}(\hel) \otimes \one_{\F} \Psi(A) \|^2 .
\end{eqnarray*}
\end{theorem}
\begin{proof}
Application of Corollary \ref{co2.5} and $H_f \Omega=0$ implies
\begin{eqnarray*}
\|[W^{(1)}_{-s},A] \go \| &\leq& c_4(0,0) (1+|s|)^{-\zeta} \|(\hel-b)^{\frac{1}{2}}
(H_f+1)^{\frac{N+1}{2}} \go \|=\\
&=& c_4 (1+|s|)^{-\zeta} |e_0-b|^{\frac{1}{2}},
\end{eqnarray*}
which shows convergence of
\[\Psi=\Psi(A)=\lim_{t \to \infty}\int\limits_{-t}^t ds [W^{(1)}_{-s},A] \go.\]
Due to Corollary \ref{co1.1} and Theorem \ref{co2.1}
\begin{eqnarray*}
\lefteqn{\int\limits_0^{t+\tau(g)} 
ds \,e^{-is(H_g-E_g)}[W^{(1)}+gW^{(2)},A_{\tau(g)-s}] \gs=}\\[-0.5cm]
&=&\int\limits_{-\infty}^{\infty} dr \, e^{-i(\tau(g)-r)(H_0-E_0)}[W^{(1)},A_r] \go
+ \tilde{\cR}(\tau(g),t), 
\end{eqnarray*}
where $\displaystyle \sup_{t \geq g^{-1}} \|\tilde{\cR}(\tau(g),t) \| \leq 
\co(1)$
as $g \searrow 0$. So for any $g >0$
\[\lim_{t \to \infty} \|F_R \tilde{\cR}(\tau(g),t) \| \leq 
\sup_{t \geq g^{-1}} \|\tilde{\cR}(\tau(g),t) \| \leq \co(1) \]
and therefore
\[\lim_{R \to \infty} \,\lim_{g \searrow 0 } \,\lim_{t \to \infty} 
\|F_R \tilde{\cR}(\tau(g),t)\|=0=
\lim_{R \to \infty} \, \lim_{g \searrow 0 } \, \lim_{t \to \infty} 
\|F_R \tilde{\cR}(\tau(g),t)\|^2.\]
Furthermore
\[\Big \|F_R \int\limits_{-\infty}^{\infty} e^{-i(\tau(g)-r)(H_0-E_0)}
[W^{(1)},A_r] \go \Big \|
\leq \Big\| \int\limits_{-\infty}^{\infty} [W^{(1)}_{-r},A] \go dr \Big\| =
\|\Psi\|,\]
so
\begin{eqnarray*}
\lefteqn{ \Big| \lim_{R \to \infty} \, \lim_{g \searrow 0} \, 
\lim_{t \to \infty}
\Re \Big \lkl F_R \int\limits_{-\infty}^{\infty} 
e^{-i(\tau(g)-r)(H_0-E_0)}[W^{(1)},A_r] \go , 
F_R \tilde{\cR}(\tau(g),t) \Big \rkl \Big|}\\
&\leq& \|\Psi\| \lim_{R \to \infty} \, \lim_{g \searrow 0} \, 
\lim_{t \to \infty} \| F_R \tilde{\cR}(\tau(g),t)\| =0 \hspace{5cm}
\end{eqnarray*}
and we conclude
\begin{eqnarray} 
Q^{(2)}(A)&=&\lim_{R \to \infty} \, \lim_{g \searrow 0} \, \lim_{t \to \infty}
\Bigg[ \Big\| F_R \int\limits_{-\infty}^{\infty} 
e^{-i(\tau(g)-r)(H_0-E_0)}[W^{(1)},A_r] \go \Big\|^2 + \nonumber \\
&&+\|F_R \tilde{\cR}(\tau(g),t)\|^2+ \nonumber\\
&&+2 \Re \Big \lkl F_R \int\limits_{-\infty}^{\infty} 
e^{-i(\tau(g)-r)(H_0-E_0)}[W^{(1)},A_r] \go , F_R \tilde{\cR}(\tau(g),t) 
\Big \rkl \Bigg]=  \nonumber\\
&=& \lim_{R \to \infty} \, \lim_{g \searrow 0} 
\Big\| F_R \int\limits_{-\infty}^{\infty} 
e^{-i(\tau(g)-r)(H_0-E_0)}[W^{(1)},A_r] \go \Big\|^2 = \nonumber \\
&=& \lim_{R \to \infty} \lim_{g \searrow 0} \|F_R e^{-i\tau(g) H_0} \Psi(A) \|^2.
\label{gl17}
\end{eqnarray}
Apart from $\tau(g)$ any other $g$ dependence has disappeared from 
(\ref{gl17}), so
\begin{equation} \label{gl19}
Q^{(2)}(A)= \lim_{R \to \infty} \lim_{\tau \to \infty} 
\|F_R e^{-i\tau H_0} \Psi(A) \|^2
\end{equation}
The algebraic tensor-product $\Hel \otimes \F$ is dense in $\hr$ and 
$\one_{pp}(\hel) \Hel$ is the closure of finite linear combinations of 
eigenfunctions of $\hel$. So for any $\varepsilon >0$, there are
$M \in \N$, $\phi_1,...,\phi_m \in \F$, $h_1,...,h_M \in \Hel$, such that
\[\Big\|\Psi-\sum_{j=1}^M h_j \otimes \phi_j\Big\| < \frac{\varepsilon}{2} \]
and furthermore $m_j \in \N$ and eigenfunctions $\eta_{j,l}$ of $\hel$
corresponding to the eigenvalues $e_{j,l}$, $j=1,...,M$, $l=1,...,m_j$, such 
that
\[\Big\|\one_{pp}(\hel) h_j -\sum_{l=1}^{m_j} \eta_{j,l} \Big\| < 
\frac{\varepsilon}{2M \|\phi_j\|}.\]
$\|F_Re^{-i\tau H_0}\| \leq 1$ so
\begin{eqnarray}
\lefteqn{\|F_Re^{-i\tau H_0} \one_{pp}(\hel) \Psi\| \leq
\Big\|F_Re^{-i\tau H_0} \sum_{j=1}^M \one_{pp}(\hel) h_j \otimes \phi_j\Big\| 
+\frac{\varepsilon}{2} \leq}\label{gl200}\\
&\leq& \sum_{j=1}^M \|\one_{\{|x| \geq R\}} e^{-i\tau \hel} \one_{pp}(\hel) h_j \|\,
\|\phi_j\| +\frac{\varepsilon}{2} \leq \nonumber\\
&\leq& \sum_{j=1}^M \|\one_{\{|x| \geq R\}} e^{-i\tau \hel} 
\sum_{l=1}^{m_j} \eta_{j,l} \|\, \|\phi_j\| +\varepsilon \leq 
 \sum_{j=1}^M \sum_{l=1}^{m_j} \|\one_{\{|x| \geq R\}} \eta_{j,l} \|\, 
\|\phi_j\| +\varepsilon  \nonumber
\end{eqnarray}
The right hand side of (\ref{gl200}) does not depend on $\tau$, so
\[\sup_{\tau \in \R} \|F_Re^{-i\tau H_0} \one_{pp}(\hel) \Psi\| \leq
\sum_{j=1}^M \sum_{l=1}^{m_j} \|\one_{\{|x| \geq R\}} \eta_{j,l} \|\, 
\|\phi_j\| +\varepsilon\]
and $\one_{\{|x| \geq R\}}$ converges strongly to $0$ for $R \to \infty$, so
\[\lim_{R \to \infty}\sup_{\tau \in \R} \|F_Re^{-i\tau H_0} \one_{pp}(\hel) \Psi\| 
\leq 
\sum_{j=1}^M \sum_{l=1}^{m_j} \lim_{R \to \infty} \|\one_{\{|x| \geq R\}} 
\eta_{j,l} \|\,  \|\phi_j\| +\varepsilon = \varepsilon,\]
hence
\begin{equation} \label{eq234}
\limsup_{R \to \infty} \sup_{\tau \in \R_+} 
\|F_R e^{-i\tau H_0} \one_{pp}(\hel) \otimes \one_{\F} \Psi \| = 0.
\end{equation}
Due to Hypothesis \ref{H-1} the singular continuous spectrum 
$\sigma_{sc}(\hel)=\emptyset$ is empty, hence 
$\one_{\Hel}=\one_{pp}(\hel)+\one_{ac}(\hel)$ and in combination with 
(\ref{eq234}) we get:
\begin{eqnarray}
\lefteqn{
\lim_{R \to \infty} \lim_{\tau \to \infty} \|F_R e^{-i\tau H_0}
\Psi\|^2= 
\lim_{R \to \infty} \lim_{\tau \to \infty} \|F_R e^{-i\tau H_0}
\one_{ac}(\hel) \otimes \one_{\F} \Psi \|^2= \label{eq235} }\\
&=& \|\one_{ac}(\hel) \otimes \one_{\F}\Psi \|^2 -
\lim_{R \to \infty} \lim_{\tau \to \infty} \|(\one-F_R) e^{-i\tau H_0}
\one_{ac}(\hel) \otimes \one_{\F} \Psi \|^2. \nonumber
\end{eqnarray}
$\D(\hel)$ is dense in $\Hel$, so for each $\varepsilon >0$ there are
$\varphi_1,...,\varphi_n \in \D(\hel)$ and
$\phi_1,...,\phi_n \in \F$, such that $\displaystyle \|\sum_{j=1}^n \varphi_j
\otimes \phi_j -\Psi\| < \varepsilon$, hence
\begin{eqnarray}
\lefteqn{
\lim_{R \to \infty} \lim_{\tau \to \infty} \| (\one-F_R) e^{-i\tau H_0}
\one_{ac}(\hel) \otimes \one_{\F} \Psi \| \leq \label{eq236} }\\
&\leq& \sum_{j=1}^n 
\lim_{R \to \infty} \lim_{\tau \to \infty} \| \one_{\{|x| < R\}} 
e^{-i\tau \hel} \one_{ac}(\hel) \varphi_j\| \, \|e^{-i \tau H_f} \phi_j\| +
\varepsilon=\varepsilon. \nonumber
\end{eqnarray}
For the last estimate we used
\begin{itemize}
\item 
$\hspace{1cm} \displaystyle  \lim_{\tau \to \infty} \| \one_{\{|x| < R\}} 
e^{-i\tau \hel} \one_{ac}(\hel) \varphi_j\| = $ \\
$\hspace{1cm} \displaystyle =
\lim_{\tau \to \infty} \| \one_{\{|x| < R\}} (\hel-b)^{-1}
e^{-i\tau \hel} \one_{ac}(\hel) (\hel-b) \varphi_j\|$
\item $\hspace{1cm} \displaystyle \one_{\{|x| < R\}} (\hel-b)^{-1}=
\one_{\{|x| < R\}} (1-\lap)^{-1} (1-\lap)(\hel-b)^{-1} $ \\
is compact: $\one_{\{|x| < R\}} (1-\lap)^{-1}$ is Hilbert-Schmidt 
(integral kernel for $(1-\lap)^{-1}$) and 
$\Ran(\hel-b)^{-1}=\D(\hel)=\D(-\lap)$ by Hypothesis \ref{H-1}, 
so $(1-\lap)(\hel-b)^{-1}$ is bounded by closed graph theorem.
\item  $\one_{ac}(\hel) (\hel-b) \varphi_j$ is a well defined element of 
$\one_{ac}(\hel) \Hel$ provided $\varphi_j \in \D(\hel)$,
\end{itemize}
so we are in a situation to apply Riemann-Lebesgue Lemma like in \cite{RS3},
XI.3, Lemma 2, which yields $\| \one_{\{|x| < R\}} 
e^{-i\tau \hel} \one_{ac}(\hel) \varphi_j\| 
\stackrel {\tau \to \infty}{\longrightarrow} 0$ 
for each $R>~0$. Putting together 
these results and with $\Psi(A)$ as in (\ref{gl18}), finally we get:
\begin{eqnarray*}
Q^{(2)}(A)=\|\one_{ac}(\hel) \otimes \one_{\F} \Psi(A) \|^2. \\[-1.5cm]
\end{eqnarray*}
\end{proof}
\vspace*{0.2cm}
\noindent Note, that in this RAGE-type theorem, we have to do the finite
rank approximations of $\Psi$ ``by hand'', because 
$F_R=\one_{\{|x| \geq R\}} \otimes \one_{\F}$ destroys relative $H_0$ compactness.
\begin{lemma} \label{l3.3}
Under the assumptions of Theorem \ref{thm3.2}
let $(\varphi_n)_{n \in \N}$ be an orthonormal family in 
$C_0^{\infty} (\R^3 \backslash \{0\})$, such that 
$\lkl \varphi_j, \varphi_l\rkl_{L^2}=\delta_{jl}$. Suppose
$(m_1,...,m_{\eta}), (n_1,...,n_{\eta}) \in \N^{\eta}$ with 
\[m_1+...+m_{\eta}+n_1+...+n_{\eta}=N,\] 
then the second order 
\[Q^{(2)}(a^*_+(\varphi_1)^{m_1} a^*_-(\varphi_1)^{n_1} \cdots
a^*_+(\varphi_{\eta})^{m_{\eta}} a^*_-(\varphi_{\eta})^{n_{\eta}}) \]
of the ionisation probability by a photon cloud
\begin{equation}
A=a^*_+(\varphi_1)^{m_1}a^*_-(\varphi_1)^{n_1} \cdots 
a^*_+(\varphi_{\eta})^{m_{\eta}}a^*_-(\varphi_{\eta})^{n_{\eta}},
\label{eq264}
\end{equation}
is given by
\begin{eqnarray}
\lefteqn{ \frac{ Q^{(2)}(a^*_+(\varphi_1)^{m_1} a^*_-(\varphi_1)^{n_1} \cdots
a^*_+(\varphi_{\eta})^{m_{\eta}} a^*_-(\varphi_{\eta})^{n_{\eta}}) }
{m_1! \cdots m_{\eta}! n_1! \cdots n_{\eta}!}= \nonumber} \\
&=& \sum_{j=1}^{\eta} \Big( n_j Q^{(2)}_-(\varphi_j)+
m_j Q^{(2)}_+(\varphi_j)\Big) \label{gl52}
\end{eqnarray}
with one photon terms
\begin{equation}
Q^{(2)}_{\lambda}(\varphi):= \Big\|\one_{ac}(\hel)
\int\limits_{-\infty}^{\infty}  ds 
\int\limits_{\R^3} dk \, e^{is(\hel-e_0)} w^{(0,1)}(k,\lambda) 
e^{-is\omega(k)} \varphi(k) \varphi_0 \Big\|^2.
\end{equation}
\end{lemma}
\begin{proof}
(\ref{eq38}) implies
\begin{eqnarray}
\lefteqn{ \|(\one_{ac} (\hel) \otimes \one_{\F}) \Psi(A) \|^2=
\Big\| (\one_{ac}(\hel) \otimes \one_{\F}) 
\int\limits_{-\infty}^{\infty}  ds\,
e^{is(H_0-e_0)} [W^{(1)},A_s] \go \Big\|^2= \nonumber}\\
&=& \hspace{-2pt} \Bigg\|(\one_{ac}(\hel) \otimes \one_{\F}) \hspace{-4pt}
\int\limits_{-\infty}^{\infty} \hspace{-4pt} ds 
\Big\{ \sum_{j=1}^{\eta} n_j
\int\limits_{\R^3} dk \, e^{is(\hel-e_0)} w^{(0,1)}(k,-) 
e^{-is\omega(k)} \varphi_j(k) \varphi_0 
\nonumber \\[-0.5cm] &&\hspace{6cm} 
\prod_{l=1}^{\eta} a^*_+(\varphi_l)^{m_l}
a^*_-(\varphi_l)^{n_l-\delta_{jl}} \Omega + \nonumber\\
&& \hspace{3.6cm}
+\sum_{j=1}^{\eta} m_j
\int\limits_{\R^3} dk \, e^{is(\hel-e_0)} w^{(0,1)}(k,+) 
e^{-is\omega(k)} \varphi_j(k) \varphi_0  \nonumber \\[-0.5cm] 
&&\hspace{6cm} 
\prod_{l=1}^{\eta} a^*_+(\varphi_l)^{m_l-\delta_{jl}}
a^*_-(\varphi_l)^{n_l} \Omega\Big\} \Bigg\|^2  \label{eq263}
\end{eqnarray}
Commuting creation and annihilation operators, the canonical
commutation relations together with the orthonormality of
$\varphi_1,...,\varphi_m$ imply
\[a_{\lambda}(\varphi_j)a^*_{\lambda'}(\varphi_l)=
a^*_{\lambda'}(\varphi_l) a_{\lambda}(\varphi_j) +
\delta_{\lambda,\lambda'}\delta_{jl}.\]
By induction
$a_{\lambda}(\varphi_j)^{q} a^*_{\lambda'}(\varphi_l)^{q} \Omega= 
\delta_{\lambda,\lambda'}\delta_{jl} q! \Omega$ for $q \in \N$, where
$a(\varphi_j) \Omega=0$ was used. As a generalisation of the last result
\begin{equation} \label{eq540} 
\lkl \prod_{l=1}^{\eta} a^*_+(\varphi_l)^{q_l}a^*_-(\varphi_l)^{r_l} 
\Omega ,
\prod_{l=1}^{\eta} a^*_+(\varphi_l)^{q_l'}a^*_-(\varphi_l)^{r_l'} \Omega \rkl=
\prod_{l=1}^{\eta} \delta_{q_lq_l'}\, \delta_{r_lr_l'} \, q_l! \, r_l!
\end{equation}
for $(q_1,...,q_{\eta}),(q_1',...,q_{\eta}'),
(r_1,...,r_{\eta}),(r_1',...,r_{\eta}') \in \N^{\eta}$.
When we use these orthogonality relations in expanding the sum under
the norm square in (\ref{eq263})
\begin{eqnarray*}
\lefteqn{ \frac{\|(\one_{ac} (\hel) \otimes \one_{\F}) \Psi(A)
\|^2}{m_1! \cdots m_{\eta}! n_1! \cdots n_{\eta}!}=}\\
&=&  \sum_{j=1}^{\eta} n_j 
\Big\|\one_{ac}(\hel) \int\limits_{-\infty}^{\infty}  ds 
\int\limits_{\R^3} dk \, e^{is(\hel-e_0)} w^{(0,1)}(k,-) 
e^{-is\omega(k)} \varphi_j(k) \varphi_0 \Big\|^2
\nonumber \\
&&+ \sum_{j=1}^{\eta} m_j \Big\|\one_{ac}(\hel) 
\int\limits_{-\infty}^{\infty}  ds 
\int\limits_{\R^3} dk \, e^{is(\hel-e_0)} w^{(0,1)}(k,+) 
e^{-is\omega(k)} \varphi_j(k) \varphi_0 \Big\|^2
\nonumber 
\end{eqnarray*}
which is the desired result.
\end{proof}
\begin{remark} \end{remark} \vspace*{-0.3cm} 
\noindent 
At that point we see, a typical situation, where the reduction of the 
photon field as
a multi-particle system to an effective one photon system is justified:
If the photon cloud is of the form (\ref{eq264}), the second
order of the ionisation probability is additive in
the photons involved and not a collective effect of the whole
system. This step is already contained in Einstein's model, where the
electron is (at least implicitly) allowed to absorb only one
photon. When we use photons of momentum $k_1,...,k_m$ in
Einstein's model, this model assumption is motivated from quantum
electrodynamics by Lemma \ref{l3.3}: 
A~photon of momentum $k_j$, i.e. with momentum distribution
$\delta(k-k_j)$ in Einstein's model, is ``approximated'' in our model 
by photons with a smooth momentum distribution $\varphi_{j,\varepsilon}$ of
compact support in $\{|k-k_j| < \varepsilon\}$ and  
$\varphi_{j,\varepsilon} \stackrel{\varepsilon \to
0}{\longrightarrow} \delta(k-k_j)$ as distributions. If $\varepsilon$ is
small enough, then $\varphi_{j,\varepsilon} \perp
\varphi_{l,\varepsilon}$ for $j \not = l$ and we may apply Lemma \ref{l3.3}.

\subsection{Expansion in generalised eigenfunctions and the ``explicit''
calculation of ${\protect{Q^{(2)}(\varphi_j)}}$} \label{sec3.3}
To see an analogon of (\ref{eqein2}) in our model, we need a more explicit
calculation of $Q^{(2)}(\varphi_j)$.
For such an ``explicit'' calculation of $Q^{(2)}(\varphi_j)$ for some 
given momentum distribution 
$\varphi_j \in C_0^{\infty}(\R^3 \backslash\{0\})$, 
we need some results from scattering theory
of the electron Hamiltonian $\hel$, in particular an expansion in 
(generalised) eigenfunctions. 
For the application of eigenfunction expansion to the calculation of
$Q^{(2)}(\varphi_j)$ we assume:
\begin{hypothesis} \label{hyp6}
The wave operators
\[ \Omega^{\pm}(-\lap, \hel):= \slim_{t \to \mp \infty} e^{it(-\lap)}
e^{-it\hel} \one_{ac}(\hel) \]
exist. For compact $K \subseteq \R^3 \backslash \{0\}$, $\alpha \in \N_0^3$,
$|\alpha| \leq \zeta$, there is some $\theta \in L^2(\R^3)$, such that
\begin{equation} \label{gl55}
\sup_{k \in K \atop \lambda \in \Z_2} \big| \lkl x \rkl^2 (\partial_k^{\alpha} 
w^{(0,1)}(k,\lambda) \varphi_0)(x)\big| \leq |\theta(x)|
\end{equation}
and for $s \in \R$, $k \in \R^3 \backslash \{0\}$ and $\lambda \in \Z_2$
\begin{equation} \label{gl56}
w^{(0,1)}_s (k,\lambda) \varphi_0 =
e^{-is(\hel-e_0)} w^{(0,1)}(k,\lambda) \varphi_0\in \D(\lkl \cdot \rkl^2) 
\end{equation}
\end{hypothesis}
\begin{theorem} \label{thm3.3}
Suppose Hypothesis \ref{H-1}, \ref{hyp5}, \ref{hyp4}, \ref{H-2}, 
\ref{hyp6}, $(\hel,1)$ and $(\hel,1,1)$ are satisfied, then there is a 
function 
$\displaystyle \begin{array}[t]{rcl} \widehat{\rho}_{\lambda} : 
\R^3 \times \R^3  &
\longrightarrow & \C \\
(p,k) & \longmapsto & \widehat{\rho}_{\lambda}(p,k) \end{array}$,
such that for $|\alpha| \leq 2$ all partial derivatives 
$\partial_k^{\alpha} \widehat{\rho}_{\lambda}$
exist on $\R^3 \backslash \{0\}$,
\[ \int\limits_{\R^3} dp \int\limits_K dk \,|\partial_k^{\alpha}
\widehat{\rho}_{\lambda}(p,k)|^2 < \infty\]
for each compact set $K \subseteq \R^3 \backslash \{0\}$ and 
for $\varphi_j \in C_0^{\infty}(\R^3 \backslash \{0\})$
the second order of the ionisation probability is
\begin{equation}
Q^{(2)}_{\lambda} (\varphi_j)= \lim_{t \to \infty} \int\limits_{\R^3} dp
\left|\, \int\limits_{-t}^t ds \int\limits_{\R^3} dk
\, e^{is(p^2-e_0-\omega(k))} \widehat{\rho}_{\lambda}(p,k) \varphi_j(k) \right|^2
\end{equation}
\end{theorem}
\begin{proof}
The absolute continuous subspace $\one_{ac}(\hel) \Hel$ of $\hel$ is a
reducing subspace for $\hel$. Due to the assumptions, the wave
operators
\[ \Omega^{\pm}(-\lap, \hel):= \slim_{t \to \mp \infty} e^{it(-\lap)}
e^{-it\hel} \one_{ac}(\hel) \]
exist. Using the intertwining properties of $\Omega^{\pm}$ and conjugation with 
Fourier transform $\fou$, we obtain 
$\hel|_{\one_{ac}(\hel) \Hel}$ as a multiplication operator with $p^2$:
\[\fou \Omega^{\pm}(-\lap, \hel)  \hel|_{\one_{ac}(\hel) \Hel}
(\fou \Omega^{\pm}(-\lap, \hel) )^* = p^2
\one_{\Ran (\fou \Omega^{\pm})}.\]
So for an application of \cite{PS} Theorem 2.2 and 2.3 we may take $H=\hel$,
$M=\one_{ac}(\hel) \Hel$, $X=\R^3$, $d\rho(p)=dp$, $h(p)=p^2$
and 
\[U:=\fou \Omega^{\pm}(-\lap, \hel): \one_{ac}(\hel) \Hel \longrightarrow 
L^2(\R^3).\] 
We further fix $z \in \C \backslash \R$ and define
$\gamma(\lambda):=(\lambda-z)^{-2}$, then $\gamma(h(p))=(p^2-z)^{-2}
\not= 0$ for each $p \in \R^3$ and $\frac{1}{\gamma(h(p))}=(p^2-z)^2$
remains bounded on each compact subset of $\R^3$. Hence the limiting arguments
in the proof of \cite{PS} Theorem 2.2 can be done for the $\sigma$-compact
space $\R^3$ with Borel measure as in the case of a $\sigma$-finite
measure space with $\gamma$ finite on sets of finite measure. 
\cite{PS} Theorem 3.6
is applicable due to the relative $-\lap$-bound of $V$ in
Hypothesis \ref{H-1}, hence it implies, that we can choose 
$T=\lkl \cdot \rkl^2$ and $S=\one-\lap$,
so that 
\[\gamma(H)T^{-1} S=(\hel-z)^{-2} \lkl \cdot \rkl^{-2} (\one-\lap) 
\subseteq \left( (\one-\lap) \lkl \cdot \rkl^{-2}
(\hel-\overline{z})^{-2} \right)^*\]
is the restriction of the Hilbert-Schmidt operator 
$\left( (\one-\lap) \lkl \cdot \rkl^{-2}
(\hel-\overline{z})^{-2} \right)^*$.
According to Hypothesis \ref{hyp6}
for any $k \in \R^3\backslash \{0\}$, $\lambda \in \Z_2$ and $s \in \R$
\[e^{is(\hel-e_0)}w^{(0,1)}(k,\lambda) \varphi_0=
w^{(0,1)}_{-s}(k,\lambda) \varphi_0 \in \D(\lkl \cdot \rkl^2),\]
hence
\[\Psi(t):= \int\limits_{-t}^t  ds 
\int\limits_{\R^3} dk \, e^{is(\hel-e_0-\omega(k))} \varphi_j(k)
w^{(0,1)}(k,\lambda) \varphi_0 \in \D(\lkl \cdot \rkl^2).\]
and by \cite{PS} Theorem 2.2 and 2.3 we get for 
$(U \one_{ac}(\hel) \Psi(t))(p)$:
\begin{eqnarray}
\lefteqn{ \|\one_{ac}(\hel) \Psi(t) \|^2=
\int\limits_{\R^3} dp |(U \one_{ac}(\hel)  \Psi(t) )(p)|^2=
\int\limits_{\R^3} dp | \lkl \varphi(p),  \Psi(t) \rkl_{\mp}|^2
\nonumber }\\
&=&\int\limits_{\R^3} dp \left| \left \lkl \varphi(p),
\int\limits_{-t}^t  ds 
\int\limits_{\R^3} dk \, e^{is(\hel-e_0-\omega(k))} \varphi_j(k)
w^{(0,1)}(k,\lambda) \varphi_0 \right\rkl_{\mp} \right|^2
\nonumber\\
&=&\int\limits_{\R^3} dp \left|\, 
\int\limits_{-t}^t ds \int\limits_{\R^3} dk e^{-is\omega(k)} \varphi_j(k)
\lkl \varphi(p), e^{is(\hel-e_0)} w^{(0,1)}(k,\lambda) \varphi_0 
\rkl_{\mp} \right|^2
\nonumber\\
&=&\int\limits_{\R^3} dp \left|\, 
\int\limits_{-t}^t ds \int\limits_{\R^3} dk e^{is(p^2-e_0-\omega(k))} 
\varphi_j(k) \lkl \varphi(p), w^{(0,1)}(k,\lambda) \varphi_0 \rkl_{\mp} 
\right|^2 \label{eq266}
\end{eqnarray}
Now we define 
\begin{equation} \label{eq267}
\widehat{\rho}_{\lambda}(p,k):=
\lkl \varphi(p), w^{(0,1)}(k,\lambda) \varphi_0 \rkl_{\mp} ,
\end{equation}
and note, that the construction of the generalised eigenfunctions $\varphi(p)$ 
in \cite{PS} and our choice of $S$ and $T$ implies
\begin{eqnarray*}
\varphi(p) \in \Ran(\lkl \cdot \rkl ^2 (\one-\lap)^{-1})
\hspace{-4pt} &=& \hspace{-4pt}
H^2_{-2}(\R^3) \equiv \\
&\equiv& \hspace{-4pt} \{f:\R^3 \longrightarrow \C \,
{\mathrm {measurable}} \,, \lkl \cdot \rkl^{-2} f \in H^2(\R^3) \},
\end{eqnarray*} 
so due to $w^{(0,1)}(k,\lambda) \varphi_0 \in \D(\lkl \cdot \rkl^2)$
and the definition of the dual pairing 
$\lkl \cdot, \cdot \rkl_{\mp}$ in \cite{PSW}, in (\ref{eq267}) it 
boils down to the following integral
\begin{equation}
\Big\lkl \varphi(p),  w^{(0,1)}(k,\lambda) \varphi_0 \Big\rkl_{\mp}=
\int\limits_{\R^3} \ol{\varphi(p,x)}(w^{(0,1)}(k,\lambda) \varphi_0)(x) dx
\end{equation}
By (\ref{eq266}) and (\ref{eq267}), we get the
same type of formula for $Q^{(2)}_{\lambda}(\varphi_j)$ as in
\cite{BKZ}:
\begin{eqnarray}
\lefteqn{ Q^{(2)}_{\lambda}(\varphi_j)=
\lim_{t \to \infty} \|\one_{ac}(\hel) \Psi(t)\|^2=
\label{eq268}}\\
&=& \lim_{t \to \infty} \int\limits_{\R^3} dp \left| \,
\int\limits_{-t}^t ds \int\limits_{\R^3} dk\,
e^{is(p^2-e_0-\omega(k))} \widehat{\rho}_{\lambda}(p,k) \varphi_j(k) \right|^2
\nonumber
\end{eqnarray}
Due to Hypothesis \ref{hyp6}
\[\big| \ol{\varphi(p,x)} (\partial_k^{\alpha} w^{(0,1)}(k,\lambda) \varphi_0)(x)
\big| \leq |\ol{\varphi(p,x)} \lkl x \rkl^{-2} | |\theta(x)|\]
is dominated by the $L^1$ function on the right hand side, so dominated 
convergence theorem implies
\begin{eqnarray}
\partial_k^{\alpha} \widehat{\rho}_{\lambda}(p,k)&=&
\partial_k^{\alpha} \int\limits_{\R^3} dx \ol{\varphi(p,x)} (w^{(0,1)}(k,\lambda)
\varphi_0)(x)=\label{gl33}\\
&=&\int\limits_{\R^3} dx \ol{\varphi(p,x)} (\partial_k^{\alpha}w^{(0,1)}(k,\lambda)
\varphi_0)(x)=
\lkl \varphi(p), \partial_k^{\alpha} w^{(0,1)}(k,\lambda) \varphi_0 \rkl_{\mp}
\nonumber
\end{eqnarray}
{For} further applications, now we check the regularity properties of
$\widehat{\rho}_{\lambda}$ and its derivatives.
$\partial_k^{\alpha} w^{(0,1)}(k,\lambda) \varphi_0 \in \D(\lkl \cdot \rkl^2)$
according to Hypothesis \ref{hyp6}, so
\begin{eqnarray*} 
\int\limits_{K} dk \int\limits_{\R^3} dp 
|\partial_k^{\alpha}\widehat{\rho}_{\lambda}(p,k)|^2
&=& \int\limits_{K} dk \int\limits_{\R^3} dp 
\Big |\lkl \varphi(p),\partial_k^{\alpha} w^{(0,1)}(k,\lambda) 
\varphi_0 \rkl_{\mp} \Big|^2 =\\
&=&\int\limits_{K} dk \int\limits_{\R^3} dp 
|U \one_{ac}(\hel)\partial_k^{\alpha} w^{(0,1)}(k,\lambda) \varphi_0|^2(p)=
\nonumber\\
&=&\int\limits_{K} dk \|\one_{ac}(\hel) \partial_k^{\alpha} 
w^{(0,1)}(k,\lambda) \varphi_0\|^2
\nonumber \leq \\
&\leq& (e_0-b) \int\limits_{K} dk 
\|\partial_k^{\alpha} w^{(0,1)}(k,\lambda)(\hel-b)^{-\frac{1}{2}}\|^2, \nonumber
\end{eqnarray*}
which is finite for any compact $K \subseteq \R^3 \backslash \{0\}$ 
and $\alpha \in \N_0^3$ with $|\alpha| \leq 2$ 
according to Hypothesis \ref{hyp4}.
\end{proof}
\begin{theorem} \label{thm3.4}
Let the assumptions of Theorem \ref{thm3.3} be satisfied and let
$\mu_r$ be the Lebesgue measure on the sphere $S^2(r):=\{k \in \R^3:
|k|=r\}$. Then 
the second order of the ionisation probability is given by
\begin{equation}
Q^{(2)}_{\lambda}(\varphi_j)
=\int\limits_{\R^3} dp \left|\; \int\limits_{S^2(p^2-e_0)} d\mu_{p^2-e_0}(k)
\widehat{\rho}_{\lambda}(p,k) \varphi_j(k) \right|^2 \label{eq279}
\end{equation}
\end{theorem}
\begin{proof}
As $\varphi_j \in C_0^{\infty}(\R^3 \backslash \{0\})$, the dispersion
$\omega$ is differentiable on the support of $\varphi_j$, hence
\begin{equation} \label{eq271}
\varphi_j(k) e^{is(p^2-e_0-\omega(k))}=
\varphi_j(k) 
\left[ \frac{i}{s |\nabla \omega|^2(k)} \sum_{l=1}^3 \frac{\partial
\omega}{\partial k_l}(k) \frac{\partial}{\partial k_l}\right]^2
e^{is(p^2-e_0-\omega(k))}. 
\end{equation}
Two times integration by parts of (\ref{eq271}) shows us, that there
are $C_0^{\infty}(\R^3 \backslash \{0\})$ functions $f_{\alpha}$,
$\alpha \in \N_0^3$, $|\alpha| \leq 2$, such that for $|s|>1$ we
obtain:
\begin{equation} \label{eq272}
\int\limits_{\R^3} dk e^{is(p^2-e_0-\omega(k))} \widehat{\rho}_{\lambda}(p,k)  
\varphi_j(k)=
\frac{1}{s^2} \hspace{-2pt} \int\limits_{\R^3} \hspace{-2pt}
dk \sum_{|\alpha| \leq 2}
(\partial_k^{\alpha} \widehat{\rho}_{\lambda})(p,k) f_{\alpha}(k)
e^{is(p^2-e_0-\omega(k))} 
\end{equation}
By (\ref{eq272}) and Schwarz inequality, we obtain:
\begin{eqnarray}
\lefteqn{\left| \,\int\limits_{-t}^t ds \int\limits_{\R^3}dk  
e^{is(p^2-e_0-\omega(k))} \widehat{\rho}_{\lambda}(p,k) \varphi_j(k)\right|^2 
\leq \label{eq273} }\\
&\leq&  \left| \, \int\limits_{-t}^{-1} ds 
\frac{1}{s^2} \int\limits_{\R^3} dk \sum_{|\alpha| \leq 2}
(\partial_k^{\alpha} \widehat{\rho}_{\lambda})(p,k) f_{\alpha}(k)
e^{is(p^2-e_0-\omega(k))} \right|^2+ \nonumber\\
&&+\left| \, \int\limits_{-1}^1 ds \int\limits_{\R^3}dk \,  
e^{is(p^2-e_0-\omega(k))} \widehat{\rho}_{\lambda}(p,k) \varphi_j(k)\right|^2+
\nonumber\\ 
&&+\left| \int\limits_{1}^t ds 
\frac{1}{s^2} \int\limits_{\R^3} dk \sum_{|\alpha| \leq 2}
(\partial_k^{\alpha} \widehat{\rho}_{\lambda})(p,k) f_{\alpha}(k)
e^{is(p^2-e_0-\omega(k))} \right|^2\leq \nonumber\\
&\leq&
\left|\, \int\limits_{-1}^1 \hspace{-2pt}  ds \hspace{-3pt}
\int\limits_{\R^3}\hspace{-2pt}dk 
|\widehat{\rho}_{\lambda}(p,k)| \, |\varphi_j(k)| \right|^2\hspace{-3pt}+
2\left| \int\limits_{1}^{\infty} 
\frac{ds}{s^2} \int\limits_{\R^3} \hspace{-2pt} dk  \hspace{-2pt}
\sum_{|\alpha| \leq 2}
\left| (\partial_k^{\alpha} \widehat{\rho}_{\lambda})(p,k) \right| \, 
|f_{\alpha}(k)|
\right|^2\nonumber\\
&\leq& \int\limits_{\R^3} dk |\varphi_j(k)|^2 
\int\limits_{\supp \varphi_j}  dk 
|\widehat{\rho}_{\lambda}(p,k)|^2 + \nonumber\\
&&+200 \max_{|\alpha| \leq 2}
\int\limits_{\R^3} dk |f_{\alpha}(k)|^2 
\max_{|\alpha| \leq 2} \int\limits_{K} dk \left|
\partial_k^{\alpha} \widehat{\rho}_{\lambda}(p,k) \right|^2, \nonumber
\end{eqnarray}
where $K:=\displaystyle \bigcup_{|\alpha| \leq 2} \supp f_{\alpha}$ is
a compact subset of $\R^3 \backslash \{0\}$. In the last estimate, we 
integrated $\displaystyle \int\limits_1^{\infty} s^{-2} ds =1$ and used, 
that there are 10 multi-indices $\alpha \in \N_0^3$ with $|\alpha| \leq 2$.
Thus by Theorem
\ref{thm3.3}, the bound on the right hand side of
(\ref{eq273}), which is uniform in $t$, is integrable in $(\R^3,dp)$, 
so by dominated convergence
\begin{eqnarray}
\lefteqn{Q^{(2)}_{\lambda}(\varphi_j)= 
\lim_{t \to \infty} \int\limits_{\R^3} dp \left| 
\,\int\limits_{-t}^t ds \int\limits_{\R^3}dk\,  
e^{is(p^2-e_0-\omega(k))} \widehat{\rho}_{\lambda}(p,k) \varphi_j(k)\right|^2 =
\label{eq274} }\\
&=& \int\limits_{\R^3} dp \left| \lim_{t \to \infty} 
\,\int\limits_{\R^3} dk  \int\limits_{-t}^t ds \, 
e^{is(p^2-e_0-\omega(k))} \widehat{\rho}_{\lambda}(p,k) \varphi_j(k)\right|^2= 
\nonumber\\
&=& \int\limits_{\R^3} dp \left| \lim_{t \to \infty} 
\,\int\limits_{\R^3} dk  \, 
\frac {e^{it(p^2-e_0-\omega(k))}- e^{-it(p^2-e_0-\omega(k))}} 
{i(p^2-e_0-\omega(k))} \widehat{\rho}_{\lambda}(p,k) \varphi_j(k)\right|^2=
\nonumber\\
&=&\int\limits_{\R^3} dp \left| \lim_{t \to \infty} 
\,\int\limits_0^{\infty} dr  \, 
\frac {e^{it(p^2-e_0-r)}- e^{-it(p^2-e_0-r)}} 
{i(p^2-e_0-r)} \int\limits_{S^2(r)} d\mu_r(k) 
\widehat{\rho}_{\lambda}(p,k) \varphi_j(k)\right|^2\hspace{-1pt}. \nonumber
\end{eqnarray}
In the last step we changed to polar coordinates for the
$k$-integration and used the Lebesgue measure $\mu_r$ on the sphere
$S^2(r):=\{k \in \R^3: |k|=r\}$, which is normalised as
$\mu_r(S^2(r))=4\pi r^2$. Passing to the new integration variable
$y:=p^2-e_0-r$ and 
\begin{equation} \label{eq275}
u_p(y):= \int\limits_{S^2(p^2-e_0-y)} d\mu_{p^2-e_0-y}(k)
\widehat{\rho}_{\lambda}(p,k) \varphi_j(k),
\end{equation}
we see, that the differentiability of $\widehat{\rho}_{\lambda}$ in $k$ and
$\varphi_j \in C_0^{\infty}(\R^3 \backslash \{0\})$ imply $u_p \in
C_0^1(\R)$, hence we can perform the $y$-integration in the limit 
$t \to \infty$ explicit:
\begin{eqnarray*}
Q^{(2)}_{\lambda}(\varphi_j)&=&
\int\limits_{\R^3} dp \left| \lim_{t \to \infty} \int\limits_{\R} dy 
\frac{2 \sin(ty) }{y} u_p(y) \right|^2=
\int\limits_{\R^3} dp \left| 2 \pi u_p(0) \right|^2=\nonumber\\
&=&\int\limits_{\R^3} dp \left|\; \int\limits_{S^2(p^2-e_0)} d\mu_{p^2-e_0}(k)
\widehat{\rho}_{\lambda}(p,k) \varphi_j(k) \right|^2. \\[-1.75cm]
\end{eqnarray*}
\end{proof}
\vspace*{0.5cm}
\begin{remark} \end{remark}
\noindent The formula (\ref{eq279}) for the second order
term $Q^{(2)}_{\lambda}(\varphi_j)$ of the ionisation probability 
produced by a single photon in an
incoming scattering state reflect just Einstein's condition:
Instead of having an electron with a momentum and a photon with one
frequency, we have an electron wavefunction (viewed in Fourier space with 
momentum as variable) and a photon wavefunction.
The integral $\int\limits_{\R^3} dp$ takes into account all possible 
electron momenta, the integrals
$\int\limits_{S^2(p^2-e_0)} \hspace{-10pt}d\mu_{p^2-e_0}(k)$ 
pose the condition $p^2-e_0-\omega(k)=0$. So this is a ``local version''
of (\ref{eqein2}) for fixed momenta $p$ and $k$ 
and for the free kinetic energy $p^2$ of the electron and for 
ionisation gap $\lap E=|e_0|$ .
%
%
\setcounter{equation}{0}
\begin{appendix}
\section{(Regularised) field energy and relative bounds for 
creation- and annihilation operators}
Let $0 \leq \tilde{r}$ be an infrared and $r > \tilde{r}$ 
be an ultraviolet regularization parameter and define the 
regularised dispersion relation
\begin{equation} \label{gla10}
\omega_{(\tilde{r},r)}(k):=\omega(k) \one_{\{\tilde{r}\leq \omega(k) \leq r\}} (k) 
\end{equation}
and the regularised free field
\begin{equation} \label{gla11}
H_{f,(\tilde{r},r)}:=\sum_{\lambda \in \Z_2} \;\int\limits_{\R^3} dk 
\omega_{(\tilde{r},r)}(k) a_{\lambda}^*(k) a_{\lambda}(k)
\end{equation}
$H_{f,(\tilde{r},r)}$ is the second quantisation $d\Gamma(\omega_{(\tilde{r},r)})$ 
of the multiplication
with $\omega_{(\tilde{r},r)}$, so as in the non-regularised case 
$\tilde{r}=0$ and $r=\infty$ the pull-through formula
\begin{eqnarray}
a_{\lambda}(k) F(H_{f,(\tilde{r},r)})&=&
F(H_{f,(\tilde{r},r)}+\omega_{(\tilde{r},r)}(k)) a_{\lambda}(k) 
\label{gla12}\\
F(H_{f,(\tilde{r},r)})a^*_{\lambda}(k)&=&
a^*_{\lambda}(k) F(H_{f,(\tilde{r},r)}+\omega_{(\tilde{r},r)}(k)) 
\label{gla13}
\end{eqnarray}
hold true and for some measurable $F:\R \to \C$. The restriction of 
$F(H_{f,(\tilde{r},r)})$ to the $n$-photon sector
$\cS_n(L^2(\R^3 \times \Z_2)^n)$ is the multiplication operator 
\[\Psi(k_1,\lambda_1,...,k_n,\lambda_n) \mapsto 
F(\omega_{(\tilde{r},r)}(k_1)+...+\omega_{(\tilde{r},r)}(k_n))
\Psi(k_1,\lambda_1,...,k_n,\lambda_n).\]
\begin{lemma} \label{coa1}
For $0\leq \tilde{s} \leq \tilde{r} <r\leq s \leq \infty$ and 
$0 \leq \beta \leq \alpha$
\begin{equation}
\| (H_{f,(\tilde{r},r)}+1+\omega_{(\tilde{r},r)}(k))^{\beta}
(H_{f,(\tilde{s},s)}+1+\omega_{(\tilde{s},s)}(k))^{-\alpha} \|
\leq  1 \label{gla14} \end{equation}
\begin{eqnarray}
\lefteqn{\| (H_{f,(\tilde{r},r)}+1+\omega_{(\tilde{r},r)}(k)+
\omega_{(\tilde{r},r)}(k'))^{\beta}
(H_{f,(\tilde{s},s)}+1+\omega_{(\tilde{s},s)}(k)+
\omega_{(\tilde{s},s)}(k''))^{-\alpha} \|\nonumber} \\
&\leq& (1+\omega_{(\tilde{r},r)}(k'))^{\beta}  \hspace{9.7cm}\label{gla15}
\end{eqnarray}
\end{lemma}
\begin{proof}
These two operators leave the $n$-photon sectors $\F^{(n)}$ invariant:
Applied to $\Psi_n \in \F^{(n)}=\cS_n(L^2(\R^3 \times \Z_2)^n)$ 
in the $n$ photon sector
the operator 
$(H_{f,(\tilde{r},r)}+1+\omega_{(\tilde{r},r)}(k))^{\beta}
(H_{f,(\tilde{s},s)}+1+\omega_{(\tilde{s},s)}(k))^{-\alpha}$ 
is just the
multiplication operator with the function
\[\frac{(\omega_{(\tilde{r},r)}(k_1)+...+\omega_{(\tilde{r},r)}(k_n)+1+
\omega_{(\tilde{r},r)}(k))^{\beta}}
{(\omega_{(\tilde{s},s)}(k_1)+...+\omega_{(\tilde{s},s)}(k_n)+1+
\omega_{(\tilde{s},s)}(k))^{\alpha}},\]
which has $L^{\infty}$ norm on $\{\tilde{s} \leq \omega(k) \leq s\}$
less or equal one due to the choices 
$0 \leq\tilde{s} \leq \tilde{r}< r \leq s \leq \infty$ and 
$0 \leq \beta \leq \alpha$.
For the second inequality, note that
\[\frac{\omega_{(\tilde{r},r)}(k_1)+...+\omega_{(\tilde{r},r)}(k_n)+1+
\omega_{(\tilde{r},r)}(k)+\omega_{(\tilde{r},r)}(k')}
{\omega_{(\tilde{s},s)}(k_1)+...+\omega_{(\tilde{s},s)}(k_n)+1+
\omega_{(\tilde{s},s)}(k)+\omega_{(\tilde{s},s)}(k'')} >0 \]
so by monotonicity of powers on $\R_+$
\begin{eqnarray*}
\lefteqn{
\frac{(\omega_{(\tilde{r},r)}(k_1)+...+\omega_{(\tilde{r},r)}(k_n)+1+
\omega_{(\tilde{r},r)}(k)+\omega_{(\tilde{r},r)}(k'))^{\beta}}
{(\omega_{(\tilde{s},s)}(k_1)+...+\omega_{(\tilde{s},s)}(k_n)+1+
\omega_{(\tilde{s},s)}(k)+\omega_{(\tilde{s},s)}(k''))^{\alpha}}\leq}\\
&\leq& 
\left(1+\frac{\omega_{(\tilde{r},r)}(k')-\omega_{(\tilde{r},r)}(k'')}
{\omega_{(\tilde{s},s)}(k_1)+...+\omega_{(\tilde{s},s)}(k_n)+1+
\omega_{(\tilde{s},s)}(k)+\omega_{(\tilde{s},s)}(k'')}\right)^{\beta} \leq\\
&\leq& (1+\omega_{(\tilde{r},r)}(k'))^{\beta},
\end{eqnarray*}
which proves, that any restriction to some $n$-photon sector has norm 
\linebreak
$\leq (1+\omega_{(\tilde{r},r)}(k'))^{\beta}$.
\end{proof}
\begin{lemma} \label{la3}
Let $0 \leq \tilde{s} \leq \tilde{r}< r \leq s \leq \infty$ and 
\[f:\{k \in \R^3:\tilde{s} \leq \omega(k) \leq s\} \times \Z_2 \to \C \] or 
\[f:\{k \in \R^3:\tilde{s} \leq \omega(k) \leq s\} \times \Z_2 \to L(\hel) \]
such that 
\[\vartheta_0:=\sum_{\lambda \in \Z_2} 
\int\limits_{\{\tilde{s} \leq \omega(k) \leq s\}} dk \,
\frac{\|f(k,\lambda)\|^2}{\omega_{(\tilde{s},s)}(k)} < \infty, \]
then for any $l,m \in \N_0$
\begin{equation}
\|(H_f+1)^{\frac{l}{2}}(H_{f,(\tilde{r},r)}+1)^{\frac{m}{2}} a_{\lambda}(f)
(H_{f,(\tilde{s},s)}+1)^{-\frac{m+1}{2}}(H_f+1)^{-\frac{l}{2}}\|
\leq \sqrt{\vartheta_0} \label{gla16}
\end{equation}
If for some $n \in \N_0$
\begin{eqnarray*}
\theta_n:&=&\sum_{\lambda \in \Z_2} \;\int\limits_{\{\tilde{s} \leq \omega(k) \leq s\}} 
dk (1+ \frac{1}{\omega_{(\tilde{s},s)}(k)}) (1+\omega_{(\tilde{r},r)}(k))^n \|f(k,\lambda)\|^2 < \infty
\end{eqnarray*}
then
\begin{equation}
\|(H_{f,(\tilde{r},r)}+1)^{\frac{n}{2}} a^*_{\lambda}(f)(H_{f,(\tilde{s},s)}+1)^{-\frac{n+1}{2}}\|
\leq \sqrt{\theta_n}. \label{gla20}
\end{equation}
If moreover $s < \infty$ and 
\[\vartheta:=\sum_{\lambda \in \Z_2} \;
\int\limits_{\{\tilde{s} \leq \omega(k) \leq s\}}
(1+\frac{1}{\omega_{(\tilde{s},s)}(k)}) \|f(k,\lambda)\|^2 < \infty, \]
then for any $m,n \in \N_0$
\begin{equation}
\|(H_f+1)^{\frac{m}{2}}(H_{f,(\tilde{r},r)}+1)^{\frac{n}{2}} a^*_{\lambda}(f)
(H_{f,(\tilde{s},s)}+1)^{-\frac{n+1}{2}}(H_f+1)^{-\frac{m}{2}}\|
\leq \sqrt{\vartheta}(1+s)^{\frac{m+n}{2}}. \label{gla50}
\end{equation}
\end{lemma}
\begin{proof}
Definition of $H_{f,(\tilde{s},s)}$ as a quadratic form and H\"older inequality 
imply
\begin{eqnarray}
\|a_{\lambda}(f) \Psi\|^2 &= &
\Big\| \int\limits_{\{\tilde{s} \leq \omega(k) \leq s\}} dk 
f(k,\lambda)^* a_{\lambda}(k) \Psi \Big\|^2 \leq
 \label{gla1}\\
&\leq & \Bigg[
\sum_{\lambda \in \Z_2} \;\int\limits_{\{\tilde{s} \leq \omega(k) \leq s\}} 
\hspace{-0.7cm} dk \frac{\|f(k,\lambda)\|^2}{\omega_{(\tilde{s},s)}(k)} \Bigg]
\Bigg[\sum_{\lambda \in \Z_2} \;\int\limits_{\{\tilde{s} \leq \omega(k) \leq s\}}  
\hspace{-0.7cm} dk
\omega_{(\tilde{s},s)}(k) \|a_{\lambda}(k) \Psi \|^2\Bigg] \nonumber\\
&\leq& \vartheta_0 \|H_{f,(\tilde{s},s)}^{\frac{1}{2}} \Psi \|^2, \nonumber
\end{eqnarray}
so 
$\|a_{\lambda}(f) (H_{f,(\tilde{s},s)}+1)^{-\frac{1}{2}} \| \leq 
\vartheta_0^{\frac{1}{2}}$.
If there are powers of $H_f+1$, $H_{f,(\tilde{r},r)}+1$ and $H_{f,(\tilde{s},s)}+1$ 
on both sides of 
$a_{\lambda}(f)$, the pull through
formula allows us to shift them to one side, rearrange them
because $[H_f,H_{f,(\tilde{s},s)}]=[H_f,H_{f,(\tilde{r},r)}]=
[H_{f,(\tilde{r},r)},H_{f,(\tilde{s},s)}]=0$
and finally use Lemma \ref{coa1}:
\begin{eqnarray}
\lefteqn{\|(H_f+1)^{\frac{l}{2}}(H_{f,(\tilde{r},r)}+1)^{\frac{m}{2}} a_{\lambda}(f) 
(H_{f,(\tilde{s},s)}+1)^{-\frac{m+1}{2}} (H_f+1)^{-\frac{l}{2}}\Psi\|= 
\nonumber}\\
&=& \Big\| \int\limits_{\{\tilde{s} \leq \omega(k) \leq s\}} dk 
(H_f+1)^{\frac{l}{2}} (H_{f,(\tilde{r},r)}+1)^{\frac{m}{2}} a_{\lambda}(k) 
f(k,\lambda)^* (H_{f,(\tilde{s},s)}+1)^{-\frac{m+1}{2}} \nonumber\\[-0.7cm]
&&\hspace{5cm}(H_f+1)^{-\frac{l}{2}} \Psi \Big\|
\nonumber\\
&=& \Big\| \int\limits_{\{\tilde{s} \leq \omega(k) \leq s\}} dk (H_{f,(\tilde{r},r)}+1)^{\frac{m}{2}} 
(H_{f,(\tilde{s},s)}+1+\omega_{(\tilde{s},s)}(k))^{-\frac{m}{2}}(H_f+1)^{\frac{l}{2}}  \nonumber\\[-0.6cm]
&& \hspace{2cm}
(H_f+1+\omega(k))^{-\frac{l}{2}} a_{\lambda}(k) 
f(k,\lambda)^* (H_{f,(\tilde{s},s)}+1)^{-\frac{1}{2}} \Psi \Big\|  \nonumber\\
&\leq & \sup_{{\{|k| \leq s\}}} \Big\| (H_{f,(\tilde{r},r)}+1)^{\frac{m}{2}} 
(H_{f,(\tilde{s},s)}+1+\omega_{(\tilde{s},s)}(k))^{-\frac{m}{2}} \Big\| \; \nonumber\\[-0.3cm]
&&\hspace{0.5cm} \sup_{{\{|k| \leq s\}}} \Big\| (H_f+1)^{\frac{l}{2}} 
(H_f+1+\omega(k))^{-\frac{l}{2}} \Big\| \;
\Big(\sum_{\lambda \in \Z_2}\; \int\limits_{\{\tilde{s} \leq \omega(k) \leq s\}} 
\hspace{-0.4cm}dk 
\frac{\|f(k,\lambda)\|^2}{\omega_{(\tilde{s},s)}(k)}\Big)^{\frac{1}{2}} \nonumber\\
&& \hspace{1cm}\Big(\sum_{\lambda \in \Z_2}\;
\int\limits_{\{\tilde{s} \leq \omega(k) \leq s\}} dk 
\omega_{(\tilde{s},s)}(k)\|a_{\lambda}(k) (H_{f,(\tilde{s},s)}+1)^{-\frac{1}{2}} 
\Psi \|^2 
\Big)^{\frac{1}{2}} = \nonumber\\
&=& \sqrt{\vartheta_0} 
\|H_{f,(\tilde{s},s)}^{\frac{1}{2}}(H_{f,(\tilde{s},s)}+1)^{-\frac{1}{2}} \Psi \|
\leq \sqrt{\vartheta_0} \|\Psi\|, \label{gla3}
\end{eqnarray}
proving (\ref{gla16}).
The canonical commutation relations allow us to convert creation into
annihilation operators plus some extra terms, so
\begin{eqnarray}
\lefteqn{\|a^*_{\lambda}(f) \Psi\|^2 = 
\Big\| \int\limits_{\{\tilde{s} \leq \omega(k) \leq s\}} dk f(k,\lambda)
 a^*_{\lambda}(k) 
\Psi \Big\|^2 = \nonumber}\\
&=& \hspace{-0.2cm}
\int\limits_{\{\tilde{s} \leq \omega(k_1) \leq s\}} \hspace{-0.5cm}
dk_1 \int\limits_{\{\tilde{s} \leq \omega(k_2) \leq s\}}  \hspace{-0.5cm} dk_2 
\lkl f(k_1,\lambda) \Psi,
(a^*_{\lambda}(k_2)a_{\lambda}(k_1) +\delta(k_1-k_2)) f(k_2,\lambda) \Psi \rkl
\nonumber\\
&=&\int\limits_{\{\tilde{s} \leq \omega(k) \leq s\}} \hspace{-0.3cm} dk 
\|f(k,\lambda) \Psi \|^2+ \nonumber\\
&&+ \int\limits_{\{\tilde{s} \leq \omega(k_1) \leq s\}} \hspace{-0.3cm} dk_1 
\hspace{-0.3cm}\int\limits_{\{\tilde{s} \leq \omega(k_2) \leq s\}} 
\hspace{-0.3cm}dk_2 
\lkl f(k_1,\lambda) a_{\lambda}(k_2) \Psi,f(k_2,\lambda) a_{\lambda}(k_1) \Psi
\rkl \nonumber\\
&\leq& \|\Psi\|^2 \int\limits_{\{\tilde{s} \leq \omega(k) \leq s\}} dk 
\|f(k,\lambda) \|^2  +
\Big( \sum_{\lambda \in \Z_2} \;\int\limits_{\{\tilde{s} \leq \omega(k) \leq s\}} dk 
\|f(k,\lambda)\| \|a_{\lambda}(k) \Psi \| \Big)^2  \nonumber\\
&\leq& \|\Psi\|^2 \hspace{-0.2cm} \int\limits_{\{\tilde{s} \leq \omega(k) \leq s\}} 
\hspace{-0.2cm}dk 
\|f(k,\lambda) \|^2  + \nonumber\\
&&+\sum_{\lambda \in \Z_2} \;\int\limits_{\{\tilde{s} \leq \omega(k) \leq s\}} \hspace{-0.2cm}dk 
\frac{\|f(k,\lambda)\|^2}{\omega_{(\tilde{s},s)}(k)}
\sum_{\lambda \in \Z_2} \;\int\limits_{\{\tilde{s} \leq \omega(k) \leq s\}} \hspace{-0.2cm}dk  
\omega_{(\tilde{s},s)}(k)\|a_{\lambda}(k) \Psi \|^2 \nonumber\\
&\leq& \theta_0 \|\Psi\|^2 + \theta_0 \|H_{f,(\tilde{s},s)}^{\frac{1}{2}} \Psi\|^2 =
\theta_0 \|(H_{f,(\tilde{s},s)}+1)^{\frac{1}{2}} \Psi \|^2. \label{gla2}
\end{eqnarray}
If there are powers of $H_{f,(\tilde{r},r)}+1$ and $H_{f,(\tilde{s},s)}+1$ 
on both sides of 
$a^*_{\lambda}(f)$, the pull through formula allows us to shift them to one 
side and by the canonical commutation relations we convert the
creation into annihilation operators and  use Lemma \ref{coa1}:
\begin{eqnarray}
\lefteqn{\|(H_{f,(\tilde{r},r)}+1)^{\frac{n}{2}} a^*_{\lambda}(f) 
(H_{f,(\tilde{s},s)}+1)^{-\frac{n+1}{2}} \Psi\|^2= \label{gla19}}\\
&=& \Big\| \int\limits_{\{\tilde{s} \leq \omega(k) \leq s\}} 
\hspace{-0.3cm}dk \;a^*_{\lambda}(k) 
f(k,\lambda) (H_{f,(\tilde{r},r)}+1+\omega_{(\tilde{r},r)}(k))^{\frac{n}{2}}
(H_{f,(\tilde{s},s)}+1)^{-\frac{n+1}{2}}\Psi\Big\|^2\nonumber\\
&=& \int\limits_{\{\tilde{s} \leq \omega(k) \leq s\}} dk
\Big\|f(k,\lambda) (H_{f,(\tilde{r},r)}+1+\omega_{(\tilde{r},r)}(k))^{\frac{n}{2}} 
(H_{f,(\tilde{s},s)}+1)^{-\frac{n+1}{2}}\Psi \Big\|^2+ \nonumber\\
&&+\hspace{-0.5cm} \int\limits_{\{\tilde{s} \leq \omega(k_1) \leq s\}} 
\hspace{-0.35cm}dk_1 \hspace{-0.35cm} 
\int\limits_{\{\tilde{s} \leq \omega(k_2) \leq s\}} \hspace{-0.3cm} dk_2 \Big\lkl
f(k_1,\lambda) (H_{f,(\tilde{r},r)}+1+\omega_{(\tilde{r},r)}(k_1)+
\omega_{(\tilde{r},r)}(k_2))^{\frac{n}{2}} 
\nonumber \\[-0.5cm]
&& \hspace{3.45cm} (H_{f,(\tilde{s},s)}+1+\omega_{(\tilde{s},s)}(k_2))^{-\frac{n}{2}}  
a_{\lambda}(k_2) (H_{f,(\tilde{s},s)}+1)^{-\frac{1}{2}}\Psi, \nonumber\\
&&\hspace{3.8cm} f(k_2,\lambda)
(H_{f,(\tilde{r},r)}+1+\omega_{(\tilde{r},r)}(k_1)+
\omega_{(\tilde{r},r)}(k_2))^{\frac{n}{2}} \nonumber\\ 
&& \hspace{3.45cm}
(H_{f,(\tilde{s},s)}+1+\omega_{(\tilde{s},s)}(k_1))^{-\frac{n}{2}} 
a_{\lambda}(k_1)(H_{f,(\tilde{s},s)}+1)^{-\frac{1}{2}}\Psi \Big\rkl \nonumber \\
&\leq& \int\limits_{\{\tilde{s} \leq \omega(k) \leq s\}} dk 
(1+\omega_{(\tilde{r},r)}(k))^n
\Big\|f(k,\lambda)  (H_{f,(\tilde{s},s)}+1)^{-\frac{1}{2}}\Psi
\Big\|^2+ \nonumber\\
&&+\hspace{-0.5cm}\int\limits_{\{\tilde{s} \leq \omega(k_1) \leq s\}}
\hspace{-0.35cm} dk_1 \hspace{-0.35cm}
\int\limits_{\{\tilde{s} \leq \omega(k_2) \leq s\}} \hspace{-0.4cm}dk_2 
(1+\omega_{(\tilde{r},r)}(k_1))^{\frac{n}{2}}
\|f(k_1,\lambda) 
a_{\lambda}(k_2) (H_{f,(\tilde{s},s)}+1)^{-\frac{1}{2}}\Psi \| \nonumber\\[-0.6cm]
&&\hspace{3.7cm} (1+\omega_{(\tilde{r},r)}(k_2))^{\frac{n}{2}}
\| f(k_2,\lambda)
a_{\lambda}(k_1)(H_{f,(\tilde{s},s)}+1)^{-\frac{1}{2}}\Psi \| \nonumber\\
&\leq& \Big\| (H_{f,(\tilde{s},s)}+1)^{-\frac{1}{2}}\Psi \Big\|^2 
\int\limits_{\{\tilde{s} \leq \omega(k) \leq s\}} dk (1+\omega_{(\tilde{r},r)}(k))^n
\|f(k,\lambda)\|^2+ \nonumber\\
&&+\Big( \int\limits_{\{\tilde{s} \leq \omega(k) \leq s\}} dk (1+\omega_{(\tilde{r},r)}(k))^{\frac{n}{2}}
\|f(k,\lambda) \| \cdot
\| a_{\lambda}(k) (H_{f,(\tilde{s},s)}+1)^{-\frac{1}{2}}\Psi \| \Big)^2 \nonumber\\
&\leq& \theta_n \Big\| (H_{f,(\tilde{s},s)}+1)^{-\frac{1}{2}}\Psi \Big\|^2 +
\sum_{\lambda \in \Z_2 } \; \int\limits_{\{\tilde{s} \leq \omega(k) \leq s\}} dk (1+\omega_{(\tilde{r},r)}(k))^n
\frac{\|f(k,\lambda) \|^2}{\omega_{(\tilde{s},s)}(k)} \nonumber\\
&& \hspace{1cm}\sum_{\lambda \in \Z_2 } \;\int\limits_{\{\tilde{s} \leq \omega(k) \leq s\}} dk 
\omega_{(\tilde{s},s)}(k) \| a_{\lambda}(k) (H_{f,(\tilde{s},s)}+1)^{-\frac{1}{2}}\Psi \|^2 \nonumber\\
&\leq& \theta_n \Big\| (H_{f,(\tilde{s},s)}+1)^{-\frac{1}{2}}\Psi \Big\|^2 +
\theta_n \|H_{f,(\tilde{s},s)}^{\frac{1}{2}}(H_{f,(\tilde{s},s)}+1)^{-\frac{1}{2}}\Psi \|^2 
=\theta_n\|\Psi\|^2  \nonumber
\end{eqnarray}
If $s < \infty$, then with the additional estimate $\omega(k) \leq s$ on
$\{|k| \leq s\}$ and along the same lines:
\begin{eqnarray}
\lefteqn{\|(H_f+1)^{\frac{m}{2}}(H_{f,(\tilde{r},r)}+1)^{\frac{n}{2}} a^*_{\lambda}(f) 
(H_{f,(\tilde{s},s)}+1)^{-\frac{n+1}{2}} (H_f+1)^{-\frac{m}{2}}\Psi\|^2= 
\label{gla52}}\\
&=& \Big\| \int\limits_{\{\tilde{s} \leq \omega(k) \leq s\}} dk a^*_{\lambda}(k) 
f(k,\lambda) (H_f+1+\omega(k))^{\frac{m}{2}} 
(H_{f,(\tilde{r},r)}+1+\omega_{(\tilde{r},r)}(k))^{\frac{n}{2}} \nonumber\\[-0.5cm]
&& \hspace{2cm}(H_{f,(\tilde{s},s)}+1)^{-\frac{n+1}{2}}(H_f+1)^{-\frac{m}{2}}\Psi
\Big\|^2=\nonumber\\
&=& \hspace{-8pt}\int\limits_{\{\tilde{s} \leq \omega(k) \leq s\}} \hspace{-10pt}dk
\Big\|f(k,\lambda) 
\Big(\frac{H_f+1+\omega(k)}{H_f+1} \Big)^{\frac{m}{2}}
(H_{f,(\tilde{r},r)}+1+\omega_{(\tilde{r},r)}(k))^{\frac{n}{2}} \nonumber\\[-0.7cm]
&&\hspace{7cm}(H_{f,(\tilde{s},s)}+1)^{-\frac{n+1}{2}}\Psi \Big\|^2 \nonumber\\
&&+\hspace{-0.3cm}\int\limits_{\{\tilde{s} \leq \omega(k_1) \leq s\}} 
\hspace{-0.3cm}dk_1 \hspace{-0.3cm} 
\int\limits_{\{\tilde{s} \leq \omega(k_2) \leq s\}} \hspace{-0.4cm} dk_2 \Big\lkl
f(k_1,\lambda) (H_{f,(\tilde{r},r)}+1+\omega_{(\tilde{r},r)}(k_1)+
\omega_{(\tilde{r},r)}(k_2))^{\frac{n}{2}} 
\nonumber \\[-0.5cm]
&& \hspace{3.8cm}
\Big(\frac{H_f+1+\omega(k_1)+\omega(k_2)}{H_f+1+\omega(k_2)}\Big)^{\frac{m}{2}}
\nonumber\\[-0.1cm]
&& \hspace{3.3cm} (H_{f,(\tilde{s},s)}+1+\omega_{(\tilde{s},s)}(k_2))^{-\frac{n}{2}}  
a_{\lambda}(k_2) (H_{f,(\tilde{s},s)}+1)^{-\frac{1}{2}}\Psi, \nonumber\\
&&\hspace{3.3cm} f(k_2,\lambda)
\Big(\frac{H_f+1+\omega(k_1)+\omega(k_2)}{H_f+1+\omega(k_1)}\Big)^{\frac{m}{2}}
\nonumber\\
&&\hspace{3.3cm}
(H_{f,(\tilde{r},r)}+1+\omega_{(\tilde{r},r)}(k_1)+
\omega_{(\tilde{r},r)}(k_2))^{\frac{n}{2}} \nonumber\\ 
&& \hspace{3.3cm}
(H_{f,(\tilde{s},s)}+1+\omega_{(\tilde{s},s)}(k_1))^{-\frac{n}{2}} 
a_{\lambda}(k_1)(H_{f,(\tilde{s},s)}+1)^{-\frac{1}{2}}\Psi \Big\rkl \nonumber\\
&\leq& \int\limits_{\{\tilde{s} \leq \omega(k) \leq s\}} dk 
(1+\omega(k))^m (1+\omega_{(\tilde{r},r)}(k))^n
\|f(k,\lambda)\|^2 \|(H_{f,(\tilde{s},s)}+1)^{-\frac{1}{2}} \Psi \|^2 \nonumber\\
&&+\int\limits_{\{\tilde{s} \leq \omega(k_1) \leq s\}} dk_1 \int\limits_{\{\tilde{s} \leq \omega(k_2) \leq s\}} dk_2
\|f(k_1,\lambda)\| (1+\omega(k_1))^{\frac{m}{2}} 
(1+\omega_{(\tilde{r},r)}(k_1))^{\frac{n}{2}} \nonumber\\[-0.2cm]
&& \hspace{3.8cm}
\|a_{\lambda}(k_2) (H_{f,(\tilde{s},s)}+1)^{-\frac{1}{2}} \Psi \|
\|f(k_2,\lambda)\| (1+\omega(k_2))^{\frac{m}{2}} \nonumber\\
&& \hspace{3.8cm}
(1+\omega_{(\tilde{r},r)}(k_2))^{\frac{n}{2}} 
\|a_{\lambda}(k_1) (H_{f,(\tilde{s},s)}+1)^{-\frac{1}{2}} \Psi \| \nonumber\\
&\leq& (1+s)^{m+n} \vartheta (\|(H_{f,(\tilde{s},s)}+1)^{-\frac{1}{2}} \Psi\|^2+
\|H_{f,(\tilde{s},s)}^{\frac{1}{2}} (H_{f,(\tilde{s},s)}+1)^{-\frac{1}{2}} \Psi\|^2)=
\nonumber\\
&=& (1+s)^{m+n} \vartheta \|\Psi\|^2. \nonumber \\[-1.3cm]\nonumber
\end{eqnarray}
\end{proof}
\vspace*{0.3cm}
\begin{corollary} \label{coa2}
Let $f_1,...,f_N \in C_0^{\infty}(\R^3 \backslash \{0\})$,
$\lambda_1,...,\lambda_N \in \Z_2$ and 
\begin{eqnarray*}
0 \leq \tilde{r} &<& \inf\{\omega(k): k \in \supp f_j: j=1,...,N\} \\
\infty > r&>&\sup\{\omega(k): k \in \supp f_j: j=1,...,N\}
\end{eqnarray*}
then for any $m,\gamma \in \N_0$ and $t \in \R$
\begin{eqnarray*}
&&(\hel-b)^{\frac{\gamma}{2}} (H_f+1)^{\frac{m}{2}} e^{-itH_0} 
a^*_{\lambda_1}(f_1) \cdots a^*_{\lambda_N}(f_N) e^{itH_0}
(H_{f,(\tilde{r},r)}+1)^{-\frac{N}{2}} \\
&&\hspace{3cm}(H_f+1)^{-\frac{m}{2}} (\hel-b)^{-\frac{\gamma}{2}}
\end{eqnarray*}
defines a bounded operator on $\hr$ and moreover
\begin{eqnarray*}
&& \sup_{t \in \R} 
\Big\|(\hel-b)^{\frac{\gamma}{2}} (H_f+1)^{\frac{m}{2}} e^{-itH_0} 
a^*_{\lambda_1}(f_1) \cdots a^*_{\lambda_N}(f_N) e^{itH_0}
(H_{f,(\tilde{r},r)}+1)^{-\frac{N}{2}} \\[-0.2cm] 
&& \hspace{1cm}(H_f+1)^{-\frac{m}{2}}(\hel-b)^{-\frac{\gamma}{2}}\Big \| 
\leq \|f_1\|_{\omega} \cdots \|f_N\|_{\omega} (1+r)^{\frac{N}{4}(2m+N-1)}
< \infty 
\end{eqnarray*}
\end{corollary}
\begin{proof}
The creation operators $a^*_{\lambda_1}(f_1),...,a^*_{\lambda_N}(f_N)$
act on the photon Fock space $\F$ and 
$e^{\pm itH_0}=e^{\pm it\hel} \otimes e^{\pm itH_f}$, so
\begin{eqnarray*}
\lefteqn{e^{-itH_0} a^*_{\lambda_1}(f_1) \cdots a^*_{\lambda_N}(f_N) e^{itH_0}=
e^{-itH_f} a^*_{\lambda_1}(f_1) \cdots a^*_{\lambda_N}(f_N) e^{itH_f}=}\\
&=&a^*_{\lambda_1}(e^{-it\omega} f_1) \cdots a^*_{\lambda_N}(e^{-it\omega}f_N).
\hspace{5cm}
\end{eqnarray*}
Now $\hel$ commutes with all other terms, so
\begin{eqnarray*}
\lefteqn{(\hel-b)^{\frac{\gamma}{2}} (H_f+1)^{\frac{m}{2}}  
a^*_{\lambda_1}(e^{-it\omega}f_1) \cdots a^*_{\lambda_N}(e^{-it\omega}f_N) 
(H_{f,(\tilde{r},r)}+1)^{-\frac{N}{2}}} \\
&& \hspace{2cm}(H_f+1)^{-\frac{m}{2}}(\hel-b)^{-\frac{\gamma}{2}}= \\
&=&(H_f+1)^{\frac{m}{2}}  
a^*_{\lambda_1}(e^{-it\omega}f_1) \cdots a^*_{\lambda_N}(e^{-it\omega}f_N) 
(H_{f,(\tilde{r},r)}+1)^{-\frac{N}{2}} (H_f+1)^{-\frac{m}{2}}
\end{eqnarray*}
Due to the choice of $f_1,...,f_N \in C_0^{\infty}(\R^3 \backslash \{0\})$ 
\[\|f_j\|_{\omega}^2=\int\limits_{\R^3}
(1+\frac{1}{\omega_{(\tilde{r},r)}(k)}) \|f_j(k,\lambda)\|^2 
< \infty, \]
Inserting identities as
$\one_{\F}=(H_{f,(\tilde{r},r)}+1)^{-\frac{j}{2}}(H_f+1)^{-\frac{m}{2}}
(H_f+1)^{\frac{m}{2}}(H_{f,(\tilde{r},r)}+1)^{\frac{j}{2}}$
and applying Lemma \ref{la3}, equation (\ref{gla50}) one gets:
\begin{eqnarray*}
\lefteqn{
\Big\|(\hel-b)^{\frac{\gamma}{2}} (H_f+1)^{\frac{m}{2}} 
a^*_{\lambda_1}(e^{-it\omega} f_1) \cdots a^*_{\lambda_N}(e^{-it\omega}f_N)
(H_{f,(\tilde{r},r)}+1)^{-\frac{N}{2}}}\\[-0.3cm]
&& \hspace{3cm}(H_f+1)^{\frac{m}{2}} (\hel-b)^{-\frac{\gamma}{2}}\Big \|\leq \\
&\leq & \prod_{j=1}^{N} \|(H_f+1)^{\frac{m}{2}}(H_{f,(\tilde{r},r)}+1)^{\frac{j-1}{2}}
a^*_{\lambda_j}(e^{-it\omega}f_j) (H_{f,(\tilde{r},r)}+1)^{-\frac{j}{2}}
(H_f+1)^{-\frac{m}{2}} \| \\
&\leq& \|f_1\|_{\omega} \cdots \|f_N\|_{\omega} (1+r)^{\frac{N}{4}(2m+N-1)}
< \infty \hspace{3cm}
\end{eqnarray*}
independent of $t$.
\end{proof}
\begin{lemma} \label{la2}
If Hypothesis \ref{H-1} is satisfied and
$\Lambda^{(1)}_{0,\gamma},\Lambda^{(1)}_{\beta,\gamma},
\widetilde{\Lambda}^{(1)}_{\beta,\gamma}< \infty$, 
then for any  $\alpha \in \N_0$ and 
$0 \leq \tilde{s} \leq \tilde{r} <r \leq s < \infty$
\begin{eqnarray*}
\lefteqn{\Big\|(H_{f,(\tilde{r},r)}+1)^{\frac{\alpha}{2}} (H_f+1)^{\frac{\beta}{2}}  
(\hel-b)^{\frac{\gamma}{2}} W^{(1)} (\hel-b)^{-\frac{\gamma+1}{2}}
(H_f+1)^{-\frac{\beta+1}{2}} }\\
&&(H_{f,(\tilde{s},s)}+1)^{-\frac{\alpha}{2}} \Big\|
\leq
\sqrt{\Lambda^{(1)}_{0,\gamma}}+(1+r)^{\frac{\alpha}{2}} 
\max\Big\{\sqrt{\Lambda^{(1)}_{\beta,\gamma}},
\sqrt{\widetilde{\Lambda}^{(1)}_{\beta,\gamma}} \Big\}
\end{eqnarray*}
\end{lemma}
\begin{proof}
The operators $H_{f,(\tilde{s},s)}$, $H_f$ and $\hel$ commute, so
for the $W^{(0,1)}$ term pull-through formula, (\ref{gla15}), 
H\"older inequality and the definition 
of $\Lambda^{(1)}_{0,\gamma}$ in (\ref{gl58}) gives:
\begin{eqnarray}
\lefteqn{ \Big\|(H_{f,(\tilde{r},r)}+1)^{\frac{\alpha}{2}} (H_f+1)^{\frac{\beta}{2}}
(\hel-b)^{\frac{\gamma}{2}}W^{(0,1)} (\hel-b)^{-\frac{\gamma+1}{2}}
(H_f+1)^{-\frac{\beta+1}{2}} \nonumber} \\[-0.2cm]
&&\hspace{5cm}(H_{f,(\tilde{s},s)}+1)^{-\frac{\alpha}{2}} \Psi \Big\|= \nonumber\\
&=& \Big\| \sum_{\lambda \in \Z_2} \int\limits_{\R^3} dk 
(H_{f,(\tilde{r},r)}+1)^{\frac{\alpha}{2}} 
(H_{f,(\tilde{s},s)}+1+\omega_{(\tilde{s},s)}(k))^{-\frac{\alpha}{2}} 
\frac{(H_f+1)^{\frac{\beta}{2}}}{(H_f+1+\omega(k))^{\frac{\beta}{2}} }
\nonumber\\[-0.2cm]
&& \hspace{2.2cm} (\hel-b)^{\frac{\gamma}{2}} w^{(0,1)}(k,\lambda) 
(\hel-b)^{-\frac{\gamma+1}{2}}
a_{\lambda}(k) (H_f+1)^{-\frac{1}{2}} \Psi \Big\|\nonumber\\
&\leq& \sum_{\lambda \in \Z_2} \int\limits_{\R^3} dk \Big\|
(\hel-b)^{\frac{\gamma}{2}} w^{(0,1)}(k,\lambda) 
(\hel-b)^{-\frac{\gamma+1}{2}}
a_{\lambda}(k) (H_f+1)^{-\frac{1}{2}} \Psi \Big\|\nonumber\\
&\leq& 
\Big( \Lambda_{0,\gamma}^{(1)} \sum_{\lambda \in \Z_2} 
\int\limits_{\R^3} dk \omega(k)
\|a_{\lambda}(k)(H_f+1)^{-\frac{1}{2}}\Psi\|^2 \Big)^{\frac{1}{2}}= \nonumber\\
&=& \|H_f^{\frac{1}{2}} (H_f+1)^{-\frac{1}{2}} \Psi\|
\sqrt{\Lambda_{0,\gamma}^{(1)}} 
\leq \|\Psi\| \sqrt{\Lambda_{0,\gamma}^{(1)}} < \infty 
\label{gla30}
\end{eqnarray}
For the next term we use pull-through plus commutation relations to bring it
in an appropriate form substituting creation operators by annihilation
operators, then along the same line as in (\ref{gla30}) we get:
\begin{eqnarray}
\lefteqn{ 
\Big\|(H_f+1)^{\frac{\beta}{2}} (H_{f,(\tilde{r},r)}+1)^{\frac{\alpha}{2}} 
(\hel-b)^{\frac{\gamma}{2}} W^{(1,0)} (\hel-b)^{-\frac{\gamma+1}{2}}
(H_{f,(\tilde{s},s)}+1)^{-\frac{\alpha}{2}} \nonumber}\\[-0.3cm]
&& \hspace{5cm} (H_f+1)^{-\frac{\beta+1}{2}} \Psi \Big\|^2=\nonumber\\
&=&\Big \| \sum_{\lambda \in \Z_2} \int\limits_{\R^3} dk  
(\hel-b)^{\frac{\gamma}{2}} w^{(1,0)}(k,\lambda) (\hel-b)^{-\frac{\gamma+1}{2}} 
a^*_{\lambda}(k) (H_f+1+\omega(k))^{\frac{\beta}{2}}\nonumber\\[-0.5cm]
&&\hspace{2cm} (H_{f,(\tilde{r},r)}+1+\omega_{(\tilde{r},r)}(k))^{\frac{\alpha}{2}}
(H_{f,(\tilde{s},s)}+1)^{-\frac{\alpha}{2}} (H_f+1)^{-\frac{\beta+1}{2}}
\Psi \Big \|^2 \nonumber\\
&=&\hspace{-0.3cm}\sum_{\lambda_1,\lambda_2 \in \Z_2} \int\limits_{\R^3} dk_1
\int\limits_{\R^3} dk_2 \Big\lkl (\hel-b)^{\frac{\gamma}{2}} 
w^{(1,0)}(k_1,\lambda_1) (\hel-b)^{-\frac{\gamma+1}{2}} \nonumber\\[-0.5cm]
&& \hspace{3.3cm}(H_{f,(\tilde{r},r)}+1+
\omega_{(\tilde{r},r)}(k_1)+\omega_{(\tilde{r},r)}(k_2))^{\frac{\alpha}{2}}\nonumber\\
&& \hspace{3.3cm}
(H_{f,(\tilde{s},s)}+1+\omega_{(\tilde{s},s)}(k_2))^{-\frac{\alpha}{2}} \nonumber\\
&& \hspace{3.3cm}\frac{(H_f+1+\omega(k_1)+\omega(k_2))^{\frac{\beta}{2}}}
{(H_f+1+\omega(k_2))^{\frac{\beta}{2}}}
a_{\lambda_2}(k_2) (H_f+1)^{-\frac{1}{2}} \Psi, \nonumber\\
&& \hspace{3.3cm} (\hel-b)^{\frac{\gamma}{2}} w^{(1,0)}(k_2,\lambda_2) 
(\hel-b)^{-\frac{\gamma+1}{2}} \nonumber\\
&& \hspace{3.3cm}(H_{f,(\tilde{r},r)}+1+
\omega_{(\tilde{r},r)}(k_1)+\omega_{(\tilde{r},r)}(k_2))^{\frac{\alpha}{2}}\nonumber\\
&& \hspace{3.3cm}
(H_{f,(\tilde{s},s)}+1+\omega_{(\tilde{s},s)}(k_1))^{-\frac{\alpha}{2}} \nonumber\\
&& \hspace{3.3cm}\frac{(H_f+1+\omega(k_1)+\omega(k_2))^{\frac{\beta}{2}}}
{(H_f+1+\omega(k_1))^{\frac{\beta}{2}}}
a_{\lambda_1}(k_1) (H_f+1)^{-\frac{1}{2}} \Psi \Big\rkl\nonumber \\
&&+\sum_{\lambda \in \Z_2} \int\limits_{\R^3} dk  
\Big\|(\hel-b)^{\frac{\gamma}{2}} w^{(1,0)}(k,\lambda) (\hel-b)^{-\frac{\gamma+1}{2}} 
\frac{(H_f+1+\omega(k))^{\frac{\beta}{2}}}{(H_f+1)^{\frac{\beta}{2}}}
\nonumber\\[-0.2cm]
&&\hspace{2.4cm} (H_{f,(\tilde{r},r)}+1+\omega_{(\tilde{r},r)}(k))^{\frac{\alpha}{2}}
(H_{f,(\tilde{s},s)}+1)^{-\frac{\alpha}{2}} (H_f+1)^{-\frac{1}{2}}
\Psi \Big \|^2  \nonumber\\
&\leq& \sum_{\lambda \in \Z_2} \int\limits_{\R^3} dk 
(1+\omega(k))^{\beta}
\frac{\|(\hel-b)^{\frac{\gamma}{2}} w^{(1,0)}(k,\lambda) 
(\hel-b)^{-\frac{\gamma+1}{2}}\|^2}{\omega(k)}
\nonumber\\[-0.4cm]
&&\hspace{2.5cm}(1+\omega_{(\tilde{r},r)}(k))^{\alpha}
\sum_{\lambda \in \Z_2} \int\limits_{\R^3} dk 
\omega(k) \|a_{\lambda}(k) (H_f+1)^{-\frac{1}{2}} \Psi \|^2 \nonumber\\
&&+\sum_{\lambda \in \Z_2} \int\limits_{\R^3} dk (1+\omega(k))^{\beta}
\|(\hel-b)^{\frac{\gamma}{2}} w^{(1,0)}(k,\lambda) 
(\hel-b)^{-\frac{\gamma+1}{2}} \|^2 \nonumber\\[-0.6cm]
&&\hspace{3cm}(1+\omega_{(\tilde{r},r)}(k))^{\alpha}
\|(H_f+1)^{-\frac{1}{2}} \Psi\|^2 \nonumber\\
&\leq& (1+r)^{\alpha} \Big(\Lambda^{(1)}_{\beta,\gamma} 
\|H_f^{\frac{1}{2}} (H_f+1)^{-\frac{1}{2}} \Psi\|^2+
\widetilde{\Lambda}^{(1)}_{\beta,\gamma} \|(H_f+1)^{-\frac{1}{2}} \Psi\|^2\Big) 
\leq \nonumber\\
&\leq& (1+r)^{\alpha} \max\{\Lambda^{(1)}_{\beta,\gamma},
\widetilde{\Lambda}^{(1)}_{\beta,\gamma} \}
\Big( 
\big\lkl (H_f+1)^{-\frac{1}{2}} \Psi, H_f (H_f+1)^{-\frac{1}{2}} \Psi \big \rkl +
\nonumber\\
&& \hspace{4.7cm}
\big\lkl (H_f+1)^{-\frac{1}{2}} \Psi, (H_f+1)^{-\frac{1}{2}} \Psi \big \rkl
\Big)= \nonumber\\
&= & (1+r)^{\alpha} \max\{\Lambda^{(1)}_{\beta,\gamma},
\widetilde{\Lambda}^{(1)}_{\beta,\gamma} \} \|\Psi\|^2. \label{gla60}
\end{eqnarray}
Adding up (\ref{gla30}) and (\ref{gla60}) finishes the proof.
\end{proof}
\begin{lemma} \label{la4}
If $\Lambda^{(2)}_{0,\gamma},\Lambda^{(2)}_{\beta,\gamma},
\widetilde{\Lambda}^{(2)}_{0,\gamma},
\widetilde{\Lambda}^{(2)}_{\frac{\beta}{2},\gamma},
\widetilde{\Lambda}^{(2)}_{\beta,\gamma}
< \infty$ and
Hypothesis \ref{H-1} and \ref{hyp5} are satisfied,
then for any  $\alpha \in \N_0$ and 
$0 \leq \tilde{s} \leq \tilde{r} <r \leq s < \infty$ 
\begin{eqnarray*}
&&\hspace{-1cm}\Big \|(H_{f,(\tilde{r},r)}+1)^{\frac{\alpha}{2}} 
(H_f+1)^{\frac{\beta}{2}}  
(\hel-b)^{\frac{\gamma}{2}} W^{(2)} (\hel-b)^{-\frac{\gamma}{2}}
(H_f+1)^{-\frac{\beta}{2}-1} \\[-0.4cm]
&&\hspace{9cm}(H_{f,(\tilde{s},s)}+1)^{-\frac{\alpha}{2}} \Big\| \\
&\leq&
\Lambda^{(2)}_{0,\gamma}+
2\sqrt{(\widetilde{\Lambda}^{(2)}_{\beta,\gamma}+
\Lambda^{(2)}_{\beta,\gamma}) \Lambda^{(2)}_{0,\gamma}} 
(1+r)^{\frac{\alpha}{2}}+
2(1+2r)^{\frac{\alpha}{2}} \max\{1,2^{\frac{\beta}{2}-1}\} \nonumber\\
&&\hspace{0.5cm} \Big[\Lambda^{(2)}_{\beta,\gamma} \Lambda^{(2)}_{0,\gamma} +
2\sqrt{\Lambda^{(2)}_{\beta,\gamma} \Lambda^{(2)}_{0,\gamma}} 
\widetilde{\Lambda}^{(2)}_{\frac{\beta}{2},\gamma}+
\Lambda^{(2)}_{\beta,\gamma} \widetilde{\Lambda}^{(2)}_{0,\gamma}+
\Lambda^{(2)}_{0,\gamma} \widetilde{\Lambda}^{(2)}_{\beta,\gamma} +
\widetilde{\Lambda}^{(2)}_{0,\gamma}
\widetilde{\Lambda}^{(2)}_{\frac{\beta}{2},\gamma} \Big] ^{\frac{1}{2}}
\nonumber
\end{eqnarray*}
\end{lemma}
\begin{proof}
The easiest two photon interaction term for the proof of this estimate
is $W^{(0,2)}$. The operators
\begin{equation} \label{gla61}
\widetilde{\fG}_{\iota}(k,\lambda)^{\#}:=(\hel-b)^{\frac{\gamma}{2}}
\fG_{\iota}(k,\lambda)^{\#}(\hel-b)^{-\frac{\gamma}{2}}
\end{equation}
define $L^2$-functions and
\[(\hel-b)^{\frac{\gamma}{2}} w^{(0,2)}(k_1,\lambda_1,k_2,\lambda_2)
(\hel-b)^{-\frac{\gamma}{2}}=
\sum_{\iota=1}^3 
\widetilde{\fG}_{\iota}(k_1,\lambda_1)\widetilde{\fG}_{\iota}(k_2,\lambda_2)\]
Using pull-through formula, we create free field terms of the form estimated 
in Lemma \ref{coa1}. Inserting this expression
for $w^{(0,2)}(k_1,\lambda_1,k_2,\lambda_2)$ we separate the two variables 
and use the definition of $H_f$ as a quadratic form:
\begin{eqnarray}
\lefteqn{ \|(H_{f,(\tilde{r},r)}+1)^{\frac{\alpha}{2}} (H_f+1)^{\frac{\beta}{2}} 
(\hel-b)^{\frac{\gamma}{2}} W^{(0,2)} (\hel-b)^{-\frac{\gamma}{2}} 
(H_f+1)^{-\frac{\beta}{2}-1}\nonumber}\\ 
&& \hspace{4cm} (H_{f,(\tilde{s},s)}+1)^{-\frac{\alpha}{2}} \Psi \|=\nonumber\\ 
&=& \hspace{-0.2cm} \Big\| \hspace{-0.2cm}
\sum_{\lambda_1,\lambda_2 \in \Z_2} \int\limits_{\R^3} \hspace{-0.1cm} dk_1
\hspace{-0.1cm}
\int\limits_{\R^3} \hspace{-0.1cm} dk_2 
(H_{f,(\tilde{r},r)}+1)^{\frac{\alpha}{2}} 
(H_{f,(\tilde{s},s)}+1+\omega_{(\tilde{s},s)}(k_1)+
\omega_{(\tilde{s},s)}(k_2))^{-\frac{\alpha}{2}} \nonumber\\[-0.5cm]
&& \hspace{3.4cm} (H_f+1)^{\frac{\beta}{2}}
(H_f+1+\omega(k_1)+\omega(k_2))^{-\frac{\beta}{2}} \nonumber\\
&& \hspace{3.4cm} (\hel-b)^{\frac{\gamma}{2}}
w^{(0,2)}(k_1,\lambda_1,k_2,\lambda_2) (\hel-b)^{\frac{\gamma}{2}}\nonumber\\
&& \hspace{3.4cm}
a_{\lambda_1}(k_1) a_{\lambda_2}(k_2) (H_f+1)^{-1} \Psi \Big\| \leq \nonumber\\
&\leq& \sum_{\lambda_2 \in \Z_2} \int\limits_{\R^3} dk_2
\Big( \sum_{\lambda_1 \in \Z_2} \int\limits_{\R^3} dk_1
\frac{\|\widetilde{\fG}(k_1,\lambda_1)\|^2}{\omega(k_1)} \Big)^{\frac{1}{2}} 
\nonumber\\
&& \hspace{1.5cm}
\Big( \sum_{\lambda_1 \in \Z_2} \int\limits_{\R^3} dk_1 \omega(k_1)
\Big\|a_{\lambda_1}(k_1) 
\widetilde{\fG}(k_2,\lambda_2)  a_{\lambda_2}(k_2) (H_f+1)^{-1} \Psi \Big\|^2
\Big)^{\frac{1}{2}}  \nonumber\\
&\leq& \sum_{\lambda_2 \in \Z_2} \int\limits_{\R^3} dk_2
\sqrt{\Lambda^{(2)}_{0,\gamma}}  \Big\|H_f^{\frac{1}{2}}
\widetilde{\fG}(k_2,\lambda_2) a_{\lambda_2}(k_2) (H_f+1)^{-1} \Psi \Big\|= 
\nonumber\\
&=& \sqrt{\Lambda^{(2)}_{0,\gamma}} \sum_{\lambda \in \Z_2} 
\int\limits_{\R^3} dk  \Big\|
H_f^{\frac{1}{2}} (H_f+1+\omega(k))^{-\frac{1}{2}} \widetilde{\fG}(k,\lambda) 
a_{\lambda}(k) (H_f+1)^{-\frac{1}{2}} \Psi \Big\| \nonumber\\
&\leq& \sqrt{\Lambda^{(2)}_{0,\gamma}} \Big( \sum_{\lambda \in \Z_2} 
\int\limits_{\R^3} dk
\frac{\|\widetilde{\fG}(k,\lambda)\|^2}{\omega(k)} \Big)^{\frac{1}{2}} 
\nonumber \\[-0.3cm]
&&\hspace{3cm}\Big( \sum_{\lambda \in \Z_2} \int\limits_{\R^3} dk \omega(k)
\|a_{\lambda}(k)(H_f+1)^{-\frac{1}{2}} \Psi \|^2 \Big)^ {\frac{1}{2}} 
\nonumber \\
&\leq& \Lambda^{(2)}_{0,\gamma} \|H_f^{\frac{1}{2}} (H_f+1)^{-\frac{1}{2}} \Psi\| 
\leq \Lambda^{(2)}_{0,\gamma} \|\Psi\| . \label{gla32}
\end{eqnarray}
For the $W^{(1,1)}$ term the change of creation into annihilation operators
by the canonical commutation relations (here we use
$a^*_{\lambda_2}(k_2) a_{\lambda_1}(k_1) a^*_{\lambda_3}(k_3) a_{\lambda_4}(k_4)=
a^*_{\lambda_3}(k_3) a^*_{\lambda_2}(k_2) a_{\lambda_1}(k_1) a_{\lambda_4}(k_4)
+\delta_{\lambda_1,\lambda_3} \delta(k_1-k_3) 
a^*_{\lambda_2}(k_2) a_{\lambda_4}(k_4)$)
 completes the program sketched above: 
\begin{eqnarray}
\lefteqn{\|(H_{f,(\tilde{r},r)}+1)^{\frac{\alpha}{2}} (H_f+1)^{\frac{\beta}{2}} 
(\hel-b)^{\frac{\gamma}{2}} W^{(1,1)}(\hel-b)^{-\frac{\gamma}{2}}
(H_f+1)^{-\frac{\beta}{2}-1}\nonumber}\\
&&\hspace{4cm} (H_{f,(\tilde{s},s)}+1)^{-\frac{\alpha}{2}}\Psi\|^2=\nonumber\\
&=& \Big\| \sum_{\lambda_1,\lambda_2 \in \Z_2} \int\limits_{\R^3} dk_1
\int\limits_{\R^3} dk_2 
(\hel-b)^{\frac{\gamma}{2}} w^{(1,1)}(k_1,\lambda_1,k_2,\lambda_2) 
(\hel-b)^{-\frac{\gamma}{2}}
a^*_{\lambda_1}(k_1) \nonumber\\[-0.5cm] 
&& \hspace{3.7cm}
(H_{f,(\tilde{r},r)}+1+\omega_{(\tilde{r},r)}(k_1))^{\frac{\alpha}{2}} (H_f+1+\omega(k_1))^{\frac{\beta}{2}} 
 \nonumber\\[-0.2cm]
&& \hspace{5cm}
a_{\lambda_2}(k_2)
(H_f+1)^{-\frac{\beta}{2}-1} (H_{f,(\tilde{s},s)}+1)^{-\frac{\alpha}{2}}
\Psi \Big\|^2 = \nonumber\\
&=&\sum_{\lambda_1,\lambda_2,\lambda_3,\lambda_4 \in \Z_2} 
\int\limits_{\R^3} dk_1 \int\limits_{\R^3} dk_2 \int\limits_{\R^3} dk_3
\int\limits_{\R^3} dk_4 \nonumber\\
&&\hspace{0.2cm} \Big\lkl 
(\hel-b)^{\frac{\gamma}{2}} w^{(1,1)}(k_1,\lambda_1,k_2,\lambda_2) 
(\hel-b)^{-\frac{\gamma}{2}}
(H_f+1+\omega(k_1)+\omega(k_3))^{\frac{\beta}{2}} \nonumber\\
&& \hspace{1.35cm}
(H_{f,(\tilde{r},r)}+1+\omega_{(\tilde{r},r)}(k_1)+
\omega_{(\tilde{r},r)}(k_3))^{\frac{\alpha}{2}}\nonumber\\
&& \hspace{1.35cm}
(H_{f,(\tilde{s},s)}+1+\omega_{(\tilde{s},s)}(k_2)
+\omega_{(\tilde{s},s)}(k_3))^{-\frac{\alpha}{2}} \nonumber\\
&& \hspace{1.35cm}(H_f+1+\omega(k_2)+\omega(k_3))^{-\frac{\beta}{2}}
a_{\lambda_3}(k_3) a_{\lambda_2}(k_2) (H_f+1)^ {-1}\Psi, \nonumber\\
&&\hspace{0.5cm} (\hel-b)^{\frac{\gamma}{2}} w^{(1,1)}(k_3,\lambda_3,k_4,\lambda_4) 
(\hel-b)^{-\frac{\gamma}{2}}
(H_f+1+\omega(k_1)+\omega(k_3))^{\frac{\beta}{2}} \nonumber\\
&& \hspace{1.35cm}
(H_{f,(\tilde{r},r)}+1+\omega_{(\tilde{r},r)}(k_1)+
\omega_{(\tilde{r},r)}(k_3))^{\frac{\alpha}{2}}\nonumber\\
&& \hspace{1.35cm}
(H_{f,(\tilde{s},s)}+1+\omega_{(\tilde{s},s)}(k_1)+
\omega_{(\tilde{s},s)}(k_4))^{-\frac{\alpha}{2}} \nonumber\\
&& \hspace{1.35cm}(H_f+1+\omega(k_1)+\omega(k_4))^{-\frac{\beta}{2}}
a_{\lambda_1}(k_1) a_{\lambda_4}(k_4) (H_f+1)^ {-1}\Psi\Big\rkl \nonumber\\
&&+\sum_{\lambda_1,\lambda_2,\lambda_4 \in \Z_2} 
\int\limits_{\R^3} dk_1 \int\limits_{\R^3} dk_2 \int\limits_{\R^3} dk_4 
\nonumber\\
&&\hspace{1cm} \Big\lkl 
(\hel-b)^{\frac{\gamma}{2}} w^{(1,1)}(k_1,\lambda_1,k_2,\lambda_2) 
(\hel-b)^{-\frac{\gamma}{2}}
(H_f+1+\omega(k_1))^{\frac{\beta}{2}} \nonumber\\[-0.2cm]
&& \hspace{1.5cm} (H_{f,(\tilde{r},r)}+1+\omega_{(\tilde{r},r)}(k_1))^{\frac{\alpha}{2}}
(H_{f,(\tilde{s},s)}+1+\omega_{(\tilde{s},s)}(k_2))^{-\frac{\alpha}{2}} 
\nonumber\\
&& \hspace{1.5cm} (H_f+1+\omega(k_2))^{-\frac{\beta}{2}}
a_{\lambda_2}(k_2) (H_f+1)^ {-1}\Psi, \nonumber\\
&&\hspace{1cm}  
(\hel-b)^{\frac{\gamma}{2}} w^{(1,1)}(k_1,\lambda_1,k_4,\lambda_4) 
(\hel-b)^{-\frac{\gamma}{2}}
(H_f+1+\omega(k_1))^{\frac{\beta}{2}} \nonumber\\
&& \hspace{1.5cm}
(H_{f,(\tilde{r},r)}+1+\omega_{(\tilde{r},r)}(k_1))^{\frac{\alpha}{2}}
(H_{f,(\tilde{s},s)}+1+\omega_{(\tilde{s},s)}(k_4))^{-\frac{\alpha}{2}}
\nonumber\\
&& \hspace{1.5cm}
(H_f+1+\omega(k_4))^{-\frac{\beta}{2}} 
a_{\lambda_4}(k_4) (H_f+1)^ {-1}\Psi\Big\rkl \nonumber\\
&\leq& \Bigg(\sum_{\lambda_2,\lambda_3, \in \Z_2} 
\int\limits_{\R^3} dk_2  \int\limits_{\R^3} dk_3
(1+\omega(k_3))^{\frac{\beta}{2}} (1+\omega_{(\tilde{r},r)}(k_3))^{\frac{\alpha}{2}}
\|\widetilde{\fG}(k_3,\lambda_3)\| \nonumber\\[-0.5cm]
&& \hspace{3.3cm}
\Big(\Big\|a_{\lambda_3}(k_3) 
\widetilde{\fG}(k_2,\lambda_2)a_{\lambda_2}(k_2) (H_f+1)^{-1} \Psi\Big\|+ 
\nonumber\\
&& \hspace{3.5cm} \Big\|a_{\lambda_3}(k_3) 
\widetilde{\fG}(k_2,\lambda_2)^*a_{\lambda_2}(k_2) (H_f+1)^{-1} 
\Psi\Big\| \Big)\Bigg)^2 \nonumber\\
&& +\sum_{\lambda_1 \in \Z_2} \int\limits_{\R^3} dk_1 
(1+\omega(k_1))^{\beta} (1+\omega_{(\tilde{r},r)}(k_1))^{\alpha} 
\|\widetilde{\fG}(k_1,\lambda_1)\|^ 2
\nonumber\\[-0.5cm]
&& \hspace{1.7cm}\Bigg[ \sum_{\lambda_2 \in \Z_2} \int\limits_{\R^3} dk_2
(\|\widetilde{\fG}(k_2,\lambda_2)a_{\lambda_2}(k_2)(H_f+1)^{-1} \Psi\|+ 
\nonumber\\[-0.7cm]
&& \hspace{3.5cm}
\|\widetilde{\fG}^*(k_2,\lambda_2)a_{\lambda_2}(k_2)(H_f+1)^{-1} \Psi\|) \Bigg]^2
\nonumber\\
&\leq& 2(1+r)^{\alpha} \Lambda_{\beta,\gamma}^{(2)} \nonumber\\
&& \Bigg\{ \sum_{\lambda \in \Z_2} \int\limits_{\R^3} dk \|H_f^{\frac{1}{2}}
(H_f+1+\omega(k))^{-\frac{1}{2}} \widetilde{\fG}(k,\lambda) a_{\lambda}(k)
(H_f+1)^{-\frac{1}{2}} \Psi \|^2+\nonumber\\
&&+
\sum_{\lambda \in \Z_2} \int\limits_{\R^3} dk \|H_f^{\frac{1}{2}}
(H_f+1+\omega(k))^{-\frac{1}{2}} \widetilde{\fG}(k,\lambda)^* a_{\lambda}(k)
(H_f+1)^{-\frac{1}{2}} \Psi \| ^2 \Bigg\}\nonumber\\
&&+4(1+r)^{\alpha} \widetilde{\Lambda}^{(2)}_{\beta,\gamma} 
\Lambda^{(2)}_{0,\gamma}
\sum_{\lambda \in \Z_2} \int\limits_{\R^3} dk \|H_f^{\frac{1}{2}}
(H_f+1)^{-1} \Psi \|^2\nonumber\\
&\leq &4(1+r)^{\alpha} (\widetilde{\Lambda}^{(2)}_{\beta,\gamma} +
\Lambda^{(2)}_{\beta,\gamma})
\Lambda^{(2)}_{0,\gamma} \|\Psi\|^2 \label{gla33}
\end{eqnarray}
For the interaction term with two creation operators the idea is the same,
but the commutation relation is more complicated:
\begin{eqnarray*}
\lefteqn{a_{\lambda_2}(k_2) a_{\lambda_1}(k_1) a^*_{\lambda_3}(k_3)
a^*_{\lambda_4}(k_4) = }\\
&=& a_{\lambda_2}(k_2) (a^*_{\lambda_3}(k_3) a_{\lambda_1}(k_1)+
\delta_{\lambda_1\lambda_3} \delta(k_1-k_3)) a^*_{\lambda_4}(k_4) =
\nonumber\\
&=& (a^*_{\lambda_3}(k_3) a_{\lambda_2}(k_2)+
\delta_{\lambda_2\lambda_3} \delta(k_2-k_3))
(a^*_{\lambda_4}(k_4) a_{\lambda_1}(k_1)+
\delta_{\lambda_1\lambda_4} \delta(k_1-k_4))+\\
&&+(a^*_{\lambda_4}(k_4) a_{\lambda_2}(k_2)+
\delta_{\lambda_2\lambda_4} \delta(k_2-k_4))
\delta_{\lambda_1\lambda_3} \delta(k_1-k_3)=\nonumber\\
&=& a^*_{\lambda_3}(k_3) a^*_{\lambda_4}(k_4) a_{\lambda_2}(k_2)
a_{\lambda_1}(k_1) +
a^*_{\lambda_3}(k_3)a_{\lambda_1}(k_1)
\delta_{\lambda_2\lambda_4} \delta(k_2-k_4) + \\
&&+a^*_{\lambda_3}(k_3)a_{\lambda_2}(k_2)
\delta_{\lambda_1\lambda_4} \delta(k_1-k_4) +
a^*_{\lambda_4}(k_4)a_{\lambda_1}(k_1)
\delta_{\lambda_2\lambda_3} \delta(k_2-k_3) \nonumber\\
&&+a^*_{\lambda_4}(k_4)a_{\lambda_2}(k_2)
\delta_{\lambda_1\lambda_3} \delta(k_1-k_3)+
\delta_{\lambda_1\lambda_4} \delta_{\lambda_2\lambda_3} 
\delta(k_1-k_4)\delta(k_2-k_3)+\\ 
&&+\delta_{\lambda_1\lambda_3} \delta_{\lambda_2\lambda_4} 
\delta(k_1-k_3)\delta(k_2-k_4)
\end{eqnarray*}
When summing up all terms with the same number of creation and 
annihilation operators renaming some indices the $W^{(2,0)}$ term yields:
\begin{eqnarray}
\lefteqn{\|(H_f+1)^{\frac{\beta}{2}}(H_{f,(\tilde{r},r)}+1)^{\frac{\alpha}{2}}
(\hel-b)^{\frac{\gamma}{2}} W^{(2,0)}(\hel-b)^{-\frac{\gamma}{2}}
(H_{f,(\tilde{s},s)}+1)^{-\frac{\alpha}{2}} \nonumber}\\
&&\hspace{3cm}(H_f+1)^{-\frac{\beta}{2}-1} \Psi\|^2= \nonumber \\
&=& \Big\| \hspace{-0.2cm}\sum_{\lambda_1,\lambda_2 \in \Z_2} \int\limits_{\R^3} dk_1
\int\limits_{\R^3} dk_2 
(\hel-b)^{\frac{\gamma}{2}} w^{(2,0)}(k_1,\lambda_1,k_2,\lambda_2) 
(\hel-b)^{-\frac{\gamma}{2}} \nonumber\\[-0.5cm]
&&\hspace{3cm}
a^*_{\lambda_1}(k_1) a^*_{\lambda_2}(k_2)
(H_f+1+\omega(k_1)+\omega(k_2))^{\frac{\beta}{2}} \nonumber\\
&&\hspace{3cm}
(H_{f,(\tilde{r},r)}+1+\omega_{(\tilde{r},r)}(k_1)+\omega_{(\tilde{r},r)}(k_2))^{\frac{\alpha}{2}} \nonumber\\
&&\hspace{3cm} (H_{f,(\tilde{s},s)}+1)^{-\frac{\alpha}{2}} 
(H_f+1)^{-\frac{\beta}{2}-1}\Psi \Big\|^2 = \nonumber\\
&=&\sum_{\lambda_1,\lambda_2,\lambda_3,\lambda_4 \in \Z_2} 
\int\limits_{\R^3} dk_1 \int\limits_{\R^3} dk_2 \int\limits_{\R^3} dk_3
\int\limits_{\R^3} dk_4 \nonumber\\
&& \Big\lkl \widetilde{\fG}(k_1,\lambda_1)^* \widetilde{\fG}(k_2,\lambda_2)^* 
(H_f+1+\omega(k_1)+\omega(k_2)+\omega(k_3)+\omega(k_4))^{\frac{\beta}{2}} 
\nonumber\\[-0.2cm]
&& \hspace{0.5cm}(H_{f,(\tilde{r},r)}+1+\omega_{(\tilde{r},r)}(k_1)+\omega_{(\tilde{r},r)}(k_2)+\omega_{(\tilde{r},r)}(k_3)+
\omega_{(\tilde{r},r)}(k_4))^{\frac{\alpha}{2}}
\nonumber\\
&& \hspace{0.5cm}(H_{f,(\tilde{s},s)}+1+\omega_{(\tilde{s},s)}(k_3)+\omega_{(\tilde{s},s)}(k_4))^{-\frac{\alpha}{2}} 
(H_f+1+\omega(k_3)+\omega(k_4))^{-\frac{\beta}{2}} \nonumber\\
&& \hspace{0.5cm} a_{\lambda_3}(k_3)
a_{\lambda_4}(k_4) (H_f+1)^{-1} \Psi,\nonumber\\
&& \hspace{0.2cm} \widetilde{\fG}(k_3,\lambda_3)^* 
\widetilde{\fG}(k_4,\lambda_4)^*
(H_f+1+\omega(k_1)+\omega(k_2)+\omega(k_3)+\omega(k_4))^{\frac{\beta}{2}} 
\nonumber\\
&& \hspace{0.5cm}(H_{f,(\tilde{r},r)}+1+\omega_{(\tilde{r},r)}(k_1)+\omega_{(\tilde{r},r)}(k_2)+\omega_{(\tilde{r},r)}(k_3)+
\omega_{(\tilde{r},r)}(k_4))^{\frac{\alpha}{2}}
\nonumber\\
&& \hspace{0.5cm}(H_{f,(\tilde{s},s)}+1+\omega_{(\tilde{s},s)}(k_1)+\omega_{(\tilde{s},s)}(k_2))^{-\frac{\alpha}{2}} 
(H_f+1+\omega(k_1)+\omega(k_2))^{-\frac{\beta}{2}} \nonumber\\
&& \hspace{0.5cm} a_{\lambda_1}(k_1)
a_{\lambda_2}(k_2) (H_f+1)^{-1} \Psi \Big\rkl\nonumber\\
&&+4\sum_{\lambda_1,\lambda_2,\lambda_3 \in \Z_2} 
\int\limits_{\R^3} dk_1 \int\limits_{\R^3} dk_2 \int\limits_{\R^3} dk_3 
\nonumber\\
&& \Big\lkl \widetilde{\fG}(k_1,\lambda_1)^* \widetilde{\fG}(k_2,\lambda_2)^* 
(H_f+1+\omega(k_1)+\omega(k_2)+\omega(k_3))^{\frac{\beta}{2}} 
\nonumber\\[-0.2cm]
&& \hspace{0.5cm}
(H_{f,(\tilde{r},r)}+1+\omega_{(\tilde{r},r)}(k_1)+\omega_{(\tilde{r},r)}(k_2)+\omega_{(\tilde{r},r)}(k_3))^{\frac{\alpha}{2}} 
\nonumber\\
&& \hspace{0.5cm}(H_{f,(\tilde{s},s)}+1+\omega_{(\tilde{s},s)}(k_3))^{-\frac{\alpha}{2}} 
(H_f+1+\omega(k_3))^{-\frac{\beta}{2}} 
a_{\lambda_3}(k_3) (H_f+1)^{-1} \Psi,\nonumber\\
&& \hspace{0.2cm} \widetilde{\fG}(k_3,\lambda_3)^* 
\widetilde{\fG}(k_2,\lambda_2)^* 
(H_f+1+\omega(k_1)+\omega(k_2)+\omega(k_3))^{\frac{\beta}{2}} \nonumber\\
&& \hspace{0.5cm}
(H_{f,(\tilde{r},r)}+1+\omega_{(\tilde{r},r)}(k_1)+\omega_{(\tilde{r},r)}(k_2)+\omega_{(\tilde{r},r)}(k_3))^{\frac{\alpha}{2}} 
\nonumber\\
&& \hspace{0.5cm}(H_{f,(\tilde{s},s)}+1+\omega_{(\tilde{s},s)}(k_1))^{-\frac{\alpha}{2}} 
(H_f+1+\omega(k_1))^{-\frac{\beta}{2}} 
a_{\lambda_1}(k_1) (H_f+1)^{-1} \Psi \Big\rkl\nonumber\\
&&+2\sum_{\lambda_1,\lambda_2 \in \Z_2} 
\int\limits_{\R^3} dk_1 \int\limits_{\R^3} dk_2 \nonumber\\
&& \Big\| \widetilde{\fG}(k_1,\lambda_1)^* \widetilde{\fG}(k_2,\lambda_2)^*
(H_f+1+\omega(k_1)+\omega(k_2))^{\frac{\beta}{2}} \nonumber\\[-0.2cm]
&&\hspace{1.5cm}
(H_{f,(\tilde{r},r)}+1+\omega_{(\tilde{r},r)}(k_1)+
\omega_{(\tilde{r},r)}(k_2))^{\frac{\alpha}{2}} \nonumber\\
&&\hspace{1.5cm}
(H_{f,(\tilde{s},s)}+1)^{-\frac{\alpha}{2}} (H_f+1)^{-\frac{\beta}{2}-1} \Psi \Big\|^2 
\nonumber\\
&\leq& \Bigg( \sum_{\lambda_1,\lambda_2 \in \Z_2} 
\int\limits_{\R^3} dk_1 \int\limits_{\R^3} dk_2 (1+2r)^{\frac{\alpha}{2}}
(1+\omega(k_1)+\omega(k_2))^{\frac{\beta}{2}} 
\|\widetilde{\fG}(k_1,\lambda_1)\|
\nonumber\\[-0.6cm]
&&\hspace{3.5cm}
\|\widetilde{\fG}(k_2,\lambda_2)\| \|a_{\lambda_1}(k_1) a_{\lambda_2}(k_2) 
(H_f+1)^{-1} \Psi \| \Bigg)^2 \nonumber\\
&&+4(1+2r)^{\alpha} \sum_{\lambda_1,\lambda_2,\lambda_3 \in \Z_2} 
\int\limits_{\R^3} dk_1 \int\limits_{\R^3} dk_2 \int\limits_{\R^3} dk_3 
(1+\omega(k_1)+\omega(k_2))^{\frac{\beta}{2}} 
\nonumber\\
&&\hspace{0.7cm} \|\widetilde{\fG}(k_1,\lambda_1)\|
\|\widetilde{\fG}(k_2,\lambda_2)\|^2 \|\widetilde{\fG}(k_3,\lambda_3)\|
(1+\omega(k_3)+\omega(k_2))^{\frac{\beta}{2}} \nonumber\\
&& \hspace{0.7cm}\|a_{\lambda_1}(k_1) (H_f+1)^{-1} \Psi \| 
\|a_{\lambda_3}(k_3) (H_f+1)^{-1} \Psi \|\nonumber\\
&&+2 \sum_{\lambda_1,\lambda_2 \in \Z_2} 
\int\limits_{\R^3} dk_1 \int\limits_{\R^3} dk_2 (1+2r)^{\alpha}
(1+\omega(k_1)+\omega(k_2))^{\beta} \|\widetilde{\fG}(k_1,\lambda_1)\|^2
\nonumber\\[-0.6cm]
&&\hspace{3.7cm}
\|\widetilde{\fG}(k_2,\lambda_2)\|^2 \|(H_f+1)^{-1} \Psi \|^2  \nonumber
\end{eqnarray}
For any $s>0$ and $a,b \geq 0$ the estimate 
$(a+b)^s \leq \max\{1,2^{s-1}\} (a^s+b^s)$ holds true, so applying this 
inequality to the $(1+\omega(k_1)+\omega(k_2))^{\frac{\beta}{2}}$ terms above,
we use H\"older inequality and get:
\begin{eqnarray}
\lefteqn{\|(H_f+1)^{\frac{\beta}{2}}(H_{f,(\tilde{r},r)}+1)^{\frac{\alpha}{2}}
(\hel-b)^{\frac{\gamma}{2}} W^{(2,0)}(\hel-b)^{-\frac{\gamma}{2}}
(H_{f,(\tilde{s},s)}+1)^{-\frac{\alpha}{2}} \nonumber}\\
&& \hspace{3cm}(H_f+1)^{-\frac{\beta}{2}-1} \Psi\|^2 \label{gla34}\\
&\leq& (1+2r)^{\alpha} \max\{1,2^{\beta-2}\} 
\Bigg[ \sum_{\lambda_1,\lambda_2 \in \Z_2} \int\limits_{\R^3} dk_1 
\int\limits_{\R^3} dk_2 
\|\widetilde{\fG}(k_1,\lambda_1)\|
\|\widetilde{\fG}(k_2,\lambda_2)\| \nonumber\\
&&\hspace{0.5cm}
\Big((1+\omega(k_1))^{\frac{\beta}{2}} + (1+\omega(k_2))^{\frac{\beta}{2}}\Big)
\|a_{\lambda_1}(k_1) a_{\lambda_2}(k_2) (H_f+1)^{-1} \Psi\| \Bigg]^2
\nonumber\\
&&+4 (1+2r)^{\alpha} \max\{1,2^{\beta-2}\}  
\sum_{\lambda_1,\lambda_2,\lambda_3 \in \Z_2} \int\limits_{\R^3} dk_1
\int\limits_{\R^3} dk_2 \int\limits_{\R^3} dk_3 \nonumber\\
&&\hspace{0.5cm}
\Big((1+\omega(k_1))^{\frac{\beta}{2}} + (1+\omega(k_2))^{\frac{\beta}{2}} \Big) 
\Big((1+\omega(k_2))^{\frac{\beta}{2}} + (1+\omega(k_3))^{\frac{\beta}{2}} \Big) 
\nonumber\\
&&\hspace{0.5cm} 
\|\widetilde{\fG}(k_1,\lambda_1)\| \|\widetilde{\fG}(k_2,\lambda_2)\|^2
\|\widetilde{\fG}(k_3,\lambda_3)\|
\|a_{\lambda_1}(k_1) (H_f+1)^{-1} \Psi \| \nonumber\\
&&\hspace{0.5cm} \|a_{\lambda_1}(k_3) (H_f+1)^{-1} \Psi \|
\nonumber\\
&&+ 2 (1+2r)^{\alpha} \max\{1,2^{\beta-1}\}
\sum_{\lambda_1,\lambda_2 \in \Z_2} \int\limits_{\R^3} dk_1 
\int\limits_{\R^3} dk_2 
\|\widetilde{\fG}(k_1,\lambda_1)\|^2 \nonumber\\
&&\hspace{0.5cm}\|\widetilde{\fG}(k_2,\lambda_2)\|^2
\Big((1+\omega(k_1))^{\frac{\beta}{2}} + (1+\omega(k_2))^{\frac{\beta}{2}} \Big) 
\|(H_f+1)^{-1} \Psi\|^2
\nonumber\\
&\leq& 
\Big[\Lambda^{(2)}_{\beta,\gamma} \Lambda^{(2)}_{0,\gamma} +
2\sqrt{\Lambda^{(2)}_{\beta,\gamma} \Lambda^{(2)}_{0,\gamma}} 
\widetilde{\Lambda}^{(2)}_{\frac{\beta}{2},\gamma}+
\Lambda^{(2)}_{\beta,\gamma} \widetilde{\Lambda}^{(2)}_{0,\gamma}+
\Lambda^{(2)}_{0,\gamma} \widetilde{\Lambda}^{(2)}_{\beta,\gamma} +
\widetilde{\Lambda}^{(2)}_{0,\gamma}
\widetilde{\Lambda}^{(2)}_{\frac{\beta}{2},\gamma} \Big] \nonumber\\
&& 4(1+2r)^{\alpha} \max\{1,2^{\beta-2}\} 
\|\Psi\|^2. \nonumber\\[-1.7cm] \nonumber
\end{eqnarray} 
\end{proof}
\vspace*{0.4cm}
\end{appendix}
\noindent{\bf Acknowledgements:} 
The author thanks Laszlo Erd\"os and Marcel Griesemer 
for various helpful discussions and comments. The introduction of $g^{-\gamma}$
in the proof of Theorem \ref{co2.1}, which simplified my former proof,
commes from a discussion with Marcel Griesemer.

\end{document}